\documentclass[reprint,superscriptaddress, amsmath,amssymb, aps,
showkeys ]{revtex4-1}

\usepackage{graphicx}
\usepackage{dcolumn}
\usepackage{bm}
\usepackage{hyperref}
\usepackage{physics}
\usepackage{mathtools}
\usepackage{float}
\usepackage{subcaption}

\usepackage{graphicx,amssymb,amsmath,amsthm,amsfonts}
\usepackage[usenames]{color}
\usepackage{mathtools}

\definecolor{blue-violet}{rgb}{0.54, 0.17, 0.89}

\begin{document}

\preprint{APS/123-QED}

\title{Electrically charged spherical matter shells in higher
dimensions: Entropy, thermodynamic stability, and the black hole
limit}


\author{Tiago V. Fernandes}
\email{tiago.vasques.fernandes@tecnico.ulisboa.pt}
\affiliation{Centro de Astrof\'isica e Gravita\c c\~ao -
CENTRA, Departamento de F\'isica,
Instituto Superior T\'{e}cnico - IST, Universidade de Lisboa
- UL,\\
Avenida Rovisco Pais 1, 1049-001 Lisboa, Portugal}
\author{Jos\'e P. S. Lemos}
\email{joselemos@ist.utl.pt}
\affiliation{Centro de Astrof\'isica e Gravita\c c\~ao -
CENTRA, Departamento de F\'isica,
Instituto Superior T\'{e}cnico - IST, Universidade de Lisboa
- UL,\\
Avenida Rovisco Pais 1, 1049-001 Lisboa, Portugal}

\begin{abstract}

We study the thermodynamic properties of a static electrically charged
spherical thin shell in $d$ dimensions by imposing the first law of
thermodynamics on the shell.  The shell is at radius $R$, inside it
the spacetime is Minkowski, and outside it the spacetime is
Reissner-Nordstr\"om.  We obtain that the shell thermodynamics is
fully described by giving two additional reduced equations of state,
one for the temperature and another for the thermodynamic
electrostatic potential.  We choose the equation of state for the
temperature as essentially a power law in the gravitational radius
$r_+$ with exponent $a$, such that the $a=1$ case gives the
temperature of a shell with black hole thermodynamic properties, and
for the electrostatic potential we choose an equation of state
characteristic of a Reissner-Nordstr\"om spacetime.  The entropy of
the shell is then found to be proportional to $A_+^a$, where $A_+$ is
the gravitational area corresponding to $r_+$, with the exponent $a$
obeying $a>0$ to have a physically reasonable entropy.  We are then
able to perform the black hole limit $R=r_+$, find the Smarr formula
for $d$-dimensional electrically charged black holes, and generically
recover the thermodynamics of a $d$-dimensional Reissner-Nordstr\"om
black hole.  We further study the intrinsic thermodynamic stability of
the shell with the chosen equations of state. We obtain that for
$0<a\leq \frac{d-3}{d-2}$ all the configurations of the shell are
thermodynamically stable, for $\frac{d-3}{d-2}<a<1$ stability depends
on the mass and electric charge, for $a=1$ the configurations are
unstable, unless the shell is at its own gravitational radius, i.e.,
at the black hole limit, in which case it is marginally stable, and
that for $1<a<\infty$ all configurations are unstable.  Rewriting the
stability conditions with variables that can be measured in the
laboratory, it is found that the sufficient condition for the
stability of these shells is when the isothermal electric
susceptibility $\chi_{p,T}$ is positive, marginal stability happening
when this quantity is infinite, and instability, and thus depart from
equilibrium, arising for configurations with a negative electric
susceptibility.

\end{abstract}

\keywords{self-gravitating thin shells, entropy,
thermodynamics, black holes}
\maketitle

\section{\label{sec:Intro}Introduction}

Self-gravitating
thin shell physics in general relativity has proved of great value in
the understanding of the many aspects involving the interaction
between gravitational and matter fields.
We mention a few of these aspects. 
First, 
the main features of gravitational collapse and black hole
formation
has been described in detail
from the dynamics of thin
shells in Schwarzschild and Reissner-Nordstr\"om spacetimes, as well
as in some of their generalizations to higher dimensions
\cite{israelshellcollapse,kuchar,kijowski,diasgaolemos,gaolemos,
eiroasimeone1}.
Second, 
the issue of the compactness of stars has
been studied
through the help of neutral and electrically charged thin shells
\cite{andreasson}. Third, 
 the understanding of stars with outer
normals to their
surface pointing to decreasing radius can be clearly formulated
through the maximum analytical extensions of shell
solutions and their
appearance in the other side of the Carter-Penrose diagram
\cite{lemosluz1}.
Fourth, 
wormhole spacetimes 
can be constructed through the support of thin shells
\cite{diaslemos,bejaranoeiroa},
the complementary of wormholes and bubble universes
is clearly displayed with the use of thin shells
\cite{lemosluz2}. Fifth, 
regular black holes can be found employing thin
shells \cite{lemoszanchin,uchikata,brandenberger}.

Self-gravitating thin shells allow a precise mathematical treatment
not only in
a dynamic context,
but also in a thermodynamic framework
embedded within general relativity,
and as such they are of great interest in the understanding of the
thermodynamics of matter in a gravitational field, as well as in the
understanding of the thermodynamics of the gravitational field itself.
Remarkably, general relativity by itself fixes the equation of state
for pressure of the matter in a static spherical shell.  On the other
hand, general relativity does not determine an equation of state for
the temperature.  To find one, the first law of thermodynamics, valid
for the matter on the shell, has to be
postulated with all thermodynamic quantities that
enter into it having a precise and correct meaning.
Then, the integrability conditions
applied to the first law of thermodynamics
restrict the form of
the temperature equation of state, leaving nonetheless some freedom
for its choice, that can be performed through
some deduced reasonable guess
or via some more fundamental description of matter, and which, in turn,
narrows the possible types for the entropy function.  For a shell with
an exterior given by a Schwarzschild spacetime this was performed
in~\cite{Martinez:1996}, also treated in \cite{bergliaffa1}, and
generalized to $d$-dimensions in \cite{Andre:2019}.  The
inclusion of electric charge means that the exterior spacetime is
Reissner-Nordstr\"om, and although one needs to take care of a further
equation of state for the thermodynamic electric potential, it also
allows an exact treatment as done in \cite{Lemos:2015a}, also studied
in \cite{bergliaffa2}, and with the extremal state being analyzed in
\cite{Lemos:2015c,lemoshellextremal12}.
One
important fact about thermodynamics of shells is that
the shell can be put to its own gravitational radius. One can then
argue that at this stage its properties should be black hole like, and
indeed they are.

The interest in thermodynamics of spacetimes, in particular in
thermodynamics of shell spacetimes, comes from black holes.  Black
holes radiate at the Hawking temperature, have the Bekenstein-Hawking
entropy, and thus, in appropriate settings,  can be described as
thermodynamic objects.  In a path integral statistical mechanics
formulation for spherical black holes, the suitable ensemble to use is
the canonical ensemble, where one fixes a temperature at a cavity of a
given radius, and from which the full thermodynamic properties of the
system can then be deduced.  For a Schwarzschild vacuum spacetime one
finds a small black hole solution which is thermodynamically unstable and
a large black hole solution, of radius near the cavity radius, which
is stable, see \cite{York:1986,can} and its $d$-dimensional
generalization \cite{andrelemos1,andrelemos2}.  In this
path integral formulation,
one can include a matter shell surrounding the black hole \cite{ym},
put electric charge into the black hole spacetime \cite{Braden:1990},
and work with AdS spacetimes \cite{pecalemos}.

Alternatively, one can find black hole thermodynamics through matter
thermodynamics via the quasiblack hole approach
\cite{Lemos:2007,Lemos:2010,Lemos:2011,lemoszaslavskii}.  In general,
to have a thermodynamic formulation, knowledge of the matter equations
of state is required.  However, using the quasiblack hole approach one
is able to skip the setting of specific equations of state.  In this
approach, one keeps the gravitational radius fixed, and changes the
proper mass and the radius of the configuration, maintaining it near
the black hole threshold.  One can then integrate the first law of
thermodynamics over this set of configurations, finding that the
result is indeed model independent, and retrieving fully the black
hole properties.

Thus, thin shells, black holes, and quasiblack holes are of importance
in the understanding of thermodynamics of spacetimes.  It is certainly
of significance to proceed with these themes.  In particular, it is of
interest to study further thermodynamic shell properties.  Here we
analyze the entropy and the thermodynamic stability of a static
electrically charged spherical thin shell in $d$ dimensions, as well
as its black hole limit, extending the analysis and the results given
in \cite{Andre:2019,Lemos:2015a}.  The study of systems in $d$
dimensions is of relevance in obtaining knowledge of features that are
permanent, independent of dimension, and is of use as several theories
of gravitation imply the existence of extra dimensions.  To make
progress in our analysis,
we impose an equation of state such that the entropy of the
shell is a power in the gravitational area, i.e.,
the area corresponding to
the gravitational radius, and find that stability requires positive
heat capacity,  positive
isothermal compressibility, and  positive
isothermal electric
susceptibility.  Putting the shell at its own gravitational radius we
find the corresponding black hole thermodynamic
properties including the Smarr formula and
its thermodynamic stability. The
essential thermodynamic formalism used
is in the book by Callen \cite{callen}.

This paper is organized as follows. In Sec.~\ref{sec:thinformalism},
we apply the thin shell formalism to obtain the internal energy and
the pressure of the shell in terms of the spacetime variables. In
Sec.~\ref{sec:firstlaw}, we apply the first law of thermodynamics to
the shell and compute its entropy after providing two equations of
state, and afterward we analyze the black hole limit. In
Sec.~\ref{sec:stability}, we study the intrinsic stability of the
shell with such equations of state and entropy, analyzing when
applicable the black hole limit.  In Sec.~\ref{sec:Intrinsic}, we
establish the connection of intrinsic stability to physical quantities
such as the heat capacity, the isothermal compressibility and the
isothermal electric susceptibility.  In Sec.~\ref{conc} we
conclude. There are several appendices that complete the paper,
including one where all the necessary plots to understand the
thermodynamic stability of the shell are displayed.

\section{Electrically
charged spherical  shell spacetime \label{sec:thinformalism}}

\subsection{Thin shell spacetime formalism}

General relativity coupled to Maxwell electromagnetism
has the equations 
\begin{align}
G_{ab}=8\pi G T_{ab}\,,
\label{eq:einstein}
\end{align}
\begin{align}
\nabla_b\hskip0.04cm{F_{a}}^b= J_{a}\,,
\label{eq:max}
\end{align}
where, $G_{ab}$ is the Einstein tensor given in terms of the metric
$g_{ab}$ and its first two derivatives, $G$ is the gravitational
constant in $d$ dimensions, the speed of light $c$ is set to one, $c=1$,
$T_{ab}$ is the stress-energy tensor, $\nabla_b$ is the
covariant derivative of the metric $g_{ab}$, $F_{ab}$ is the Maxwell
tensor, $J_{a}$ is the electric current, and indices $a,b$, are
$d$-dimensional indices, running from 0 to $d-1$.  The Maxwell tensor
$F_{ab}$ also obeys the internal equations $\nabla_{[c}F_{ab]}=0$,
where brackets in indices means total antisymetrization, and allow us 
to
write $F_{ab}$ in terms of an electromagnetic potential vector $A_a$
as ${F}_{ab}=\frac{\partial A_b}{\partial x^a}- \frac{\partial
A_a}{\partial x^b}$.
For an electrovacuum spacetime the stress-energy tensor 
$T_{ab}$ is given by 
\begin{align}
T_{ab}=\varepsilon
\left(
{F_a}^cF_{bc}-\frac14 g_{ab}F^{cd}F_{cd}
\right)
\,,
\label{eq:stressenergytensormaxwell}
\end{align}
where  the parameter
$\varepsilon$ is defined as
$\varepsilon=\epsilon\frac{(d-3)}{\Omega}$,
with $\epsilon$ being the electromagnetic
coupling constant, and 
$\Omega =
\frac{
\pi^{\frac{d-2}{2}}}
{\Gamma(\frac{d}{2})}$
is the area of a $d-2$ unit sphere,
which 
in four dimensions yields the usual $4\pi$.
In a thin shell spacetime, one has an interior
region, $\mathcal{V}_i$
say, that obeys Eqs.~\eqref{eq:einstein} and \eqref{eq:max},
an exterior region, $\mathcal{V}_e$,
that also obeys Eqs.~\eqref{eq:einstein} and
\eqref{eq:max}, and a boundary surface, i.e., a thin shell,
in-between these two regions that has properties found by appropriate
junction conditions to match the two different spacetime regions.

The interior $\mathcal{V}_i$ has coordinates $x_i^a$ assigned to it
and a metric ${{g}_i}_{ab}$.  We denote ${{n}_i}_{a}$ as the covector
orthogonal to a hypersurface.
An important quantity is given by the way
in which ${{n}_i}_{a}$ changes along the hypersurface, i.e.,
${{\nabla_i}}_{a}{{n_i}}_{b}$, where ${{\nabla_i}}_{a}$ is the
covariant derivative
in the interior region.  If the coordinates of the hypersurface are
denoted by $y^\alpha$, where Greek
indices $\alpha,\beta$, are $d-1$-dimensional indices
and run from 0 to
$d-2$,  then the tangent vectors at the hypersurface are
$(e_i)^a_{\alpha} = \frac{\partial x_i^a}{\partial y^\alpha}$.  The
interior $\mathcal{V}_i$ is assumed to have an electromagnetic vector
potential ${{A}_i}_{a}$ and a
corresponding field strength
${{F}_i}_{ab}=\frac{\partial {A_i}_{{b}}}{\partial {x_i}^{{a}}}-
\frac{\partial {A_i}_{{a}}}{\partial {x_i}^{{b}}} $.

The exterior $\mathcal{V}_e$ has coordinates $x_e^a$ assigned to it
and a metric ${{g}_e}_{ab}$.  We denote ${{n}_e}_{a}$ as the covector
orthogonal to a hypersurface. An important quantity is related to
how ${{n}_e}_{a}$ changes along the hypersurface,
${{\nabla_e}}_{a}{{n_e}}_{b}$, where ${{\nabla_e}}_{a}$ is the
covariant derivative
in the exterior region.  If the coordinates of the hypersurface are
denoted by $y^\alpha$ then the tangent vectors at the hypersurface are
$(e_e)^a_{\alpha} = \frac{\partial x_e^a}{\partial y^\alpha}$.  The
exterior $\mathcal{V}_e$ is assumed to have an electromagnetic vector
potential ${{A}_e}_{a}$ and a
corresponding field strength
${{F}_e}_{ab}=\frac{\partial {A_e}_{b}}{\partial {x_e}^{a}}-
\frac{\partial {A_e}_{a}}{\partial {x_e}^{b}} $.

The boundary hypersurface,
with coordinates $y^\alpha$ assigned to it, 
and which can be
a thin shell, is assumed to be 
timelike
and common to $\mathcal{V}_i$ and $\mathcal{V}_e$.
The pull-back of a
covariant tensorial quantity in each region allows
the definition of the
covariant tensorial quantity at this boundary hypersurface.
Then, the junction conditions give
that tensorial quantity 
uniquely at the
common boundary hypersurface.
For the interior, the metric ${g}_i$ has the pull-back $(\phi^*
{g_i})_{\alpha\beta} = {{g_i}}_{ab}
(e_i)^{a}_{\alpha}(e_i)^{b}_{\beta} \equiv {{h_i}}_{\alpha\beta}$,
the quantity
$\nabla_i n_i$ has a pull-back yielding the
hypersurface extrinsic curvature
${K_i}_{\alpha\beta}$, namely, 
$(\phi^* \nabla_i n_i)_{\alpha \beta} = {{\nabla_i}}_{a}{{n_i}}_{b}
e^{a}_{\alpha} e^{b}_{\beta} ={{\nabla_i}}_{\alpha}{{n_i}}_{\beta}
\equiv {K_i}_{\alpha\beta} $, ${A_i}$ has the pull-back $(\phi^*
{A_i})_{\alpha} = {{A_i}}_{a} (e_i)^{a}_{\alpha}
\equiv {{A_i}}_{\alpha}$,
${\bm F}_i$ has the pull-back $(\phi^* {\bm F_i})_{\alpha\beta} =
{{F_i}}_{ab} (e_i)^{a}_{\alpha}(e_i)^{b}_{\beta} \equiv
{{F_i}}_{\alpha\beta}$, and ${F}_i$ defined such that ${{F}_i}_a\equiv
{{F_i}}_{ab}{n_i}^b $ has the pull-back $(\phi^* {F_i})_{\alpha} =
{{F_i}}_{ab}{n_i}^b (e_i)^{a}_{\alpha} \equiv {{F_i}}_{\alpha}$.
For the exterior, the metric ${g}_e$ has the pull-back
$(\phi^* {g_e})_{\alpha\beta} = {{g_e}}_{ab}
(e_e)^{a}_{\alpha}(e_e)^{b}_{\beta} \equiv {{h_e}}_{\alpha\beta}$,
the quantity
$\nabla_e n_e$ has a pull-back yielding the extrinsic curvature
${K_e}_{\alpha\beta}$, namely, 
$(\phi^* \nabla_e n_e)_{\alpha \beta} = {{\nabla_e}}_{a}{{n_e}}_{b}
e^{a}_{\alpha} e^{b}_{\beta} ={{\nabla_e}}_{\alpha}{{n_e}}_{\beta}
\equiv {K_e}_{\alpha\beta}$,
${A_e}$ has the pull-back $(\phi^* {A_e})_{\alpha} = {{A_e}}_{a}
(e_e)^{a}_{\alpha} \equiv
{{A_e}}_{\alpha}$, ${\bm F}_e$ has the pull-back
$(\phi^* {\bm F_e})_{\alpha\beta} = {{F_e}}_{ab}
(e_e)^{a}_{\alpha}(e_e)^{b}_{\beta} \equiv {{F_e}}_{\alpha\beta}$, and
${F}_e$ defined such that ${{F}_e}_a\equiv {{F_e}}_{ab}{n_e}^b $ has
the pull-back $(\phi^* {F_e})_{\alpha} = {{F_e}}_{ab}{n_e}^b
(e_e)^{a}_{\alpha} \equiv {{F_e}}_{\alpha}$.
The first junction condition is the continuity of the
metric 
\begin{align}
    [h_{\alpha \beta}] = 0\,, 
\label{eq:Curvaturecondition1}
\end{align}
where $[h_{\alpha \beta}]$ means $[h_{\alpha
\beta}]={h_e}_{\alpha\beta}-
{h_i}_{\alpha\beta}$, and the
same for
any other quantity. Equation~\eqref{eq:Curvaturecondition1}
means that one can define a metric $h_{\alpha \beta}$
at the boundary surface with coordinates $y^\alpha$
and of course such that
it obeys $h_{\alpha \beta}={h_i}_{\alpha\beta}=
{h_e}_{\alpha\beta}$.
The second junction condition is 
\begin{align}
-\Big([K_{\alpha\beta}] -
    [K]h_{\alpha\beta} \Big)=
    {8\pi G}S_{\alpha\beta}  \,,
    \label{eq:Curvaturecondition}
\end{align}
where 
 $K$ is the trace of the extrinsic curvature $K_{\alpha \beta}$,
 and $S_{\alpha\beta}$ is the stress-energy tensor for
matter in
the shell. We consider
that the thin shell is made of a perfect fluid having the
stress-energy tensor
\begin{align}
    S_{\alpha \beta} = (\sigma + p)u_\alpha u_\beta +
    p h_{\alpha \beta}\,,\label{eq:Sab}
\end{align}
where $\sigma$ is the energy density, $p$ is the pressure and
$u^\alpha$ is the velocity of the fluid on the boundary.

There are also junction
conditions for the pull backs
of the covector potential 
and of the electromagnetic field strength. They are
given by
\begin{align}
    [A_\alpha] = 0\,,
    \label{eq:Vectorcondition1}
\end{align}
\begin{align}
   [F_{\alpha \beta}] = 0\,\,,\,\,\,
    [F_{\alpha}] = j_\alpha\,,
    \label{eq:Vectorcondition2}
\end{align}
where $F_{\alpha}=F_{a b  }n^be^a_{\alpha}$
is defined on each side of the boundary
hypersurface, i.e, from the interior
and exterior region sides, 
and the electric current $j_\alpha$
is given by 
\begin{align}
j_\alpha=
{\zeta}
\sigma_e u_\alpha,
    \label{eq:ja}
\end{align}
with
$\zeta$ being defined as 
$\zeta=\frac{\Omega }{\epsilon_q}$,
$\epsilon_q$ being the electric permittivity,
and with $\sigma_e$ being the electric charge density.

Now we apply this formalism to a particular
$d$-dimentional
spacetime, namely, a Minkowski interior, a
thin shell, and a
$d$-dimensional
Reissner-Nordstr\"om, also called
Reissner-Nordstr\"om-Tangherlini,
exterior.
We use the
convention
that
the coupling constant
$\epsilon$ appearing implicitly in
Eq.~\eqref{eq:stressenergytensormaxwell}
and the electric permittivity
 $\epsilon_q$ appearing implicitly in Eq.~\eqref{eq:ja}
are set to one,
$\epsilon=1$ and $ \epsilon_q = 1$,
and the $d$-dimensional constant of gravitation
$G$ is not set to one, it is left generic.

\subsection{The spacetime solution}

\subsubsection{The interior}

The interior
region
$\mathcal{V}_i$ is a vacuum
$d$-dimensional
spherically symmetric Minkowski
region with
spherical coordinates $x_i^a$ assigned to it
such that $x_i^a = (t_i,r_i,\theta_i^A)$
with~$A \in \{1,...,d-2\}$,
and 
with line element
\begin{align}
    ds_i^2 = - dt_i^2 + dr^2 + r^2 d\Omega^2\,,
    \quad 0\leq r_i\leq R_i\,,
    \label{in1}
\end{align}
where
we have put $r\equiv r_i$,
$d\Omega^2$ is the line element
of a~$(d-2)$ unit sphere, and $R_i$
is the radius of the shell as measured from the interior.

The Maxwell potential covector $A_{a}$ is given in the interior region
as
\begin{align}
(A_i)_{t_i} = A_i\,,
\label{covi}
\end{align}
where $A_i$ is a constant and the other components are
zero.

\subsubsection{The exterior}

The exterior 
region
$\mathcal{V}_e$ is a vacuum
$d$-dimensional
spherically symmetric  Reissner-Nordstr\"om-Tangherlini
region with
spherical coordinates $x_e^a$ assigned to it
such that $x_e^a = (t_e,r_e,\theta_e^A)$,
with~$A \in \{1,...,d-2\}$, and
with
line element
\begin{align}
    ds_e^2 = - f(r)\, dt_e^2 + f(r)^{-1} dr^2 +
    r^2 d\Omega^2\,,
    \quad R_e\leq r_e\leq \infty\,,
    \label{ex1}
    \end{align}
where
we have put $r\equiv r_e$, a redefinition
that can be done,
$R_e$
is the radius of the shell as measured from the exterior,
and
\begin{align}
f(r) = 1 - \frac{2\mu m}{r^{d-3}} +
    \frac{q Q^2}{r^{2(d-3)}}\,,
\label{fofr1}
\end{align}
where $m$ is the spacetime, also called ADM,
mass, and
$Q$ is the total electric charge,
and 
with $\mu$ and $q$
being
given by 
\begin{align}
\mu = \frac{8\pi G}{(d-2)\Omega}\,,
\quad\quad q = \frac{8\pi G
}{(d-2)\Omega}\,,
\label{muq}
\end{align}
with $G$ being the $d$-dimensional
gravitational constant, and again
$\Omega = \pi^{\frac{d-2}{2}}\Gamma(\frac{d}{2})^{-1}$
is the area of a $(d-2)$ unit sphere.
Thus, $\mu=q$.
Note that without putting
$\epsilon$
and
$\epsilon_q$ to one, $q$ in Eq.~\eqref{muq}
is
$ q = \frac{8\pi G
{\epsilon}}{(d-2)\Omega{\epsilon_q^2}}
$.

The Reissner-Nordstr\"om metric has
its gravitational radius and Cauchy radius at its
coordinate singularities given by
\begin{align}
    r_{\pm}^{d-3} = \mu m \pm \sqrt{\mu^2 m^2 - q Q^2}
    \,,\label{eq:rpm}
\end{align}
where $r_+$ corresponds to the gravitational
radius and $r_-$ to the Cauchy
radius. Note that 
the gravitational
radius and  the Cauchy
radius
in general are not
horizon radii, 
they are only horizon radii
for a full electrovacuum solution of the
Einstein-Maxwell system of equations,
in which case they are
the event horizon radius and the Cauchy horizon radius.
From Eq.~\eqref{eq:rpm}, one sees that the extremal case,
defined as $r_+=r_-$ yields
a mass to charge relation
given by $\sqrt{\mu} m = Q$, which from
Eq.~\eqref{muq} yields in four dimensions
$\sqrt{G} m = Q$. One can put
$G=1$ to give in four dimensions
$ m = Q$, but we keep the $d$-dimensional $G$
in our calculations to not fall into awkward
units along the calculations.
The area  $A_+$ corresponding
to the gravitational radius $r_+$
is an important quantity,
defined as the gravitational area,
and 
given by
\begin{align}
    A_+ = \Omega r_+^{d-2}\,.
    \label{eq:A+}
\end{align}
One can invert
Eq.~\eqref{eq:rpm}, giving
\begin{align}
    &m = \frac{1}{2\mu}(r_+^{d-3} + r_-^{d-3})
    \,,\quad\quad
    Q = \frac{(r_+ r_-)^{\frac{d-3}{2}}}{\sqrt{q}}\,.
\label{masscharge}
\end{align}
Note that $q$ in Eq.~\eqref{masscharge} can be swapped
for $\mu$ due to Eq.~\eqref{muq}, but we
stick to $q$ whenever the coefficient
is associated to the
electric charge $Q$.
Now, with the two characteristic radii $r_\pm$
defined in Eq.~\eqref{eq:rpm}
we can rewrite Eq.~\eqref{fofr1} as
\begin{align}
f(r) = 
\left(1-\left(\frac{r_+}{r}\right)^{d-3}\right)
    \left(1-\left(\frac{r_-}{r}\right)^{d-3}\right)
\,.
\label{fofr2}
\end{align}
The Maxwell potential covector $A_{a}$ is given in the exterior region
as
\begin{align}
(A_e)_{t_e} = - \frac{Q}{ (d-3) r_e^{d-3}} \,,
    \label{cove}
\end{align}
where we have set
without loss of generality
a constant of integration
$A_e$ to zero, $A_e=0$, and the other components are
zero.
Note than that the outer electric field is $(F_e)^{t_e r} =
\frac{Q}{r_e^{d-2}}$.
If we do not set $\epsilon_q=1$
then Eq.~\eqref{cove} is
$(A_e)_{t_e} =  -\frac{Q}{{(d-3)\epsilon_q} r_e^{d-3}} +
    A_e$ and
the outer electric field is $(F_e)^{t_e r} =
\frac{Q}{\epsilon_qr_e^{d-2}}$.

\subsubsection{The thin shell}

The boundary hypersurface $\Sigma$
is spherically symmetric and has in principle
a  thin
shell in it and it is useful to
give to it an intrinsic metric $h_{\alpha\beta}$
such that its line element $ds^2=
h_{\alpha\beta}dx^\alpha dx^\beta$
can be written as
\begin{align}
    ds_\Sigma^2 = -d\tau^2 + R(\tau)^2 d\Omega^2\,,
    \label{sigma1}
    \end{align}
where the coordinate system $y^\alpha = (\tau, \theta^A)$ has been
chosen, with $A \in \{1,..., d-2\}$, the coordinate $\tau$ is the
proper time of the shell, and
$R(\tau)$ is the radius of the shell.
The Maxwell potential covector $A_{\alpha}$ is given at the thin shell
as
\begin{align}
(A_\Sigma)_{\tau} = A_\Sigma\,,
\label{bsigma}
\end{align}
respectively,
where $A_\Sigma$ is a constant and the other components are
zero.
Recall that we use the latin indices to designate
quantities in the regions $\mathcal{V}_e$ and $\mathcal{V}_i$ whereas
greek indices designate quantities at the hypersurface.
The pull-back of the metric in the region $\mathcal{V}_i$ on the
hypersurface $\Sigma$,
$(\phi^* \bm{g}_i)_{\alpha \beta}$, assumes the
line element form 
\begin{equation}
    ds_{i,\Sigma}^2 = \left(-  \Dot{t}_i^2 +
     \Dot{R}_i^2\right) d\tau^2 +
    R_i(\tau)^2 d\Omega^2\,,
    \label{eq:metricinpullb}
\end{equation}
where the boundary hypersurface $\Sigma$ has
an history defined in the interior
by $R_i(t_i)$
and $\Dot{\,\,} = \dv{\,\,}{\tau}$.
The pull-back of the metric in the region $\mathcal{V}_e$ on the
hypersurface $\Sigma$,
$(\phi^* \bm{g}_e)_{\alpha \beta}$, is given by the
line element
\begin{align}
    ds_{e,\Sigma}^2 =& \Big[- f(R_e(\tau)) \Dot{t}_e^2 +
    f(R_e(\tau))^{-1} \Dot{R}_e^2\Big] d\tau^2 \notag\\&+
    R_e(\tau)^2 d\Omega^2\,,
    \label{eq:metricout}
\end{align}
where the boundary hypersurface $\Sigma$ has
an history defined in the exterior
by $R_e(t_e)$.
Now, we apply the first junction condition,
Eq.~\eqref{eq:Curvaturecondition1}.
i.e., $ [h_{\alpha \beta}] = (\phi^* \bm{g}_e)_{\alpha \beta} -
    (\phi^* \bm{g}_i)_{\alpha \beta}=0$
    to Eqs.~\eqref{in1},~\eqref{ex1}, and~\eqref{sigma1},
    to give two equations,
\begin{equation}
R_e(\tau) = R_i(\tau) = R(\tau)\,,
    \label{eq:conditioncontinuitymetric}
\end{equation}
\begin{equation}
- \Dot{t}_i^2 + \Dot{R}^2 =-
    f(R) \Dot{t}_e^2 + f(R)^{-1} \Dot{R}^2
    =-1\,.
    \label{eq:conditioncontinuitymetric2}
\end{equation}
Clearly, Eq.~\eqref{eq:conditioncontinuitymetric}
entitles us to use the same radial
coordinate $r$ for the interior, the boundary
surface, and the exterior, as we did a priori, 
and permits
to define a unique area $A$ to the shell, 
\begin{align}
    A = \Omega R^{d-2}\,.
    \label{eq:A}
\end{align}
The extrinsic curvature of the hypersurface in both regions can be
computed
from 
$K_{\alpha \beta} = (\phi^*\nabla n)_{\alpha \beta} =
    \nabla_a n_b e^{a}_{\alpha} e^{b}_{\beta}$,
where $n$ is
the unit outward normal covector to the hypersurface.
For the interior,
$n_i$ is given by 
$n_i = 
\left(1 -
\left(\dv{R}{t_i}\right)^2\right)^{-\frac12}
\left( - \dv{R}{t_i} dt_i+dr_i \right)$.
In the hypersurface, it is useful to write these components in terms
of $\tau$, so using Eq.~\eqref{eq:conditioncontinuitymetric2},
we have
$\eval{
\left(1 -
\left(\dv{R}{t_i}\right)^2\right)^{-\frac12}
}_\Sigma = \sqrt{1 +
    \Dot{R}^2}$
and 
$\eval{\dv{R}{t_i}}_\Sigma = \frac{ \Dot{R}}{\sqrt{1
    + \Dot{R}^2}}$, so that
$n_i = \left(-\Dot{R},\sqrt{1+ \Dot{R}^2},0,0\right)$.
For the exterior,
$n_e$ is given by 
$n_e =\
\sqrt{f(r_e)}\left(
f(r_e)^2 -
\left(\dv{R}{t_e}\right)^2\right)^{-\frac12}
\left(- \dv{R}{t_e} dt_e+ dr_e \right)$.
In the hypersurface,  it is useful to write these components in terms
of $\tau$, so using Eq.~\eqref{eq:conditioncontinuitymetric2}
we have
$\sqrt{f(r_e)}\eval{
\left(
f(r_e)^2 -
\left(\dv{R}{t_e}\right)^2\right)^{-\frac12}
}_\Sigma = \frac{\sqrt{f(R) +
    \Dot{R}^2}}{f(R)}
$
and
$\eval{\dv{R}{t_e}}_\Sigma = \frac{f(R) \Dot{R}}{\sqrt{f(R)
    + \Dot{R}^2}}$,
so that
$n_e = \left(-\Dot{R},  \frac{\sqrt{f(R) +
    \Dot{R}^2}}{f(R)},0,0\right)$,
    where from Eq.~\eqref{fofr1} one has
$
f(R) = 1 - \frac{2\mu m}{R^{d-3}} +
    \frac{q Q^2}{R^{2(d-3)}}
$.
Then, for the interior
the nonzero components of the extrinsic curvature are 
\begin{align}
(K_i)^{\tau}_{\,\,\,\tau} =
    \frac{\Ddot{R}}{\sqrt{1 + \Dot{R}^2}}
\,,\quad
(K_i)^{\theta^A}_{\,\,\,\theta^A}
    = \frac{\sqrt{1 + \Dot{R}^2}}{R}\,,
\label{eq:ExtrinsicCurv}
\end{align}
and for the exterior are
\begin{align}
(K_e)^{\tau}_{\,\,\,\tau} = \frac{\Ddot{R} +
    \frac{\partial_{R} f(R)}{2}}{\sqrt{f(R) + \Dot{R}^2}}
\,,\quad
(K_e)^{\theta^A}_{\,\,\,\theta^A} = \frac{\sqrt{f(R) +
    \Dot{R}^2}}{R}\,.
\label{eq:ExtrinsicCurv2}
\end{align}
We assume that the shell is static and in equilibrium,
thus $\Dot{R} = 0$, $\Ddot{R} =
0$, and $u^\alpha = (1,0,0)$. From
Eqs.~\eqref{eq:Curvaturecondition} and~\eqref{eq:Sab},
and Eqs.~\eqref{eq:ExtrinsicCurv}-\eqref{eq:ExtrinsicCurv2}, 
the energy density and the pressure of
the shell are obtained as
$\sigma = \frac{d-2}{8 \pi G R}\left(1 - \sqrt{f(R)}
    \right)
$
and
$p = \frac{d-3}{8\pi G R}\left(\sqrt{f(R)} -
    1\right) + \frac{ f'(R)}{16\pi G \sqrt{f(R)}}$,
    respectively,
where ${}'=\frac{\partial\;}{\partial R}$.
These two expressions for $\sigma$ and $p$
can be put in the form 
\begin{align}
    &\sigma = \frac{1 - k}{\mu\Omega R}\,,\label{eq:sig}\\
    &p = \frac{1}{2\mu\Omega R^{2d-5}k} \frac{d-3}{d-2}
    \Bigg[(1-k)^2 R^{2(d-3)}-q
    Q^2\Bigg]\,,
    \label{eq:p}
\end{align}
where here 
$k$ is the redshift function 
evaluated at 
 $R$, i.e.,
\begin{align}
k = \sqrt{f(R)}
\,.\label{eq:reds}
\end{align}
Note also that $q$ in Eq.~\eqref{eq:p} could be swapped
for $\mu$ due to Eq.~\eqref{muq}, but again, we
keep $q$ whenever the coefficient
is associated to the
electric charge $Q$.
It is useful to define the rest mass of the shell
by 
\begin{align}
    M =  \Omega R^{d-2} \sigma\,.\label{eq:M}
\end{align}
Then, knowing the energy density $\sigma$ from
Eq.~\eqref{eq:sig} and
using Eq.~\eqref{eq:M}
one gets $M =  \frac{R^{d-3}}{\mu}
\left(1-k\right)$.
This in turn can be manipulated to give a relation
for the spacetime mass $m$, namely,
\begin{align}
    m = M - \frac{\mu M^2}{2 R^{d-3}} +
    \frac{Q^2}{2
    R^{d-3}}\,,\label{eq:m}
\end{align}
where Eq.~\eqref{eq:reds}
in the form
$k(M,R,Q) =\sqrt{ 1 - \frac{2\mu m}{R^{d-3}} +
    \frac{q Q^2}{R^{2(d-3)}}
    }$ has been used, see Eq.~\eqref{fofr1}.
We see that the total energy $m$ of the shell is given by the rest
mass $M$ plus a second and third terms that represent the
gravitational potential energy and the electric potential energy,
respectively.
For generic $\epsilon$ and $\epsilon_q$, i.e., not set to one,
Eq.~\eqref{eq:m} is
$m = M - \frac{\mu M^2}{2 R^{d-3}} +
    \frac{{\epsilon}Q^2}{2{\epsilon_q^2}
    R^{d-3}}$.

Since the shell is static,
the junction condition for the covector field $A_a$ given in
Eq.~\eqref{eq:Vectorcondition1}, i.e., $[A_\tau] =0$,
together with Eqs.~\eqref{covi}
and~\eqref{cove} 
yield
$- \frac{Q}{(d-3) R^{d-3}k} 
   - A_i=0$, or
\begin{align}
 A_i=-\frac{Q}{(d-3) R^{d-3}k} \,.
 \label{Aijunc}
\end{align}
For generic $\epsilon$ and $\epsilon_q$,
Eq.~\eqref{Aijunc} is
$A_i=-\frac{Q}{(d-3){\epsilon_q} R^{d-3}k}$
So, Eq.~\eqref{Aijunc}
sets the constant value of the potential in the region
$\mathcal{V}_i$ as a function
of the electric
charge, the position of the shell
and the electric potential at infinity which 
we have put to zero, $A_e=0$,
and moreover gives
the value of the electric potential at the hypersurface, see
Eq.~\eqref{bsigma},
$A_\Sigma=A_i$.
In Eq.~\eqref{eq:Vectorcondition2}, the relevant condition for the
Faraday tensor
is
$[F_{\alpha}] = j_\alpha$,
which upon using
Eq.~\eqref{eq:ja}
becomes 
$-(F_e)_{tr}k\Dot{t}_e =\zeta \sigma_e$, 
with 
$\zeta = \frac{\Omega}{\epsilon_q}$, see
Eq.~\eqref{eq:ja}, and since
we are putting $\epsilon_q=1$,
$\zeta=\Omega$.
Then, from 
${F}_{ab}=\frac{\partial A_b}{\partial x^a}- \frac{\partial
A_a}{\partial x^b}$
together with Eqs.~\eqref{covi}
and~\eqref{cove} 
implies
\begin{align}
Q = R^{d-2} \sigma_e\,.
\end{align}
This is the relation between  the total
electric charge and  its corresponding charge
density.

\section{Entropy of the charged thin shell}
\label{sec:firstlaw}

\subsection{The first law of thermodynamics}

The analysis of the thermodynamics
of an electrically charged thin shell
is performed  by
imposing the first law of thermodynamics
on the shell.
Noting that the 
internal energy of the
shell is its rest mass, the first law of
thermodynamics is
\begin{align}
    TdS =  dM +  p dA -  \Phi dQ\,,
\label{1stlaw1}
\end{align}
where $T$ is the local temperature
throughout the shell,
$S$ is the entropy of the shell, 
$M$ is the rest mass of the shell
calculated in the last subsection, 
$p$ is the pressure on the shell
calculated in the last subsection, 
$A$ is the area of the shell,
$\Phi$ is the thermodynamic electric potential
at the shell, 
and $Q$ is the electric charge of the shell.
Defining the
inverse temperature $\beta$ as 
\begin{align}
   \beta=\frac1T \,,
\label{inverseT}
\end{align}
the first law
can be written as
\begin{align}
dS = \beta dM + \beta p dA - \beta \Phi dQ\,.
\label{1stlaw2}
\end{align}
To solve the first law, three
equations of state have to be
provided. 
A first equation of state is for the
inverse temperature in terms of $M$, $A$, and $Q$,
\begin{align}
\beta=\beta(M,A,Q)\,.
\label{es1}
\end{align}
A second equation of state is for the
pressure in terms of $M$, $A$, and $Q$,
\begin{align}
p = p(M,A,Q)\,.
    \label{es2}
\end{align}
This equation of state has already
been found.
Indeed, Eq.~\eqref{eq:p} is a dynamic as
well as a thermodynamic equation,
it is
$
p(M,A,Q)
= \frac{1}{2\mu\Omega R^{2d-5}k} \frac{d-3}{d-2}
    \Big[(1-k)^2 R^{2(d-3)}-q
    Q^2\Big]$,
    as $k$ is a function of $M$, $A$, and $Q$, and $R$
can be swapped for $A$.
A third equation of state is for the
thermodynamic electric potential
$\Phi$ in terms of $M$, $A$, and $Q$, 
\begin{align}
\Phi=\Phi(M,A,Q)\,.
\label{es3}
\end{align}
Note that $\beta$ and $\Phi$ are restricted
by the integrability conditions but otherwise free,
and $p$ is fixed by the equations
of motion, showing that Einstein
equations have already thermodynamics
in-built into them.
Given the functions
$\beta(M,A,Q)$, $p(M,A,Q)$, and $\Phi(M,A,Q)$
of Eqs.~\eqref{es1}-\eqref{es3}
we are interested
in calculating the entropy $S$ as
a function of $M$, $A$, and $Q$, i.e., 
$S(M,A,Q)$, through the first law of thermodynamics
given in
Eq.~\eqref{1stlaw2}.

\subsection{Integrability conditions}
The entropy $S$ is a function of the thermodynamic parameters
$(M,A,Q)$ and its differential is exact by definition. This places the
condition that the Hessian matrix of $S$ needs to be symmetric. Since
from Eq.~\eqref{1stlaw2} the first derivatives of $S(M,A,Q)$ are
\begin{align}
    &\Big(\pdv{S}{M}\Big)_{A,Q} = \beta\,,\notag\\
    &\Big(\pdv{S}{A}\Big)_{M,Q} = \beta p\,,\notag\\
    &\Big(\pdv{S}{Q}\Big)_{M,A} = - \beta \Phi\,,
\end{align}
the condition on the Hessian of $S$ is explicitly
\begin{align}
    &\Big(\pdv{\beta}{A}\Big)_{M,Q} =
    \Big(\pdv{\beta p}{M}\Big)_{A,Q}\,,\notag\\
    &\Big(\pdv{\beta}{Q}\Big)_{M,A} = -
    \Big(\pdv{\beta \Phi}{M}\Big)_{A,Q}\,,\notag\\
    &\Big(\pdv{\beta p}{Q}\Big)_{M,A} = -
    \Big(\pdv{\beta \Phi}{A}\Big)_{M,Q}\,.
    \label{eq:ExactConditions}
\end{align}
These are the integrability conditions, necessary
to have the entropy $S$ as an exact differential.

\subsection{Parameter transformation and the entropy}

In order to compute $S$, we can make parameter transformations to
simplify the differential. The parameter space ($M,A,Q$) can be easily
transformed into ($M,R,Q$) since $A$ depends solely on $R$
through Eq.~\eqref{eq:A}, i.e., $ A = \Omega R^{d-2}$.
We can also express $S$ in the parameters ($r_+,r_-,R$), which will be
more convenient.
This transformation can be performed
by using Eq.~\eqref{masscharge}.
Then, we
can
use Eq.~\eqref{eq:M} together with~\eqref{eq:sig}
and with
the redshift function 
$k=\sqrt{f(R)}$ of Eq.~\eqref{eq:reds} put in the form 
$k(r_+,r_-,R) =
\sqrt{\left(1-\left(\frac{r_+}{R}\right)^{d-3}\right)
    \left(1-\left(\frac{r_-}{R}\right)^{d-3}\right)}$,
see Eq.~\eqref{fofr2}.
The derivatives of the entropy $S(r_+,r_-,R)$ can be found by the
chain rule, so that
$\left(\pdv{S}{R}\right)_{r_+,r_-} = \beta
    \left(\pdv{M}{R}\right)_{r_+,r_-} + \beta p
    \left(\pdv{A}{R}\right)_{r_-,r_+}$
and
$\left(\pdv{S}{r_\pm}\right)_{r_\mp,R} = \beta
    \left(\pdv{M}{r_\pm}\right)_{r_\mp,R} - \beta \Phi
    \left(\pdv{Q}{r_\pm}\right)_{r_\mp,R}$.
Moreover, from Eqs.~\eqref{eq:A},
\eqref{eq:p}, and \eqref{eq:M}, we can find that
$\left(\pdv{M}{R}\right)_{r_+,r_-} = - p
    \left(\pdv{A}{R}\right)_{r_+,r_-}$.
So, clearly, the partial derivative in $R$
vanishes, $\left(\pdv{S}{R}\right)_{r_+,r_-}=0$.
This means that the entropy that in general
is a function $S=S(r_+,r_-,R)$ in this case has no dependence
on $R$, only in $r_+$ and $r_-$, and so
\begin{align}
S=S(r_+,r_-)\,,
\label{entropysimplified}
\end{align}
i.e., the entropy is independent of $R$ in this
parameter space.

\subsection{The temperature and the electric potential}

The integrability conditions or Euler relations impose restrictions on
the expressions of $\beta$ and $\Phi$, that can be worked out in the
parameters $(r_+,r_-,R)$.

Beginning with $\beta$,
we can calculate by the chain rule its derivative
with respect to
$R$ with $r_+$ and $r_-$ fixed, i.e.,
$\left(\pdv{\beta}{R}\right)_{r_+,r_-} =
    \left(\pdv{\beta}{M}\right)_{A,Q}
    \left(\pdv{M}{R}\right)_{r_+,r_-}
    +\left(\pdv{\beta}{A}\right)_{M,Q}
    \left(\pdv{A}{R}\right)_{r_+,r_-}$.
Using
the first
equation in
Eq.~\eqref{eq:ExactConditions} and that
$\left(\pdv{M}{R}\right)_{r_+,r_-} = - p
    \left(\pdv{A}{R}\right)_{r_+,r_-}$,
    which comes from Eqs.~\eqref{eq:A},
\eqref{eq:p}, and \eqref{eq:M},
and
that $\left(\pdv{p}{M}\right)_{A,Q} =
\frac{1}{(d-2)\Omega k
    R^{(d-3)}}\left(\pdv{k}{R}\right)_{r_+,r_-}$,
which comes from
Eq.~\eqref{eq:p},
we find
$\left(\pdv{\beta}{R}\right)_{r_+,r_-} = \frac{\beta}{k}
\left(\pdv{k}{R}\right)_{r_+,r_-}$,
which upon integration gives
\begin{align}
\beta(r_+,r_-,R) = b(r_+,r_-) k\,,
\label{eq:beta}
\end{align}
where $b(r_+,r_-)$
is a reduced equation of state, an intrinsic quantity
that will depend
solely on the nature
of the matter in the shell.
From Eq.~\eqref{eq:beta}
one sees that 
$b(r_+,r_-) = \beta(r_+,r_-, \infty)$.
The redshift function 
$k=\sqrt{f(R)}$ of Eq.~\eqref{eq:reds} here 
is put in the form 
$k(r_+,r_-,R) =
\sqrt{\left(1-\left(\frac{r_+}{R}\right)^{d-3}\right)
    \left(1-\left(\frac{r_-}{R}\right)^{d-3}\right)}$,
see Eq.~\eqref{fofr2}.
The dependence of $\beta$
on $k$ just found is in agreement with the
Tolman's formula for the temperature in a static
gravitational field.

Now, for the case of $\Phi(r_+,r_-,R)$, the chain rule for the
derivative in $R$ together with $\left(\pdv{M}{R}\right)_{r_+,r_-} = -
p \left(\pdv{A}{R}\right)_{r_+,r_-}$
gives
$\left(\pdv{\Phi}{R}\right)_{r_+,r_-} =
\left(\pdv{A}{R}\right)_{r_+,r_-}\left(\pdv{\Phi}{A}\right)_{Q,M}
- p
\left(\pdv{A}{R}\right)_{r_+,r_-}\left(\pdv{\Phi}{M}\right)_{A,Q}$.
The three equations in Eq.~\eqref{eq:ExactConditions} can be
rearranged to substitute the right-hand side of
the latter equation  into
$\left(\pdv{\Phi}{R}\right)_{r_+,r_-} =
    -\left(\pdv{A}{R}\right)_{r_+,r_-}\left(\pdv{p}{Q}\right)_{M,A}
    - \Phi
    \left(\pdv{A}{R}\right)_{r_+,r_-}\left(\pdv{p}{M}\right)_{A,Q}$.
Then, we can use Eqs.~\eqref{eq:A} and
the expressions of the derivatives of the pressure, i.e.,
$\left(\pdv{p}{M}\right)_{A,Q} = \frac{1}{(d-2)\Omega k
    R^{(d-3)}}\left(\pdv{k}{R}\right)_{r_+,r_-}$
    and $\left(\pdv{p}{Q}\right)_{M,A} =- \frac{Q(d-3)}{(d-2)\Omega k
    R^{2d-5}}$,
to find explicitly the equation for $\Phi(r_+,r_-,R)$, which becomes
$\left(\pdv{k\Phi}{R}\right)_{r_+,r_-} = \frac{(d-3)Q}{R^{d-2}}$.
Upon integration one finds
\begin{align}
    &\Phi(r_+,r_-,R) = \frac{Q}{k}
\Big[ c(r_+,r_-) -
    \frac{1}{R^{d-3}}
    \Big]\,,\label{eq:Phi}
\end{align}
where $c(r_+,r_-)$
is a reduced equation of state, an intrinsic quantity
that will depend
solely on the nature
of the matter in the shell. From Eq.~\eqref{eq:Phi}
one sees that 
 $c(r_+,r_-) = \frac{\Phi(r_+,r_-,\infty)}{Q}$.

\subsection{The differential of the entropy}

The differential $dS$ in the parameters $(r_+,r_-,R)$ can be written
considering Eqs.~\eqref{eq:beta} and~\eqref{eq:Phi}. It follows
that
\begin{align}
    dS =& \frac{{(d-3)}b(r_+,r_-)}{2\mu}\Big[ \Big(1 - r_-^{d-3}
    c(r_+,r_-)\Big)r_+^{d-4}dr_+ \notag\\&+ \Big(1-
    r_+^{d-3}c(r_+,r_-)\Big)r_-^{d-4}dr_-\Big]\,.
    \label{entropydiff}
\end{align}
To ensure the integrability of the differential, we apply once again
the symmetric characteristic of the Hessian matrix in these
coordinates, which gives the condition
\begin{align}
&\pdv{b}{r_-}\,(1- c\, r_-^{d-3})r_+^{d-4} - \pdv{b}{r_+}\,(1-c
\,r_+^{d-3})r_-^{d-4} \notag\\&= \pdv{c}{r_-}b r_-^{d-3}r_+^{d-4} -
\pdv{c}{r_+}b r_+^{d-3}r_-^{d-4}\,.
\label{eq:inteC}
\end{align}
Hence, the entropy $S(r_+,r_-)$ will depend on two functions
$b(r_+,r_-)$ and $c(r_+,r_-)$ that are related by a partial
differential equation, Eq.~\eqref{eq:inteC}. These functions cannot be
specified by the first law of thermodynamics together with general
relativity as they depend on the class of matter that composes the
shell.  To make progress we need to specifiy the two reduced equations
of state for $b(r_+,r_-)$ and $c(r_+,r_-)$.

It is also
interesting to
notice that the differential for the entropy can be rewritten in a
simpler form as
$dS = \frac{b}{2\mu}
d(r_+^{d-3}+r_-^{d-3})
-\frac{b}{2\mu}c\,d[(r_+ r_-)^{d-3}]$, and thus
from Eq.~\eqref{masscharge}
one has 
$dS =bdm- b\phi dQ$, where we 
have defined the electric potential $\phi=Q\,c$,
and have used 
our convention $q=\mu$.
So the entropy and its derivatives
are functions of the ADM mass $m$
and the modulus of the electric charge of the shell,
i.e., $S=S(m,Q)$,
and, as well, the equations of state
will be functions of $m$ and $Q$, namely,
$b=b(m,Q)$ and 
$c=c(m,Q)$,
where $Q$ here means
the modulus of the electric charge. 
This shows that the
dependence on
the rest mass and on the pressure and 
the area in the first law of thermodynamics
as given in Eq.~\eqref{1stlaw2}, i.e.,
$\beta dM+\beta pdA$, 
comes from the ADM mass $m$, since
in this version these terms
are compressed to $b\,dm$, and is
aligned with the fact that
rest mass and pressure are
forms of energy in general relativity.

\subsection{The reduced equations of state: Specific choice}

To proceed, we have now to give the two reduced equations of
state, one
for $b(r_+,r_-)$ and the other for $c(r_+,r_-)$.

We choose the reduced equation of state for the temperature
of the shell, or
better, for its inverse temperature as
\begin{align}
      b(r_+,r_-) = \frac{a \gamma \Omega^{a-1}}{d-3}
      \frac{r_+^{a(d-2)}}{r_+^{d-3}-r_-^{d-3}}\,\,,
      \label{eq:tempeqstate}
\end{align}
where $a$ is a free exponent and $\gamma$ is a free parameter.
The reduced equation of state given in Eq.~\eqref{eq:tempeqstate}
imposes the restriction $r_-\leq r_+$, i.e., $r_+$ and $r_-$ have real
values.  This means that the shell can be undercharged or, in the
limit, extremely charged, but not overcharged. Thus, this reduced
equation of state cannot be applied to overcharged shells.

Inserting the $b(r_+,r_-)$ given in Eq.~\eqref{eq:tempeqstate} into
the integrability condition given in Eq.~\eqref{eq:inteC}, one finds
that one of the solutions for $c(r_+,r_-)$ is
\begin{align}
c(r_+,r_-) = \frac{1}{r_+^{d-3}}\,,
\label{eq:electeqstate}    
\end{align}
which yields the typical Reissner-Nordstr\"om equation of state for
the electric potential, and so we choose it as the second equation of
state.

The equation of state for the temperature, Eq.~\eqref{eq:tempeqstate},
introduces two free parameters, namely, $a$ and $\gamma$, which can be
chosen at will as long as the choice is physically reasonable, with
the power law exponent $a$ being the more relevant.  Envisaging the
equation of state for the temperature as arising from quantum effects
in the matter of the shell, it implies that the Planck constant
$\hbar$ appears intrinsically in the formula for $b$, as well as
Boltzmann constant $k_B$ since the whole setup involves
thermodynamics.  In this case $b$ is a length scale, a thermal one.
We set $\hbar=1$ and $k_B=1$, so the parameter $\gamma$ has units of
length to the power $(d-2)(1-a)$, and in this case, incidentally, the
$d$-dimensional constant of gravitation $G$ has units of length to the
power $d-2$.  The equation of state for the electric potential,
Eq.~\eqref{eq:electeqstate}, has no new free parameters. Both
equations of state have another parameter which is the dimension of
spacetime $d$. Being a special parameter, it can nonetheless be
treated as a free one, if one wishes. As long as the dimension $d$ is
finite one can put the desired dimension, be it $d=4$, $d=11$, or any
other finite $d$, into the formulas to obtain the corresponding
expressions for the physical quantities for that dimension. The
infinite $d$ limit is a specific case that depends on how the limit is
taken and so requires special attention. Here, we are just interested
in finite $d$ however large it is.

\subsection{The entropy formula}
\label{entropyformula}

The
differential equation for the entropy
given in Eq.~\eqref{entropydiff}
can then be integrated to yield 
$S = \frac{\gamma}{2\mu(d-2)\Omega}(\Omega r_+^{d-2})^a$
where we have chosen the constant
of integration
to be zero, and using
Eq.~\eqref{eq:A+}, it gives,
$S =\frac{\gamma}{2\mu(d-2)\Omega}A_+^a$,
or restoring the constant of gravitation $G$
from Eq.~\eqref{muq}, one has
\begin{align}
    S = \    \frac{\gamma}{16\pi G}A_+^a\,,
    \label{eq:suggestedentropy}
\end{align}
i.e., the entropy of the shell, a dimensionless quantity, is
proportional to a power of the gravitational area $A_+$.  Due to the
chosen equations of state, namely Eqs.~\eqref{eq:tempeqstate} and
\eqref{eq:electeqstate}, the generic dependence of $S$ on $r_+$ and
$r_-$, $S=S(r_+,r_-)$, see Eq.~\eqref{entropysimplified}, is now
reduced to a dependence on $r_+$ alone, $S=S(r_+)$ or, adopting the
gravitational area instead of the gravitational radius as the variable
for the entropy, one has $S=S(A_+)$.  Furthermore, in
Eq.~\eqref{eq:suggestedentropy}, we should perhaps impose that $a > 0$
so that the entropy does not diverge in the no black hole limit
$r_+\rightarrow 0$.  Note that here $A_+$ is not the event horizon
area since there is no event horizon, there is no black hole, there is
only the spacetime gravitational radius $r_+$.

We can now see the motivation for the choice of the reduced equations
of state given in Eqs.~\eqref{eq:tempeqstate}
and~\eqref{eq:electeqstate}.  It is twofold. First, power laws in
thermodynamics and statistical mechanics are ubiquitous, so it is
natural to take for the reduced equations of state $b(r_+,r_-)$ and
$c(r_+,r_-)$ power laws in $r_+$ and $r_-$, which themselves are
functions of $M$, $A$, and $Q$. Second, there is the motivation that
by choosing such equations of state they give the possibility of
taking the black hole limit $R= r_+$.  Thus, $b(r_+,r_-)$ given in
Eq.~\eqref{eq:tempeqstate} has that, for $a=1$, one gets a functional
dependence equal to the Hawking temperature of the black hole.  The
equation for the reduced potential Eq.~\eqref{eq:electeqstate}, is
simply the same as the corresponding black hole.  These two choices
yield in the $a=1$ case $S =\frac{\gamma}{16\pi G}A_+$, see
Eq.~\eqref{eq:suggestedentropy}, i.e., an entropy for the shell
proportional to the gravitational radius, which has the same
functional dependence as the Bekenstein-Hawking black hole
entropy. Note that other power laws could be chosen. For instance, one
could choose a power of Eq.~\eqref{eq:tempeqstate} itself and another
different power of Eq.~\eqref{eq:electeqstate}, and these equations
would still yield black hole features for the appropriate choice of
the exponents.  Yet a different equation of state for the reduced
inverse temperature $b$, is to choose $b$ as a power law in the ADM
mass, in which case it permits to treat not only undercharged and
extremal charged shells, but also overcharged shells, see
Appendix~\ref{eos} for such a choice. Of course, other choices with
physical meaning can be thought of.

Another important point brought about by
Eq.~\eqref{eq:suggestedentropy} is that as long as $r_+$ is fixed, the
entropy is the same for any radius $R$ of the shell.  To understand
the process involved we use Eq.~\eqref{eq:M}, or better, the equation
before it, namely,
$M = \frac{R^{d-3}}{\mu} \left(1-k\right)$, in full, $M =
\frac{R^{d-3}}{\mu} \left(1- \sqrt{
\left(1-\left(\frac{r_+}{r}\right)^{d-3}\right)
    \left(1-\left(\frac{r_-}{r}\right)^{d-3}\right)}
    \right)
$.
To simplify the discussion, put $d=4$ and $r_-=0$, i.e., $Q=0$.  Then,
$M = R \left(1- \sqrt{1 - \frac{r_+}{R}}\right)$. For $r_+$ fixed we
see that for $R=r_+$ one has $M=R=r_+$, and for $R\to\infty$ one has
$M=\frac12\,r_+$, plus the derivative of $M$ in $R$ is strictly
negative.  So, for fixed $r_+$, as $R$ increases the rest mass $M$ of
the shell decreases.  In this process of changing the radius of the
shell maintaining $r_+$ fixed, one has, from
Eq.~\eqref{eq:suggestedentropy}, that the entropy does not
change. Since the size and the energy of the system change, one
increases, the other decreases, or vice versa, but the entropy does
not change, one is in the presence of an isentropic process.

\subsection{Euler theorem}

According to Eq.~\eqref{eq:sig}
together with~\eqref{eq:M}, the rest mass $M$ can be written
in terms of $r_+$, $r_-$, and $R$, see also Eq.~\eqref{eq:m}.
Moreover, using Eqs.~\eqref{eq:A+} and
\eqref{eq:suggestedentropy}
the gravitational radius, $r_+$,
can be written in terms of $S$
as
$r_+ = \frac{1}{\Omega^{\frac{1}{d-2}}}\left(\frac{16
\pi G S }{\gamma}\right)^{\frac{1}{a(d-2)}}$, then 
using Eq.~\eqref{masscharge}
the Cauchy radius, $r_-$,
can be written in terms of $S$ and $Q$
as
$r_- = \frac{(q Q^2)^\frac{1}{d-3}
\Omega^{\frac{1}{d-2}}}{\left(\frac{16
\pi G S}{\gamma}\right)^\frac{1}{a(d-2)}}$,
and finally using
 Eq.~\eqref{eq:A}
$R$ can be written in terms of $A$ as 
$R = \left(\frac{A}{\Omega}\right)^{\frac{1}{d-2}}$.
Substituting these latter three results into 
Eq.~\eqref{eq:M}
together with~\eqref{eq:sig}, i.e., $M =  \frac{R^{d-3}}{\mu}
\left(1-k\right)$,
one has that the rest mass $M$ seen
as a function of $S$, $A$, and $Q$, i.e.
$M(S,A,Q)$, is given by
$M = \frac{1}{\mu}\left(\frac{A}{\Omega}
\right)^{\frac{d-3}{d-2}} \left[1 -
\sqrt{\left(1 - s_1\right) \left(1 -
s_2\right)} \right]$, where
we have defined
$s_1= \left(\frac{16\pi G
S}{\gamma A^a}\right)^{\frac{d-3}{a(d-2)}}$
and
$s_2 = \frac{q Q^2 \Omega^{2\frac{d-3}{d-2}}}{
\left(\frac{16\pi G
S A^a}{\gamma}\right)^{\frac{d-3}{a(d-2)}}}$.
Now, from this expression for $M(S,A,Q)$ one can
see
that $M\left(\lambda S^{\frac{1}{a}}, \lambda A,
\lambda Q^{\frac{d-2}{d-3}}\right) = \lambda^{
\frac{d-3}{d-2}} M
\left(S^{\frac{1}{a}}, A, Q^{\frac{d-2}{d-3}}\right)$,
for some arbitrary factor $\lambda$.
Since the derivatives of $M$ are described by the differential
$dM = TdS - pdA + \Phi dQ$, i.e., the first law of thermodynamics,
one obtains by the Euler
relation theorem for homogeneous functions that
\begin{align}
    \frac{d-3}{d-2}M = a T S - p A +
    \frac{d-3}{d-2} \Phi Q\,\,.\label{eq:EulerRel}
\end{align}
This relation is an integrated version of the first law of
thermodynamics for the thin shell with the
specific entropy given in Eq.~\eqref{eq:suggestedentropy}.

\subsection{Shell with black hole features,
the black hole limit, and Smarr formula}

\subsubsection{Shell with black hole features}

To get a shell with
black hole features we see
that taking $a=1$ we
obtain from Eq.~\eqref{eq:tempeqstate}
that $b(r_+,r_-) = \frac{ \gamma }{d-3}
\frac{r_+^{(d-2)}}{r_+^{d-3}-r_-^{d-3}}$, so that $T_0$
defined as $T_0=\dfrac1b$
is given by 
$T_0(r_+,r_-) = \frac{ d-3}{\gamma } \frac{r_+^{d-3}-r_-^{d-3}}
{r_+^{d-2}}$.  With the Planck length defined as
$\ell_p=(\frac{\hbar G}{c^3})^{\frac1{d-2}}$, and
since here we have $c=1$ and $\hbar=1$,
we have
$\ell_p=G^{\frac1{d-2}}$.
Putting in addition
$\gamma = 4\pi$ one gets $T_0(r_+,r_-) = \frac{ d-3}{4\pi}
\frac{r_+^{d-3}-r_-^{d-3}} {r_+^{d-2}}$, which is the Hawking
temperature $T_+$ for the
matter on the shell.
The reduced electric potential is still given by
Eq.~\eqref{eq:electeqstate}, $c(r_+,r_-) = \frac{1}{r_+^{d-3}}$.
Thus, for the shell with
black hole features,
one has that the
reduced inverse temperature and electric
potential are given by 
\begin{align}
\hskip - 0.4cm
b_+(r_+,r_-)\hskip - 0.05cm = \hskip - 0.05cm\frac{4\pi}{d-3}
\frac{r_+^{d-2}}{r_+^{d-3}-r_-^{d-3}}   ,\;
c_+(r_+,r_-)  \hskip - 0.05cm
=  \hskip - 0.05cm\frac{1}{r_+^{d-3}},\hskip -0.3cm
    \label{tempelectshellbhfinal}
\end{align}
where a subscript +
are for quantities
characteristic of black holes.
Then, the  entropy of the shell given
in Eq.~\eqref{eq:suggestedentropy}
turns into
\begin{align}
    S_+ = \frac{1}{4}\frac{A_+}{A_p}\,,
    \label{entropyshellfinal}
\end{align}
where $A_p$ is the
Planck area defined as
$A_p=l_p^{d-2}=G$.
Thus, for the shell's matter equations of state
given in 
Eqs.~\eqref{eq:tempeqstate} and~\eqref{eq:electeqstate} 
with
in addition
$a=1$ and $\gamma = 4\pi $,
one finds that the shell
at radius $R$
and with area $A$,
has black hole features,
it has
precisely the Bekenstein-Hawking entropy,
as given in Eq.~\eqref{entropyshellfinal}.
So, thermodynamically,
this spacetime being not a black hole
spacetime, rather it is a shell spacetime,
actually
mimics thermodynamically the corresponding
black hole spacetime, i.e., the
black hole
that has
the same gravitational radius $r_+$.
Indeed, for any radius $R$ greater than
the shell's gravitational radius $r_+$, $R\geq r_+$, 
the shell's entropy is always the same,
it is the Bekenstein-Hawking entropy.

\subsubsection{Black hole limit and Smarr formula}

To get a shell which not only has black hole
features but is almost a black hole,
i.e., a quasiblack hole,
we have to take the precise limit of the shell radius
$R$ going into the shell gravitational radius $r_+$,
$R
\rightarrow r_+$. 
In this case, in order to
not have divergences in the quantum state
of the matter and
to maintain thermal equilibrium,
the temperature of the shell must be precisely
the Hawking temperature, $T_+(r_+,r_-) = \frac{ d-3}{4\pi}
\frac{r_+^{d-3}-r_-^{d-3}} {r_+^{d-2}}$,
in which case the entropy of the shell
at its own gravitational radius 
has to be Bekenstein-Hawking entropy,
$S_+ = \frac{1}{4}\frac{A_+}{A_p}$,
see Eq.~\eqref{entropyshellfinal}.
When the shell is at its own gravitational radius, the shell
spacetime
is in a quasiblack hole state, the 
gravitational radius being now a quasihorizon radius.
This limit can be thought of as a sequence of
quasistatic thermodynamic equilibrium states of
the shell that reach the equilibrium state of the black hole.
Note that the pressure in Eq.~\eqref{eq:p}
diverges in this limit. In a sense this means that all degrees of
freedom are excited in this limit and the entropy is maximal.
Clearly the shell formalism, that provides an
exact solution for its dynamics and its thermodynamics,
yields in the appropriate limit
the black hole features, notably,
the Bekenstein-Hawking entropy of a black hole.
The quasiblack hole
formalism, different from the shell formalism
and with some correspondence to the
membrane paradigm formalism, 
deals with generic matter systems on the verge
of becoming a black hole and
is also able to bring
out
all the thermodynamic properties of black holes
\cite{Lemos:2010,Lemos:2011,lemoszaslavskii}.

The extremal electrically charged black hole merits
a complete investigation. Here,
we mention
some important points
connected to
the
entropy and thermodynamics of an
extremal Reissner-Nordstr\"om shell solution in $d$-dimensions and
the corresponding Reissner-Nordstr\"om black hole.
The extremal Reissner-Nordstr\"om spacetime
obeys the relation
$r_+ = r_-$, and so
for a reduced equation of state of the form given in 
Eq.~\eqref{eq:tempeqstate}
one has 
that the extremal charged shell case
has zero temperature,
whereas the reduced electric potential
still has the form given also
in Eq.~\eqref{eq:electeqstate},
and so both are well defined in the
extremal case. 
On the other hand,
the entropy of such a shell is
a subtle issue. If from a nonextremal shell,
with $R>r_+$ 
we take the limit $r_+ =
r_-$, then one obtains by continuity directly that the entropy for the
shell is given by Eq.~\eqref{eq:suggestedentropy}.  On the other hand,
if we start with an extremal shell a priori then the entropy of the
shell is some function of $A_+$, $S(A_+)$, 
that is not specified, i.e., one is free to choose it
\cite{Lemos:2015c}.
At the black hole limit in the extremal case,
i.e.,
when the radius $R$ of the shell approaches its gravitational radius
$r_+=r_-$,
and the reduced equations of state are given in 
Eq.~\eqref{tempelectshellbhfinal}, the situation is even more subtle.
Besides the two possible cases
similar to the two shell cases just mentioned,
namely, the shell is nonextremal and
is then put to its gravitational radius, and the shell is extremal
and is then put to its extremal radius, there is a third case when the
shell is turning to being
extremal and simultaneously it is approaching its own
gravitational radius \cite{lemoshellextremal12}.  The first case gives
the Bekenstein-Hawking entropy for the shell,
$S_+ = \frac{1}{4}\frac{A_+}{A_p}$,
the second case gives
that the entropy is some unspecified function of $A_+$,
$S_+=S_+\left(\frac{A_+}{A_p}\right)$, and
the third case gives again the Bekenstein-Hawking entropy for the
shell,
$S_+ = \frac{1}{4}\frac{A_+}{A_p}$.
Thus, if we take the entropy of an extremal shell at its own
gravitational radius as representative of the entropy of an extremal
black hole, then the entropy of an extremal black hole depends on 
its past, specifically, on the way it was formed, see
also~\cite{Lemos:2011} where the quasiblack hole approach is applied.

Now, we turn to the Smarr formula. It will be
derived from the Euler
relation for the shell, Eq.~\eqref{eq:EulerRel},
in the black hole limit.  The shell with black
hole properties has $a=1$ and
is given by Eqs.~\eqref{tempelectshellbhfinal}
and~\eqref{entropyshellfinal}.
Multiplying the Euler
relation, Eq.~\eqref{eq:EulerRel}, by 
the factor $k$, one obtains for a shell
with black hole properties 
$\frac{d-3}{d-2}k M  = T_+ S_+ -k p  A + \frac{d-3}{d-2}k \Phi_+  Q$,
where $\Phi_+$ is $\Phi$ defined
in Eq.~\eqref{eq:Phi} with black hole characteristics, i.e.,
$\Phi_+(r_+,r_-,R) = Q \frac{ r_+^{-(d-3)} -
    R^{-(d-3)}}{k}$.
One can now take the black hole limit, $R=r_+$.
Then the redshift function is zero,
$k=0$. This means that
$kM=0$. One also has 
$k\Phi_+(r_+,r_-,R) = Q \left({ r_+^{-(d-3)} -
    R^{-(d-3)}}\right)_{R=r_+}=0$.
So, the nonzero terms
    are $T_+ S_+$ and $-k p  A$. Then using 
Eqs.~\eqref{eq:A}
and \eqref{eq:p} for $-k p  A$, we obtain
putting $R=r_+$,
$0 =  T_+ S_+ -
\frac{1}{2\mu}\frac{d-3}{d-2}
(r_+^{d-3} - r_-^{d-3})$ or,
upon rearrangements, 
$0 =  T_+ S_+ -
\frac{1}{2\mu}\frac{d-3}{d-2}
(r_+^{d-3} + r_-^{d-3})
+
\frac{1}{\mu}\frac{d-3}{d-2} r_-^{d-3}
$. From Eqs.~\eqref{muq} and~\eqref{masscharge}
one has 
$\frac{1}{2\mu}\left(r_+^{d-3} + r_-^{d-3}\right)
=m$.
The last term can be written as
$\frac{1}{\mu}\frac{d-3}{d-2} r_+^{-(d-3)}
(r_+r_-)^{d-3}$, and from Eq.~\eqref{masscharge}
this is 
$\frac{d-3}{d-2} r_+^{-(d-3)}Q^2=\frac{d-3}{d-2}
[Qr_+^{-(d-3)}]Q=\frac{d-3}{d-2}
\phi_+ Q
$, where
the black hole potential $\phi_+$ is naturally
defined as
$\phi_+=Qr_+^{-(d-3)}$.
Then, the
Euler relation becomes
$0=T_+ S_+ -
\frac{d-3}{d-2}
m +\frac{d-3}{d-2}\phi_+Q$, i.e.,
\begin{align}
    m = \frac{d-2}{d-3}T_+ S_+ + \phi_+ Q\,\,,
\label{smarreissnernordstromtangherlini}
\end{align}
which is 
the Smarr formula for a
$d$-dimensional 
Reissner-Nordstr\"om,
i.e., a Reissner-Nordstr\"om-Tangherlini
black hole.
This relation is
the integral version of the first law of
thermodynamics for black holes, which can be
picked up from the first law formula
$dS =bdm- b\phi dQ$ or, swapping places,
$dm=T_0dS+\phi dQ$,
with $T_0=\frac1b$, found after Eq.~\eqref{entropydiff}
for thin shells,
when applied to black holes.
For the extremal case, $r_+=r_-$, 
one has $T_+=0$ and 
$\phi_+ Q=Qr_+^{-(d-3)}\,Q=\frac{Q}{\sqrt q}=
\frac{Q}{\sqrt \mu}$, where 
Eq.~\eqref{masscharge} has been used,
and the equality $\mu=q$ in our convention of units
has been applied.
Thus, for the extremal case,
the Smarr formula
given in Eq.~\eqref{smarreissnernordstromtangherlini}
turns into 
$\sqrt{\mu}m=Q$ as it should be. 
Now, in four dimensions, $d=4$,
Eq.~\eqref{smarreissnernordstromtangherlini}
gives
$m = 2T_+ S_+ + \phi_+ Q$
which is the original Smarr formula
for a four-dimensional Reissner-Nordstr\"om black hole.
Still in 
$d=4$, the extremal case gives 
$\sqrt{G}m=q$,
or $m=q$ if one puts $G=1$,
which is the usual
mass formula for an
extremal
four-dimensional Reissner-Nordstr\"om black hole.

\section{Intrinsic thermodynamic stability for the given equations of
state
\label{sec:stability}}

\subsection{Stability conditions}

The shell with  generic  equations of state given in
Eqs.~\eqref{es1}-\eqref{es3} will have its
thermodynamic equilibrium state
for some entropy $S(M,A,Q)$ found from the
first law of thermodynamics Eq.~\eqref{1stlaw1}.
We now look at the intrinsic
thermodynamic stability of the shell, see
Callen's thermodynamics book  for the
formalism.

In general, a system in thermodynamic equilibrium is susceptible to
perturbations. Let the system with entropy $S$ be split into two
subsystems. Then, the fluctuations of the matter in the boundary
between the subsystems will allow exchanges in the thermodynamic
variables, in this case $(M,A,Q)$. The entropy of the system after
those exchanges, $S+\Delta S$, will be the sum of the entropy of
the two subsystems. By the second law of thermodynamics, if
$S+\Delta S\leq S$, i.e., $\Delta S\leq 0$,
then the system will stay in equilibrium,
hence the system is stable. Otherwise, the system will evolve away
from equilibrium, building up inhomogeneities, and therefore the
equilibrium is unstable. For very small fluctuations, the conditions
of intrinsic stability are given by $dS(M,A,Q) = 0$ and $d^2S(M,A,Q)
\leq 0$, i.e., $S$ is a maximum with the Hessian of $S$ being
seminegative definite. Notice that in general the quantities
$(M,A,Q)$ do not have a relation between themselves.
However, in our case there
is a relation between those quantities since first, $S$ is solely a
function of $r_+$, and so the equilibrium configurations are given by
$r_+(M,A,Q)$, and second, since the condition $dS(M,A,Q) = 0$ holds it
implies that $r_+(M,A,Q)= {\rm constant}$, so $(M,A,Q)$ are tied by a
relation between themselves.

The stability conditions with respect to the second
derivatives of $S$,
denominated by
$S_{h_i h_j} = \frac{\partial^2
S}{\partial h_i \partial h_j}$, are
\begin{align}
    & S_{MM} \leq 0 \,,\,\,S_{AA} \leq 0 \,,\,\,S_{QQ} \leq 0
    \,,\notag\\
    & S_{MM}S_{AA} - S_{MA}^2 \geq 0 \,,\notag\\& S_{MM}S_{QQ} -
    S_{MQ}^2 \geq 0 \,,\notag\\& S_{QQ}S_{AA} - S_{QA}^2 \geq
    0\,,\notag\\
    & (S_{MM}S_{AQ} - S_{MA}S_{MQ})^2 \notag\\&- (S_{AA}S_{MM} -
    S_{AM}^2)(S_{QQ}S_{MM} - S_{QM}^2 )\leq 0\,,
    \label{eq:StabCond}
\end{align}
which have to be employed with care for each appropriate physical
situation as it is detailed below.  Note that there is a freedom on the
choice of sufficient conditions for each physical situation, which
depends on the order of the variables that one chooses. Here, we are
choosing the order $h_1 = M$, $h_2 = A$ and $h_3 = Q$. The derivation
of these conditions and the explanation of the redundancy of these
conditions are present in the Appendix~\ref{app:Stability}.

\subsection{Entropy and equations of state}

Now, we apply the formalism above.
For that, we rewrite
Eq.~\eqref{eq:suggestedentropy} for the entropy as
\begin{align}
    S(M,A,Q) =     \frac{\gamma}{16\pi G}A_+^a\,,
    \label{eq:suggestedentropystability}
\end{align}
to emphasize that we are dealing with the variables
$M$, $A$, and $Q$.
Clearly, $A_+$ is a function of 
$M$, $A$, and $Q$, since $A_+$ is a function of $r_+$,
see Eq.~\eqref{eq:A+}, which is a function of $M$, $A$, and $Q$
through Eqs.~\eqref{eq:rpm},~\eqref{eq:m}, and~\eqref{eq:A}.

There are three
equations of state that must be provided, one for the temperature,
one for the pressure, and one for the
electric potential. These already have been found in the
previous section.

For the temperature, or better, for
the inverse temperature, one has Eq.~\eqref{eq:beta},
and using the specific choice of
the reduced
equation of state given in 
Eq.~\eqref{eq:tempeqstate},
one finds the explicit
form
of the generic equation given in  Eq.~\eqref{es1}, namely,
\begin{align}
      \beta(M,A,Q) = \frac{a \gamma \Omega^{a-1}}{d-3}
      \frac{r_+^{a(d-2)}}{r_+^{d-3}-r_-^{d-3}}k\,,
      \label{eq:tempeqstatestability}
\end{align}
where clearly it is a function of 
$M$, $A$, and $Q$.

For the thermodynamic
pressure, the equation of state is
given by Einstein equations, so it is also
a dynamic pressure. Then,
Eq.~\eqref{eq:p} is indeed
the explicit
form
of the generic equation given in  Eq.~\eqref{es2}, namely,
\begin{align}
p(M,A,Q)  = 
\frac{1}{2\mu\Omega } \frac{d-3}{d-2}
    \Bigg[(1-k)^2 R^{2(d-3)}-q
    Q^2\Bigg]\frac1{ R^{2d-5}k},
    \label{eq:pstability}
\end{align}
where clearly it is a function of 
$M$, $A$, and $Q$.

For the potential,
one has Eq.~\eqref{eq:Phi},
and using the specific choice for
the  reduced
equation of state  given in 
Eq.~\eqref{eq:electeqstate}    
one finds the explicit
form
of the generic equation given in  Eq.~\eqref{es3}, namely,
\begin{align}
    &\Phi(M,A,Q)  = Q\left(
\frac{1}{r_+^{d-3}} -\frac{1}{R^{d-3}}
    \right)\frac1k\,,
    \label{eq:Phistability}
\end{align}
where clearly it is  a function of 
$M$, $A$, and $Q$.

\subsection{First and second derivatives of the entropy}

For the equations of state we use, the final form
of the entropy of the shell is
given in Eq.~\eqref{eq:suggestedentropystability}.
Then, the first derivatives of the entropy can be computed
either directly, or more easily through
the first law given in Eq.~\eqref{1stlaw1} together with
Eqs.~\eqref{eq:tempeqstatestability}-\eqref{eq:Phistability}.
They are
\begin{align}
    &S_M = \frac{a \gamma \Omega^{a-1}}{d-3}
    \frac{r_+^{a(d-2)}}{r_+^{d-3} - r_-^{d-3}}k\,,\notag\\
    &S_A = \frac{a \gamma \Omega^{a-2}r_+^{a(d-2)}\Big[(1-k)^2
    R^{2(d-3)}-q Q^2 \Big]}{2\mu(d-2)
    R^{2d-5}(r_+^{d-3}-r_-^{d-3})}\,,\notag\\
    &S_Q = -\frac{a \gamma \Omega^{a-1} Q}{{(d-3)}}\left(
    \frac{r_+^{3-d} - R^{3-d}}{r_+^{d-3} -
    r_-^{d-3}}\right) r_+^{a(d-2)}\,.
\end{align}
For the calculation of the second derivatives
of the entropy, it is useful to
consider that 
$\pdv{r_{\pm}}{M} = \pm2\mu\frac{r_\pm\, k}{(d-3)(r_+^{d-3} -
    r_-^{d-3})}$,
$\pdv{r_{\pm}}{R} = \pm \mu\frac{r_\pm}{r_+^{d-3} -
    r_-^{d-3}}\frac{\mu M^2 -
    Q^2}{R^{d-2}}$, and 
$\pdv{r_{\pm}}{Q} = \mp\frac{2 q Q r_\pm
    \left( r_\pm^{3-d} - R^{3-d}\right)
    }{(d-3) (r_+^{d-3} -
    r_-^{d-3})}$.
The components of the Hessian are
then
\begin{align}
    &S_{MM} = \frac{a \gamma \Omega^{a-2} 8\pi G
    r_+^{a(d-2)}}{(d-3)(d-2)(r_+^{d-3} - r^{d-3}_-)
    R^{d-3}}\,S_1,\notag\\
&S_{AA} = \frac{a \gamma \Omega^{a-3}r_+^{a(d-2)}}{
         2\mu(d-2)^2(r_+^{d-3}-r_-^{d-3})R^{d-1}}\,S_2\,,\notag\\
&S_{QQ} = \frac{a\gamma
    \Omega^{a-1}r_+^{a(d-2)}(1-x)}{{(d-3)}(r_+^{d-3}-r_-^{d-3})
    r_+^{d-3}}\,S_3\,,\notag\\
    &S_{MA} = \frac{a\gamma
    \Omega^{a-2}r_+^{a(d-2)}}{(d-2)(r_+^{d-3} -
    r_-^{d-3})R^{d-2}}\,S_{12}\,,\notag\\
&S_{MQ} = - \frac{2\mu a \gamma \Omega^{a-1} r_+^{a(d-2)} Q k
    }{(d-3)^{2} (r_+^{d-3} -
    r_-^{d-3})r_+^{2d-6}}\,S_{13}\,,\notag\\
&S_{AQ} = -\frac{a\gamma \Omega^{a-2}r_+^{a(d-2)}Q
    }{{(d-2)}(r_+^{d-3}-r_-^{d-3})r_+^{d-3}R^{d-2}}\,S_{23}\,,
    \label{secondderivatives}
\end{align}
where
\begin{align}
    &S_1 = \frac{2 k^2 \mathcal{G}}{(d-3)x} - 1\,,\notag\\
    &S_2 = \mathcal{F}\left[\frac{\mathcal{F} \mathcal{G}}{x} - 2d +
    5\right]\,,\notag\\
    &S_3 = -1 + \frac{2y}{d-3}\left[\mathcal{G}(1-x) -
    \frac{2(d-3)}{1-y}\right]\,,\notag\\
    &S_{12} = 1 - k + \frac{k \mathcal{G}}{x(d-3)}\mathcal{F}
    \,,\notag\\
    &S_{13} = \mathcal{G}(1-x) - \frac{(d-3)}{1-y}\,,\notag\\
    &S_{23} = x + \frac{\mathcal{F}}{x(d-3)}\Bigg[ \mathcal{G}(1-x) -
    \frac{(d-3)}{1-y}\Bigg]\,,\notag\\
    \label{eq:Sfunctions}
\end{align}
with the auxiliary functions $\mathcal{G}$, $\mathcal{F}$,
and $k$, being given by
\begin{align}
&\mathcal{G} = \frac{1}{1-y}\Bigg[ a(d-2)-(d-3)\frac{1+y}{1-y}
    \Bigg]\,,\notag\\
    &\mathcal{F} = 2 - 2k{-} x(1-y)\,,\notag\\
    &k = \sqrt{(1-x)(1-xy)}\,,
    \label{eq:Sfunctionsauxiliary}
\end{align}
and we have made use of the definitions
\begin{align}
x=\frac{r_+^{d-3}}{R^{d-3}}\,,\quad\quad
y=\frac{r_-^{d-3}}{r_+^{d-3}}\,.
\label{xy}
\end{align}

The set of inequalities in Eq.~\eqref{eq:StabCond} with the
entropy equation given in Eq.~\eqref{eq:suggestedentropystability}
and the
equations
of state given in
Eqs.~\eqref{eq:tempeqstatestability}-\eqref{eq:Phistability}
can be written as restricting conditions in
terms of the functions given in Eq.~\eqref{eq:Sfunctions}. The
conditions will then restrict the parameter space described by the
points $(d,a,x,y)$ constrained by 
\begin{align}
d\geq 4\,,\quad
a > 0\,,\quad0<x<1\,,\quad 0<y<1\,.
\end{align}
Here,  $d \geq 4$ since for lower $d$ there is
no proper Reissner-Nordstr\"om solution,
$a>0$ because  
in the
no black hole 
limit, $A_+= 0$, i.e., $r_+= 0$,
the entropy expression,
Eq.~\eqref{eq:suggestedentropystability},
should not diverge,
$0<x<1$ because the shell has to be in the limits
between no shell, $x=0$, and the black hole state,
$x=1$,
$0<y<1$ because the electric charge state of
the shell considered here
can run 
from an uncharged
one, $y=0$, to an extremally charged one, $y=1$,
overcharged shells are not treated here since
the equations of state,
Eqs.~\eqref{eq:tempeqstatestability}-\eqref{eq:Phistability},
do not apply to overcharged shells.

In what follows we deal with the algebraic conditions that arise from
the conditions given in Eq.~\eqref{secondderivatives} together with
the auxiliary functions Eqs.~\eqref{eq:Sfunctions}-\eqref{xy}.  In the
Appendix~\ref{Graphics} we make plots to help in the understanding of
these conditions.

\subsection{Mass fluctuations only}
\label{massfluconly}

A shell with only mass fluctuations will have the stability condition
given by $S_{MM} \leq 0$, see Eq.~\eqref{eq:StabCond}.  For the
equations of state we are using, and with the help of
Eq.~\eqref{secondderivatives}, one has that $S_{MM} \leq 0$ can be
written as
\begin{align}
    S_1 \leq 0\,.
    \label{eq:S1cond}
\end{align}
Then, from Eq.~\eqref{eq:Sfunctions} this inequality can be rearranged
as
\begin{align}
    a \leq \frac{x(d-3)(1-y)}{2(d-2)k^2} +
    \frac{(d-3)}{(d-2)}\frac{(1+y)}{(1-y)}\,,\label{eq:SMMa}
\end{align}
where Eqs.~\eqref{eq:Sfunctionsauxiliary} and \eqref{xy} have been
used.  From a quick analysis, the right-hand side tends to infinity at
the points $x=1$ or $y=1$. It has its minimum value at $(x,y) =
(0,0)$, corresponding to $a = \frac{d-3}{d-2}$.  A detailed analysis
of Eq.~\eqref{eq:SMMa} can be seen in Fig.~\ref{fig:S1} which is
itself split into four plots (a), (b), (c), and (d).  It is
interesting to comment on the case of the shell with thermodynamic
black hole features, i.e., the case with $a=1$.  For an uncharged
shell, $y=0$, the range of $x$ for thermodynamic stability is given by
$\frac{2}{d-1}<x<1$, in agreement with \cite{Lemos:2010}. Increasing
the value of $y$ will also increase the range of $x$ for thermodynamic
stable configurations, i.e., if the shell has more electric charge
then a higher radius $R$ is allowed for stability.  The stability is
guaranteed in the full range of $x$ if $y\geq\frac{1}{2d-5}$, see also
Fig.~\ref{fig:S1}(d) top for this $a=1$ case.
It is also interesting to see the stability with respect to the
variables $\frac{M}{R^{d-3}}$ and $\frac{Q}{R^{d-3}}$. We do this below
and one can also refer to Fig.~\ref{fig:S1}(d) bottom.

\subsection{Area fluctuations only}
\label{areafluconly}

A shell with only area fluctuations will have the stability
condition given by
$S_{AA} \leq 0$, see Eq.~\eqref{eq:StabCond}.
For the equations of state we are using,
and with the help of
Eq.~\eqref{secondderivatives},
one has that $S_{AA} \leq 0$ can be written as
\begin{align}
    S_2\leq 0\,.\label{eq:S2cond}
\end{align}
Then, from Eq.~\eqref{eq:Sfunctions} this inequality can be rearranged
as
\begin{align}
    a \leq \frac{(2d-5)x (1-y)}{(d-2)\mathcal{F}} +
    \frac{(d-3)}{(d-2)}\frac{(1+y)}{(1-y)}\,,
    \label{eq:SAAa}
\end{align}
where Eqs.~\eqref{eq:Sfunctionsauxiliary} and \eqref{xy} have been
used,
and employed the fact
that the multiplication factor $\mathcal{F}$
is always
positive for $0<x<1$ and $0<y<1$, it is also proportional to
$M-m$. 
The right-hand side
of Eq.~\eqref{eq:SAAa}
has the minimum at $(x=1,y=0)$, with the value $a
= 3- \frac{2}{d-2}$. The function then increases in the direction of
$x\rightarrow 0$ or $y\rightarrow 1$, where it tends to infinity.
 A detailed analysis of
Eq.~\eqref{eq:SAAa}
can be seen in Fig.~\ref{fig:S2} which is itself
split into three plots (a), (b), and (c).
The case of the shell with thermodynamic black hole
features, i.e., the case with $a=1$, does not need a
more detailed analysis since 
all the
configurations with $a = 1$ are below the surface of
marginal stability, therefore they are stable
to these thermodynamic perturbations.

\subsection{Charge fluctuations only}
\label{chargefluconly}

A shell with only electric charge
fluctuations will have the stability
condition given by
$S_{QQ} \leq 0$, see Eq.~\eqref{eq:StabCond}.
For the equations of state we are using,
and with the help of
Eq.~\eqref{secondderivatives},
one has that $S_{QQ} \leq 0$ can be written as
\begin{align}
    S_3\leq 0\,.
    \label{eq:S3cond}
\end{align}
Then, from Eq.~\eqref{eq:Sfunctions} this inequality can be rearranged
as
\begin{align}
    &a\leq \frac{(d-3)(1-y)}{2(d-2)y(1-x)} +
    \frac{2(d-3)}{(d-2)(1-x)} \nonumber\\&+
    \frac{(d-3)}{(d-2)}\frac{(1+y)}{(1-y)}\,,
    \label{eq:SQQa}
\end{align} 
where Eqs.~\eqref{eq:Sfunctionsauxiliary} and \eqref{xy} have been
used.
The right-hand side of Eq.~\eqref{eq:SQQa}
describes a concave surface, faced to
$a\rightarrow +\infty$. The minimum, restricted to the parameter space,
resides in $\left(x=0,y=\frac{1}{3}\right)$, where its value is $a =
5\frac{d-3}{d-2}$. It diverges to infinity at the axes $x=1$,
$y=0$ and $y=1$.
A detailed analysis of
Eq.~\eqref{eq:SQQa}
can be seen in Fig.~\ref{fig:S3} which is itself
split into three plots (a), (b), and (c).
The case of the shell with thermodynamic black hole
features, i.e., the case with $a=1$, does not need a
more detailed analysis since 
all the
configurations with $a = 1$ are below the surface of
marginal stability, therefore they are stable
to these thermodynamic perturbations.

\subsection{Mass and area fluctuations together}
\label{massandareafluctogether}

A shell with mass and area fluctuations will have the stability
conditions given by
$S_{MM} \leq 0$, $S_{AA} \leq 0$,
and  $S_{MM}S_{AA} - S_{MA}^2 \geq 0$, see Eq.~\eqref{eq:StabCond}.
Note, however, that there is redundancy on this system of
inequations, see Appendix~\ref{app:Stability}.  The sufficient
conditions can be chosen to be $S_{MM}\leq 0$ and $S_{MM}S_{AA} -
S_{MA}^2 \geq 0$.
For the equations of state we are using,
one has that $S_{MM} \leq 0$ yields 
Eq.~\eqref{eq:SMMa}, 
and with the help of
Eq.~\eqref{secondderivatives} one has that
$S_{MM}S_{AA} - S_{MA}^2 \geq 0$
can be written as 
\begin{align}
    S_4 = -\frac{1}{2(d-3)}S_1 S_2 + S_{12}^2 \leq 0\,.
    \label{eq:S4cond}
\end{align}
Then, from Eq.~\eqref{eq:Sfunctions} this inequality can be rearranged
as
\begin{align}
    &a \leq \frac{(1-y)x \Big((d-\frac{5}{2})\mathcal{F} -
    (d-3)(1-k)^2\Big)}{(d-2)\mathcal{F}\Big(\frac{k^2}{d-3} +
    2 k + \frac{\mathcal{F}}{2}\Big)} \nonumber\\&+
    \frac{(d-3)}{(d-2)}\frac{(1+y)}{(1-y)}\,,
    \label{eq:SMMAAa}
\end{align}
where Eqs.~\eqref{eq:Sfunctionsauxiliary} and \eqref{xy} have been
used.
The right-hand side  of
Eq.~\eqref{eq:SMMAAa} is minimum at $x=1$, where $a = 1$. From there
toward $x=0$, the function will bend into $a
=\frac{d-3}{d-2}\frac{1+y}{1-y}$. At $y=1$, it will tend to
infinity. When compared with Eq.~\eqref{eq:SMMa},
the right-hand side of Eq.~\eqref{eq:SMMAAa} is always lower and thus
Eq.~\eqref{eq:SMMAAa} is sufficient to describe the stability region,
in this case.  A detailed analysis of Eq.~\eqref{eq:SMMAAa} can be
seen in Fig.~\ref{fig:S4} which is itself split into four plots
(a), (b), (c), and (d).  The case of the shell with thermodynamic
black hole features, i.e., the case with $a=1$, shows that
increasing the value of $y$ will decrease the range of $x$ for
thermodynamic stable configurations, i.e., if the shell has more
electric
charge then it needs to have lower $R$ for stability, see also
Fig.~\ref{fig:S1}(d) for this $a=1$ case.

\subsection{Mass and charge fluctuations together}
\label{massandchargefluctogether}

A shell with mass and charge fluctuations will have the stability
conditions given by
$S_{MM} \leq 0$, $S_{QQ}\leq 0$,
and  $S_{MM}S_{QQ} - S_{MQ}^2 \geq 0$, see Eq.~\eqref{eq:StabCond}.
Note, however, that there is redundancy on this system of
inequations, see Appendix~\ref{app:Stability}.  The sufficient
conditions can be chosen to be $S_{MM}\leq 0$ and $S_{MM}S_{QQ} -
S_{MQ}^2 \geq 0$.
For the equations of state we are using,
one has that $S_{MM} \leq 0$ yields 
Eq.~\eqref{eq:SMMa}, 
and with the help of
Eq.~\eqref{secondderivatives} one has that
$S_{MM}S_{QQ} - S_{MQ}^2 \geq 0$
can be written as 
\begin{align}
    S_5 = - x(1-x)S_1S_3 +
    \frac{4 y k^2}{(d-3)^2}S_{13}^2\leq 0\,.
    \label{eq:S5cond}
\end{align}
Then, from Eq.~\eqref{eq:Sfunctions} this inequality can be rearranged
as
\begin{align}
   a \leq \frac{(d-3)}{2(d-2)}\,
    \frac{2 - x(1+y)}{1-x}\,,
    \label{eq:SMMQQa}
\end{align}
where Eqs.~\eqref{eq:Sfunctionsauxiliary} and \eqref{xy} have been
used.
The condition given in Eq.~\eqref{eq:SMMQQa}
is sufficient to describe the stability, since the right
hand side of it is lower than the condition
given by Eq.~\eqref{eq:SMMa}, in the respective parameter space. The inequality
is quite simple enough for analytical treatment. The function
set by the right-hand side
at $x=0$
or $y=1$ takes the value $a= \frac{d-3}{d-2}$. At $x=1$, the function
diverges to infinity. Thus, the function bends from
a constant value
to $a = \frac{d-3}{2(d-2)}\,\frac{2-x}{1-x}$, going from $y=1$ to
$y=0$.
A detailed analysis of Eq.~\eqref{eq:SMMQQa} can be seen in
Fig.~\ref{fig:S5} which is itself split into four plots (a), (b),
(c), and (d).  The case of the shell with thermodynamic black hole
features, i.e., the case with $a=1$, shows that increasing the value
of $y$ will decrease the range of $x$ for thermodynamic
stable configurations, i.e., if the shell has more
electric charge, then
it must
have lower radius $R$ for stability, see also Fig.~\ref{fig:S1}(d) for
this $a=1$ case.

\subsection{Area and charge fluctuations together}
\label{areaandchargefluctogether}

A shell with area and charge fluctuations will have the stability
conditions given by $S_{AA} \leq 0$, $S_{QQ} \leq 0$, and
$S_{AA}S_{QQ} - S_{AQ}^2 \geq 0$, see Eq.~\eqref{eq:StabCond}.
Note, however, that there is redundancy on this system of
inequations, see Appendix~\ref{app:Stability}.  The sufficient
conditions can be chosen to be $S_{AA}\leq 0$ and $S_{AA}S_{QQ} -
S_{AQ}^2 \geq 0$.
For the equations of state we are using,
one has that $S_{AA} \leq 0$ yields 
Eq.~\eqref{eq:SAAa}, 
and with the help of
Eq.~\eqref{secondderivatives} one has that
$S_{AA}S_{QQ} - S_{AQ}^2 \geq 0$
can be written as 
\begin{align}
    S_6 = -\frac{(1-x)}{2(d-3)}S_2S_3 + xy S_{23}\leq 0\,.
    \label{eq:S6cond}
\end{align}
Then, from Eq.~\eqref{eq:Sfunctions} this inequality can be rearranged
as
\begin{align}
&a \leq \frac{\frac{(1-x)\mathcal{F}(2d-5)}{2(d-3)}(1+3y) -
x^3 y(1-y) + 2 \mathcal{F}xy -
\frac{y \mathcal{F}^2}{x(1-y)}}{(d-2)(1-x)
\Big(\frac{\mathcal{F}^2}{2 x (d-3)} +
\frac{2d -5}{(d-3)^2}y(1-x)\mathcal{F} +
\frac{2 \mathcal{F} x y}{(d-3)} \Big)} \nonumber\\&+
\frac{(d-3)}{(d-2)}\,\frac{1+y}{1-y}\,.
\label{eq:SAAQQa}
\end{align}
where Eqs.~\eqref{eq:Sfunctionsauxiliary} and \eqref{xy} have been
used.
The condition given in Eq.~\eqref{eq:SAAQQa}
is sufficient to describe the stability, since the right
hand side of it is lower than the condition
given by Eq.~\eqref{eq:SAAa}, in the respective parameter space.
At $y=0$, the function set by the right-hand side intersects
$S_2$. The function then grows without bound at $(x=0,y=0)$ or
$y=1$. In the limit of $x\rightarrow 1$, the function approaches the
value of $a = \frac{8 + 6y - 3d (1+y)}{(d-2)(1+3y)}$. At $x=0$, the
right-hand side approaches $S_3$ from below.  A detailed analysis of
Eq.~\eqref{eq:SAAQQa}
can be seen in Fig.~\ref{fig:S6} which is itself
split into three plots (a), (b), and (c).  The case of the shell
with thermodynamic black hole features, i.e., the case with $a=1$,
does not need a more detailed analysis since all the configurations
with $a = 1$ are below the surface of marginal stability, therefore
they are stable to these thermodynamic perturbations.

\subsection{Mass, area, and charge fluctuations altogether}
\label{massareaandchargeflucaltogether}

A shell with mass, area, and charge fluctuations will have the
stability conditions given by all the inequalities in
Eq.~\eqref{eq:StabCond}.
Note, however, that there is redundancy on this system of
inequations, see Appendix~\ref{app:Stability}.  The sufficient
conditions can be chosen to be $S_{MM} \leq 0$, $S_{MM}S_{AA} -
S_{MA}^2 \geq 0$, and $ (S_{MM}S_{AQ} - S_{MA}S_{MQ})^2 -
(S_{AA}S_{MM} - S_{AM}^2)(S_{QQ}S_{MM} - S_{QM}^2 )\leq 0$.
For the equations of state we are using,
one has that $S_{MM} \leq 0$ yields 
Eq.~\eqref{eq:SMMa},
$S_{MM}S_{AA} - S_{MA}^2 \geq 0$
yields 
Eq.~\eqref{eq:SMMAAa},
and with the help of
Eq.~\eqref{secondderivatives} one has that
$
 (S_{MM}S_{AQ} - S_{MA}S_{MQ})^2 - (S_{AA}S_{MM} -
    S_{AM}^2)(S_{QQ}S_{MM} - S_{QM}^2 )\leq 0$
can be written as 
\begin{align}
    S_7 = \left(x S_1 S_{23} -
    \frac{2 k}{d-3}S_{12}S_{13}\right)^2y -
    S_4 S_5 \leq 0\,.
    \label{eq:S7cond}
\end{align}
Then, even though Eq.~\eqref{eq:S7cond} appears to be a polynomial on $a$ of
degree 4,
from Eq.~\eqref{eq:Sfunctions} this inequality can be rearranged
as
\begin{align}
    a\leq \frac{d-3}{d-2}\left(
    \frac{4 - 4 k + x^2 (d(1-y)^2 +
    C)}{4 - 4k + x^2 d (1-y)^2 + x D}\right)
    \,,\label{eq:SMMAAQQa}
\end{align}
where
$C = 2
    x(1+y) (k-2) - 2 -2(y-4)y$,
and
$D = 4 k - 2y - 6 - x
    \left( 1 + y (3y-8)\right)$,
    and
Eqs.~\eqref{eq:Sfunctionsauxiliary} and \eqref{xy} have been
used.
The right-hand side of Eq.~\eqref{eq:SMMAAQQa}
when compared with the conditions in Eqs.~\eqref{eq:SMMa}
and~\eqref{eq:SMMAAa}
assumes always lower values, in the respective parameter space,
therefore Eq.~\eqref{eq:SMMAAQQa} is the sufficient condition of
stability.
The equality in the condition given in Eq.~\eqref{eq:SMMAAQQa} has its
lowest value of $a = \frac{d-3}{d-2}$ at $x=0$, for every $y$. It then
increases toward $x=1$, where the limit gives $a=1$. At the limit of
$y=1$, the equality is given by the lowest value of $a =
\frac{d-3}{d-2}$ for every $x$ except $x=1$ where the limit gives $a =
1$.  Thus, the condition for stability in Eq.~\eqref{eq:SMMAAQQa}
implies that every configuration with $a\leq \frac{d-3}{d-2}$ is
stable.  On the other hand, for $\frac{d-3}{d-2} < a < 1$ the
stability region decreases with increasing $y$, being zero in the
limit of $y=1$. This means that shells with more electric charge will
have less configurations of stability.
The space of stable
configurations in the $a-d$ plane can also be made and is similar to
the analysis made for the uncharged case in~\cite{Andre:2019}.  A
detailed analysis of Eq.~\eqref{eq:SMMAAQQa} can be seen in
Fig.~\ref{fig:S7} which is itself split into three plots (a), (b), and
(c).  The case of the shell with thermodynamic black hole features,
i.e., the case with $a=1$, does not need a more detailed analysis
since all the configurations with $a = 1$ are above the surface of
marginal stability, hence unstable, except for the points with $x = 1$
which lie on the limit of the surface, hence marginally stable.  In
the black hole limit, i.e., not only $a=1$ but also $x=1$, the
configurations for every value of $y$ are marginally stable.

\subsection{Further comments on the behavior of intrinsic stability
with $a$}
\label{furthercomments}

We now make some important comments, leftovers from the previous
sections.

When discussing mass fluctuations only, Sec.~\ref{massfluconly}, we
mentioned that one can make a corresponding stability analysis in
terms of the variables $\frac{M}{R^{d-3}}$ and $\frac{Q}{R^{d-3}}$,
instead of $x=\frac{r_+^{d-3}}{R^{d-3}}$ and
$y=\frac{r_-^{d-3}}{r_+^{d-3}}$ of Eq.~\eqref{xy}.  The analysis in
$\frac{M}{R^{d-3}}$ and $\frac{Q}{R^{d-3}}$ yields some interesting
insight.  The condition given in Eq.~\eqref{eq:SMMa} becomes
\begin{align} &a \leq
    \frac{\mu}2\frac{(d-3)}{(d-2)}\frac{\left(\frac{Q^2}{R^{2(d-3)}} +
    \frac{\mu M^2}{R^{2(d-3)}} - \frac{2M}{R^{d-3}}\right)}{\left(1 -
    \frac{\mu M}{R^{d-3}}\right)^2 }\nonumber \\
    &\times
    \frac{\left(\frac{2\mu M}{R^{d-3}} + \frac{\mu (Q^2 -
     M^2 \mu^2)}{R^{2(d-3)}} -2\right)}
     {\sqrt{ \frac{\mu(\mu M^2 -
     Q^2)}{R^{2(d-3)}} \left(
     \left(2 - \frac{\mu M}{R^{d-3}}\right)^2 -\frac{\mu Q^2}
     {R^{2(d-3)}}\right) }}\,.
     \label{eq:Sa1MQ}
\end{align}
The possible physical values of $\frac{M}{R^{d-3}}$ and
$\frac{Q}{R^{d-3}}$ are restricted by the condition of subextremality,
i.e., $\sqrt{\mu} M > Q$, and the condition of no trapped surfaces,
i.e., $\frac{r_+}{R}<1$.  One finds from Eq.~\eqref{eq:Sa1MQ} that for
small values of $\frac{M}{R^{d-3}}$, the shell needs some minimum
charge $Q$, or correspondingly a minimum value of $\frac{Q}{R^{d-3}}$,
for it to be stable.  When $\frac{M}{R^{d-3}}$ has a value that
corresponds to $x=\frac{2}{d-1}$, the minimum charge for the shell to
be stable reaches zero, which means $y=0$.  For higher
$\frac{M}{R^{d-3}}$, the region of stability is restricted by the
physically possible values, namely, $\frac{\sqrt{\mu} M}{R^{d-3}} >
\frac{Q}{R^{d-3}}$ and $\frac{r_+}{R}<1$.  Thus, in brief, for
$\frac{M}{R^{d-3}}$ small, thermodynamic stability exists only for
sufficiently large electric charge.  For $\frac{M}{R^{d-3}}$ having a
value such that $x=\frac{2}{d-1}$ when $y=0$, i.e., $Q=0$, the shell
is marginally stable.  Here, it is important to note that the value of
$x=\frac{2}{d-1}$ for $Q=0$ means that the shell is at the photonic
orbit .  For higher values of $x$, maintaining $Q=0$, the shell is
inside the photonic orbit, and it is stable. This means that for
$\frac{M}{R^{d-3}}$ yielding values higher than $x=\frac{2}{d-1}$ when
$y=0$, i.e, $Q=0$, the shell is thermodynamically stable, see also
\cite{Andre:2019}.  This latter behavior, i.e., the behavior for
$\frac{M}{R^{d-3}}$ yielding values of $x$ equal or higher than
$x=\frac{2}{d-1}$, is precisely the same behavior of the large black
hole in the canonical ensemble found by York \cite{York:1986,can} and
generalized to $d$ dimensions in \cite{andrelemos1,andrelemos2}.  For
higher values of $\frac{M}{R^{d-3}}$ increasing the electrically
charge $Q$, and so essentially increasing $\frac{Q}{R^{d-3}}$, does
not alter the stability, the solutions are all thermodynamically
stable.  The result can be interpreted heuristically. To understand
it, note that the reduced inverse temperature $b$, can be envisaged as
a length scale, a thermal one. The inverse
temperature $b$ here is the one given in
Eq.~\eqref{eq:tempeqstate}.  For small $\frac{M}{R^{d-3}}$ and $Q=0$
one has
that the shells have radii higher than the photonic orbit
and are thermodynamically unstable. What happens is that the
thermal length $b$ being proportional to $M$, still in the uncharged
case, is smaller than
or of the order of the radius of the shell, and thus the
shell loses energy and mass along these thermal lengths.  Losing mass, means
that the thermal length $b$ decreases, and so the process is a runaway
process and thus unstable.  If the charge $Q$ increases,
or more correctly if the ratio
$\frac{Q}{R^{d-3}}$ increases, the thermal length $b$ gets
correspondingly higher, and for a certain sufficiently high electric
charge $Q$, or better for a sufficiently high
$\frac{Q}{R^{d-3}}$,
$b$ is now sufficiently greater than the radius of the shell,
so that it is not possible to lose energy anymore. Thus, the electric
shell is stable for charges higher than this minimum electric
charge. For higher electric charge, i.e., higher 
$\frac{Q}{R^{d-3}}$, such that one is
near the extremal limit, one has that $b$ is
proportional to $\frac{1}{\sqrt{M-Q}}$ and so it is indeed divergingly
larger than $R$.  For $Q=0$ and a value of $\frac{M}{R^{d-3}}$ such
that $x=\frac{2}{d-1}$, one has a shell with radius equal to the
photonic orbit. In this case the thermal length $b$, as the
calculations show, is barely sufficiently to not allow thermal loss
from the shell, and so maintain the shell in thermodynamic
equilibrium.  For higher $\frac{M}{R^{d-3}}$ and $Q=0$,
the shell is inside the photonic orbit, and the thermal
length $b$ is now sufficiently large relatively to the radius of the
shell to not allow thermal loss from the shell, and this holds even
truer for higher $Q$, i.e., higher
$\frac{Q}{R^{d-3}}$, 
where $b$ gets even larger and thus the shell in all these cases is
thermodynamically stable. The comments made here for generic
dimensions $d$, apply to the $d=4$ electric charged case studied in
\cite{Lemos:2015a} and are exemplified for $d=5$ in
Fig.~\ref{fig:S1}(d) bottom.

Another important point, is that in
Sec.~\ref{massareaandchargeflucaltogether} we have pointed out that
for mass, area, and charge fluctuations altogether, shells with more
electric charge will have less configurations of stability.  This
behavior differs from the case of mass fluctuations only of
Sec.~\ref{massfluconly} that was also commented in the previous
paragraph, where, for certain configurations, more electric charge
aids to the stability.  There is no contradiction between the two
cases.  The mass, area, and charge fluctuations altogether is much
more restrictive than the mass fluctuations only case, in the sense
that stable points in the former fluctuations are also stable points
in the latter fluctuations, but the converse is not true.

There is still another point worth noting.  In the case of one or two
fixed quantities,
Secs.~\ref{massfluconly}-\ref{areaandchargefluctogether}, there are
shell configurations with $a\geq 1$ that are stable.  But one notices
that the higher the $a$ the higher the entropy $S$ since it goes with
a power of $a$.  For instance, for the area fluctuations only of
Sec.~\ref{areafluconly}, we have seen that Eq.~\eqref{eq:SAAa} has its
minimum at $\frac{r_+}{R}=1$ for zero charge, i.e., $(x=1,y=0)$, with
the value for $a$ given by $a = 3- \frac{2}{d-2}$.  Since values of $a
= 3 - \frac{2}{d-2}$ are always greater than one, this could mean that
a shell with lower $a$ would tend to settle into a shell with higher
$a$ since the latter would have higher entropy.  One could think that
a change of $a$ could be achieved by some rearrangement of the
material on the shells and in this way higher entropies could be
attained.  However, the stability analysis performed is for fixed $a$,
since the very exponent $a$ gives a precise temperature equation of
state for the matter, and to treat changes in the exponent $a$ one
would have perhaps to envisage some type of phase transition.

Having worked out the thermodynamic stability criterion for all types
of fluctuation in
Secs.~\ref{massfluconly}-\ref{massareaandchargeflucaltogether} through
the parameter $a$, it begs now the question of what is the physical
reason for the behavior of the intrinsic stability with $a$ itself. We
now turn into this point.

\section{Intrinsic thermodynamic
stability in laboratory variables}
\label{sec:Intrinsic}

\subsection{The rational to introduce laboratory variables}

It begs now the question of
what is the physical reason for the behavior of intrinsic stability
with $a$.
In order to understand the physical meaning of the intrinsic
thermodynamic stability associated to this self-gravitating thin
shell, we rewrite the stability conditions with variables that can be
measured in the laboratory. One of these variables
that we are going to
define gives a good example of the way the stability conditions get
clearer when written in terms of thermodynamic coefficients. The heat
capacity at constant area and charge, $C_{A,Q}$, can be defined as
$C_{A,Q}^{-1} = \left(\frac{\partial T}{\partial M}\right)$. This
variable is important since we also have
$ S_{MM} = -\beta^2 C_{A,Q}^{-1} \leq 0$,
and so the stability condition for changes in
proper mass only is that the
heat capacity $C_{A,Q}$ is positive. The aim is to
generalize this reasoning for two types of fluctuations
which seem the most
interesting, namely, the mass and charge fluctuations studied in
Sec.~\ref{massandchargefluctogether}, and mass, area, and charge
fluctuations studied in Sec.~\ref{massareaandchargeflucaltogether}.

\subsection{Laboratory variables for mass and charge fluctuations
together}
\label{sec:LvarMassCharge}

Here we discuss the new laboratory variables for mass and charge
fluctuations, see Sec.~\ref{massandchargefluctogether}.
It emerges that the heat capacities at fixed area
play an important role
when treating mass and charge
fluctuations together. There are two
such heat capacities, namely, the heat capacity
at constant area and electric charge
$C_{A,Q}$, and the heat capacity
at constant area and electric potential $C_{A,\Phi}$.

The three equations of state
$T(M,A,Q)$, $p(M,A,Q)$,
and $\Phi(M,A,Q)$, given
in Eqs.~\eqref{eq:tempeqstatestability}-\eqref{eq:Phistability}
are to be rewritten
in laboratory variables, which are also called
thermodynamic coefficients.
In fact, for
mass and charge fluctuations, 
one only
needs
two equations of state, the ones
for the temperature 
$T(M,A,Q)$ and for the
thermodynamic electric 
potential
$\Phi(M,A,Q)$.
Since the area is kept fixed here we do not need to use
the equation of state for the pressure, $p(M,A,Q)$. 
The equations for 
$T(M,A,Q)$ and
$\Phi(M,A,Q)$
will allow us to establish the stability
conditions  for mass and charge
fluctuations in the new variables.

For the 
equation of state for temperature,
$T(M,A,Q)$,
we want to define the laboratory variables in terms
of the derivatives of $S(T,A,Q)$.  For that, note that one is able to
write the differential $dS(T,A,Q)$ as
$dS
= \left(\frac{\partial S}{\partial T}\right)_{A,Q}dT
+\left(\frac{\partial S}{\partial A} \right)_{T,Q}dA+
\left(\frac{\partial
    S}{\partial Q}\right)_{T,A}dQ
$. Now, the heat capacity $C_{A,Q}$ is defined as $C_{A,Q}^{-1} =
\left(\frac{\partial T}{\partial M}\right)_{A,Q}$ which is equivalent
to $
\frac1T C_{A,Q}=\left(\frac{\partial S}{\partial T}\right)_{A,Q}$.
The latent heat capacity at constant temperature and charge,
$\lambda_{T,Q}$, is defined as $\lambda_{T,Q}=\left(\frac{\partial
S}{\partial A} \right)_{T,Q}$.  The latent heat capacity at constant
temperature and area, $\lambda_{T,A}$, is defined as
$\lambda_{T,A}=\left(\frac{\partial S}{\partial Q}\right)_{T,A} $.
One can then change the equality for $dS$ into an equality for $dT$,
such that $T=T(S,A,Q)$,
to obtain
$dT = \frac{T}{C_{A,Q}}dS -
    T\frac{\lambda_{T,Q}}{C_{A,Q}} dA
    - \frac{T \lambda_{T,A}}{C_{A,Q}} dQ$.
One can now use the first law,
Eq.~\eqref{1stlaw1}, i.e., $TdS=dM+pdA-\Phi dQ$, substitute for the
variation of the
entropy $dS$ above and put
the equation found as an equality for $dT$, namely,
\begin{align}
    dT = \frac{1}{C_{A,Q}}dM -
    \frac{T\lambda_{T,Q} - p}{C_{A,Q}} dA-
    \frac{T \lambda_{T,A}+\Phi}{C_{A,Q}} dQ\,,
    \label{eq:dTMassCharge}
\end{align}
So, $dT$ is written in terms of the laboratory variables,
namely, the
heat capacity $C_{A,Q}$, the latent heat capacity at constant
temperature and charge $\lambda_{T,Q}$,
and the latent heat capacity
at constant temperature and area $\lambda_{T,A}$.
For the 
equation of state for the
thermodynamic electric 
potential
$\Phi(M,A,Q)$, we define the laboratory variables with
respect to $\Phi(S,A,Q)$ so that, with the aid of $S(M,A,Q)$, we
obtain $\Phi(M,A,Q)$.  Now, note that one is able to write the
differential $d\Phi(S,A,Q)$ as
$d\Phi
= \left(\frac{\partial \Phi}{\partial S}\right)_{A,Q}dS
+\left(\frac{\partial \Phi}{\partial A} \right)_{S,Q}dA+
\left(\frac{\partial
    \Phi}{\partial Q}\right)_{S,A}dQ
$. We
define the
adiabatic electric susceptibility, $\chi_{S,A}$, as
$\frac{1}{\chi_{S,A}}= \left(\frac{\partial \Phi}{\partial Q} \right)_{S,A}
$, and define the electric pressure at constant entropy and charge,
$P_{S,Q}$, as $P_{S,Q} =\left( \frac{\partial \Phi}{\partial
A}\right)_{S,Q} $.  The remaining derivative of $\Phi$ is given by the
Maxwell relation $\left(\frac{\partial \Phi}{\partial S}\right)_{A,Q}
= \left(\frac{\partial T}{\partial Q} \right)_{S,A} = - \frac{T
\lambda_{T,A}}{C_{A,Q}}$, which was calculated using the definitions
given above for $C_{A,Q}$, 
$\lambda_{T,Q}$, and $\lambda_{T,A}$, and
swapping the equality for $dS(T,A,Q)$ into
an equality for $d\Phi(S,A,Q)$, i.e., 
$d\Phi = - \frac{T \lambda_{T,A}}{C_{A,Q}} dS + P_{S,Q}
    dA +
    \frac{1}{\chi_{S,A}}dQ$.
One can now use the first law, Eq.~\eqref{1stlaw1},
i.e., $TdS=dM+pdA-\Phi dQ$, substitute for the
variation of the
entropy
$dS$ above and the equation found
as an 
equality for $d\Phi$, namely,
\begin{align}
&d\Phi = - \frac{\lambda_{T,A}}{C_{A,Q}} dM +
\left(
P_{S,Q}
-
p\frac{\lambda_{T,A}}{C_{A,Q}}\right)dA
\nonumber\\&
+
\left(\frac{1}{\chi_{S,A}} + \frac{\Phi
\lambda_{T,A}}{C_{A,Q}}\right)dQ 
\,.
\label{eq:dphiMassCharge}
\end{align}
So, $d\Phi$ is written in terms of the laboratory variables, namely,
the heat capacity, $C_{A,Q}$, the latent heat capacity at constant
temperature and area, $\lambda_{T,A}$, the electric pressure at
constant entropy and charge, $P_{S,Q}$, and the
adiabatic electric
susceptibility, $\chi_{S,A}$.  Finally, it is useful to define also
the heat capacity at constant area and constant electric potential,
$C_{A,\Phi}$, defined by $C_{A,\Phi} = T \left(\frac{\partial
S}{\partial T}\right)_{A,\Phi}$ which can be written in terms of the
coefficients in Eqs.~\eqref{eq:dTMassCharge}
and~\eqref{eq:dphiMassCharge} as $ C_{A,\Phi}= C_{A,Q} \left(1 -
\frac{T \lambda_{T,A}^2}{C_{A,Q}}\chi_{S,A} \right)^{\hskip -0.15cm
-1}$.  

The intrinsic thermodynamic stability of a thin shell for mass and
charge fluctuations together can be determined by considering the
two sufficient stability
conditions which
can be taken from Eq.~\eqref{eq:StabCond}, yielding
$S_{MM}\leq 0$ and $S_{MM}S_{QQ} - S_{MQ}^2 \geq 0$.
The first condition is almost
immediate since with the definition in Eq.~\eqref{eq:dTMassCharge} we
obtain $S_{MM} = -\beta^2 \frac1{C_{A,Q}}$ and so
it implies
that $C_{A,Q} \geq 0$.
The second condition
requires some more care and some 
more algebra, yielding in the end
$S_{MM}S_{QQ} - S_{MQ}^2 = \
\beta^2\frac{1}{C_{A,\Phi}\chi_{S,A}}$,
 and so
it implies
$C_{A,\Phi}
\,\,\chi_{S,A} \geq 0$.
The stability conditions
in the laboratory variables
can then be written as
\begin{align}\label{eq:LabMQ} 
& C_{A,Q} \geq 0 \,,\nonumber\\
& C_{A,\Phi}
\,\,\chi_{S,A} \geq 0\,.
\end{align}

For the specific equations of state we used,
Eqs.~\eqref{eq:tempeqstatestability}-\eqref{eq:Phistability}, the
coefficient $\chi_{S,A}$ is always positive. Hence, from the two
equations given in Eq.~\eqref{eq:LabMQ} , the stability conditions
become $C_{A,Q} \geq 0$ and $C_{A,\Phi} \geq 0$, respectively.
Moreover, we have found in Sec.~\ref{massandchargefluctogether} that
for these equations of state the condition $S_{MM}S_{QQ} - S_{MQ}^2
\geq 0$ is the sufficient condition for stability. Therefore, the thin
shell considered
is thermodynamic stable for mass and charge fluctuations
together if
\begin{align}\label{eq:SMMQQa2} 
C_{A,\Phi} \geq 0 \,.
\end{align}
This occurs precisely when Eq.~\eqref{eq:SMMQQa} is satisfied, 
i.e., Eq.~\eqref{eq:SMMQQa2} is equivalent to
Eq.~\eqref{eq:SMMQQa},
as for a thin shell with the equations of
state given in
Eqs.~\eqref{eq:tempeqstatestability}-\eqref{eq:Phistability}.  Note
that, when there is equality in Eq.~\eqref{eq:SMMQQa}, one must be
careful in regard to the value of the heat capacity at constant area
and constant electric potential, $C_{A,\Phi}$. If one performs the
limit to the equality as a succession of stable configurations, by
starting from a configuration with the exponent $a$ satisfying
Eq.~\eqref{eq:SMMQQa2} and then increasing $a$, the coefficient
$C_{A,\Phi}$ becomes infinite and positive. If, on the contrary, one
performs the limit to the equality as a succession of unstable
configurations, by starting from a configuration with the exponent $a$
not satisfying Eq.~\eqref{eq:SMMQQa2} and then decreasing $a$, the
coefficient $C_{A,\Phi}$ becomes infinite and negative.

The details of all the calculations presented in this
section are shown in
Appendix~\ref{app:masscharge}.

\subsection{Laboratory variables for mass, area, and
charge fluctuations altogether}
\label{mac}

Here we discuss the new
laboratory variables for mass, area, and
charge fluctuations, see
Sec.~\ref{massareaandchargeflucaltogether}.
The analysis in Sec.~\ref{sec:LvarMassCharge} highlights the
importance of the specific heat capacities $C_{A,Q}$ and $C_{A,\Phi}$
in the intrinsic stability of thermodynamic systems with fixed
area. Nevertheless, there are other important quantities playing a
role in the intrinsic stability for the case of mass, area and charge
fluctuations. In this case
the important laboratory quantities 
are the heat capacity
at constant area and electric charge
$C_{A,Q}$ again,
the expansion coefficient at constant temperature
and electric charge
$\kappa_{T,Q}$,
and the electric susceptibility  at constant pressure
and temperature
$\chi_{p,T}$.

The three equations of state
$T(M,A,Q)$, $p(M,A,Q)$,
and $\Phi(M,A,Q)$ are to be rewritten
    in laboratory variables.
This will allow us to establish the
stability conditions in these new
variables for mass, area, and charge fluctuations, and so
no fixed quantities.
We now want to
define the laboratory variables in terms of the derivatives of
$S(T,p,Q)$, $A(T,p,Q)$, and $\Phi(T,p,Q)$, for convenience, since these
variables will simplify the considered stability conditions. Notice
that the three functions $S(T,p,Q)$, $A(T,p,Q)$ and $\Phi(T,p,Q)$ are
the derivatives of the Gibbs potential, i.e., $dG = - SdT + Adp + \Phi
dQ$.
Let us start with the area $dA$
for which the coefficients have
a direct physical meaning.
We write
$dA=
\left(\frac{\partial A}{\partial T}\right)_{p,Q}dT+
\left( \frac{\partial A}{\partial p}\right)_{T,Q}dp+
\left( \frac{\partial A}{\partial Q}\right)_{T,p} dQ$,
such that 
$\alpha_{p,Q}=\frac1A\left(\frac{\partial A}{\partial T}\right)_{p,Q} 
$ is the expansion coefficient, 
$\kappa_{T,Q}=-\frac1A
\left( \frac{\partial A}{\partial p}\right)_{T,Q}$
is the isothermal compressibility,
and 
$\kappa_{p,T}=-\frac1A
\left( \frac{\partial A}{\partial Q}\right)_{T,p}
$
is the electric
compressibility.
Now, $dS$ is written as
$dS=
\left(\frac{\partial S}{\partial T}\right)_{p,Q}dT+
\left( \frac{\partial S}{\partial p}\right)_{T,Q}dp+
\left( \frac{\partial S}{\partial Q}\right)_{T,p} dQ$,
where $\left(\frac{\partial S}{\partial T}\right)_{p,Q}$
can be written in terms of previously defined
coefficients, specifically, 
$\left(\frac{\partial S}{\partial T}\right)_{p,Q} =
    \frac{C_{A,Q}}{T} + A \frac{\alpha_{p,Q}^2}{\kappa_{T,Q}}$,
$\left(\frac{\partial S}{\partial p}\right)_{T,Q} $
can be written
using the Maxwell relation $\left(\frac{\partial S}{\partial
p}\right)_{T,Q} = -\left(\frac{\partial A}{\partial T}\right)_{T,Q}
=
 A\alpha_{p,Q}
$, and
$\lambda_{p,T}=
  \left(\frac{\partial S}{\partial Q}\right)_{p,T}$
  is a new coefficient, the latent heat capacity.
Finally, $d\Phi$ is written as
$d\Phi=
\left(\frac{\partial \Phi}{\partial T}\right)_{p,Q}dT+
\left( \frac{\partial \Phi}{\partial p}\right)_{T,Q}dp+
\left( \frac{\partial \Phi}{\partial Q}\right)_{T,p} dQ$,
where
two of the derivatives are written using Maxwell relations, i.e.,
$\left(\frac{\partial \Phi}{\partial T}\right)_{p,Q} =
-\left(\frac{\partial S}{\partial Q}\right)_{p,T}=\lambda_{p,T}$
and
$\left(\frac{\partial \Phi}{\partial p}\right)_{T,Q}
= \left(\frac{\partial A}{\partial Q}\right)_{p,T}
=-A\kappa_{p,T}
$ as defined above, 
and  $\frac1{\chi_{p,T}}=
\left(\frac{\partial \Phi}{\partial Q}\right)_{p,T}$
is a new coefficient, the
isothermal electric susceptibility.
With the differentials $dA(T,p,Q)$, $dS(T,p,Q)$ defined
above in terms of physical coefficients, we are able
to invert the system composed by these two differentials in order to
obtain $dT(S,A,Q)$ and $dp(S,A,Q)$. Then, using Eq.~\eqref{1stlaw1},
i.e.,
$TdS=dM+pdA-\Phi dQ$, we
are able to obtain the differentials of the two equations of state in
the desired form, i.e., $dT(M,A,Q)$ and $dp(M,A,Q)$. Inserting
$dT(M,A,Q)$ and $dp(M,A,Q)$ into $d\Phi(T,p,Q)$, we find the
differential of the remaining equation of state, $d\Phi(M,A,Q)$.
Thus, these
differentials are written in terms of the defined laboratory variables
and the differentials of $dM$, $dA$, and $dQ$,
as
\begin{align}
    dT =& \frac{dM}{C_{A,Q}} + \left(\frac{p}{C_{A,Q}} -
    T\frac{\alpha_{p,Q}}{C_{A,Q}\kappa_{T,Q}}\right)dA \nonumber\\&-
    \left(\frac{\Phi}{C_{A,Q}} + T\frac{\lambda_{p,T}}{C_{A,Q}} + A
    \frac{\alpha_{p,T} \kappa_{p,T}}{\kappa_{T,Q} C_{A,Q}}
    \right)dQ\,,\label{eq:dTMAQ}
\end{align}
\begin{align}
    dp =& \frac{\alpha_{p,Q}}{C_{A,Q}\kappa_{T,Q}} dM\nonumber\\& -
    \left[ \frac{1}{A \kappa_{T,Q}} - \frac{\alpha_{p,Q}}{C_{A,Q}
    \kappa_{T,Q}}\left( p - T \frac{\alpha_{p,Q}}{\kappa_{T,Q}}\right)
    \right] dA \nonumber\\& - \left(\frac{\kappa_{p,T}}{\kappa_{T,Q}}
    - \frac{\alpha_{p,Q}}{C_{A,Q} \kappa_{T,Q}}\mathcal{C}\right)dQ\,,
    \label{eq:dpMAQ}
\end{align}
\begin{align}
    d\Phi = & - \mathcal{B} dM +
    \left[\frac{\kappa_{p,T}}{\kappa_{T,Q}} - \left(p -
    T\frac{\alpha_{p,Q}}{\kappa_{T,Q}}\right)\mathcal{B}
    \right]dA\nonumber\\& + \left(\mathcal{B}\mathcal{C} +
    \frac{1}{\chi_{p,T}} + A\frac{\kappa_{p,T}^2}{\kappa_{T,Q}}
    \right)dQ\,,\label{eq:dPhiMAQ}
\end{align}
where $\mathcal{B}$ is defined as
$\mathcal{B} = A\frac{\kappa_{p,T} \alpha_{p,Q}}{C_{A,Q}
\kappa_{T,Q}} + \frac{\lambda_{p,T}}{C_{A,Q}}$,
and $\mathcal{C}$ is defined as $\mathcal{C} =
T A
\frac{\kappa_{p,T}\alpha_{p,Q}}{\kappa_{T,Q}} + T\lambda_{p,T} +
\Phi$.
With the differentials $dT(M,A,Q)$, $dp(M,A,Q)$,
and $d\Phi(M,A,Q)$ in
Eqs.~\eqref{eq:dTMAQ}-\eqref{eq:dPhiMAQ}, and the first
law of thermodynamics, Eq.~\eqref{1stlaw1}, i.e.,
$TdS=dM+pdA-\Phi dQ$, the second derivatives
of
the entropy that enter into the thermodynamic
stability problem
and are given in Eq.~\eqref{secondderivatives}
can be calculated directly.

The intrinsic thermodynamic stability of a thin shell
for generic mass, area, and charge fluctuations,
is given
by
the three sufficient stability
conditions which
can be taken from Eq.~\eqref{eq:StabCond}, yielding
$S_{MM}\leq 0 $,
$S_{MM}S_{AA} - S_{MA}^2 \geq 0$, and 
$(S_{MM}S_{AQ} - S_{MA}S_{MQ})^2
-(S_{AA}S_{MM} - S_{AM}^2)(S_{QQ}S_{MM} - S_{QM}^2 ) 
\leq 0$.
Now, having the second derivatives of the entropy written in terms of
the laboratory variables defined in this section, one finds that
$S_{MM}\leq 0 $ is equivalent to
$\beta^2\frac{1}{C_{A,Q}} \geq 0$, 
$S_{MM}S_{AA} - S_{MA}^2 \geq 0$ is equivalent to 
 $\beta^3\frac{1}{A \kappa_{T,Q}
    C_{A,Q}} \geq 0$, and 
$(S_{MM}S_{AQ} - S_{MA}S_{MQ})^2
-(S_{AA}S_{MM} - S_{AM}^2)(S_{QQ}S_{MM} - S_{QM}^2 ) 
\leq 0$
is equivalent to 
$\beta^6\frac{1}{A C_{A,Q}^2 \kappa_{T,Q} \,\,\chi_{p,T}} \geq
    0$. 
We then have that the stability conditions
for generic mass, area, and charge fluctuations are given
by the following three equations,
\begin{align}
    &C_{A,Q}\geq 0,\,\nonumber\\
    &\kappa_{T,Q}\geq 0\,,\nonumber\\
    & \chi_{p,T}\geq 0\,,\label{eq:stabcondbetter}
\end{align}
i.e.,
all three laboratory quantities have to be positive,
specifically, 
the heat capacity $C_{A,Q}$ which is related to changes in
temperature, the isothermal compressibility $\kappa_{T,Q}$ which is
related to changes in the pressure, and the
isothermal electric
susceptibility $\chi_{p,T}$ which is related to changes in the
electric charge, have to be positive,
with the case of marginal stability
corresponding to these physical
variables going to infinity.

For the specific equations of state
we use, 
Eqs.~\eqref{eq:tempeqstatestability}-\eqref{eq:Phistability},
for $T$, $p$, and
$\Phi$, respectively, 
one finds that the most
restrictive condition is for $\chi_{p,T}$, 
\begin{align}\label{eq:SMAQlab} 
\chi_{p,T}\geq 0 \,,
\end{align}
so the positivity of the
isothermal
electric susceptibility $\chi_{p,T}$
is the sufficient condition in the case
of the thin shell we are considering, 
with marginal stability happening when this
quantity is infinite,
and with
the unstable configurations
having a negative electric susceptibility and thus  departing from
equilibrium.
Making the connection to
Sec.~\ref{massareaandchargeflucaltogether},
Eq.~\eqref{eq:SMAQlab} is equivalent to Eq.~\eqref{eq:SMMAAQQa}
and one finds that
for
$a<\frac{d-3}{d-2}$
the
isothermal electric
susceptibility $\chi_{p,T}$
will always be positive, for
$\frac{d-3}{d-2}<a<1$
it will be positive for some
values of $(r_+, r_-,R)$, for $a=1$ and $R>r_+$ 
it will be negative with the case $R=r_+$
having to be treated with care, 
and for $a>1$ it will always be negative.
The
shell with black hole features, namely,
$a=1$ and $R=r_+$,
is thermodynamic unstable if
the shell approaches its own gravitational radius $r_+$,
$R\to r_+$, since in this case $\chi_{p,T}
\rightarrow -\infty$.
But there is the possibility, of
having a configuration
with $R=r_+$ that is created
from the start, i.e.,
a configuration not belonging
to a sequence of quasistatic configurations
that has its radius $R$ decreased
up to $r_+$. In this case
the stability depends on
whether  the exponent $a$ of the equation of state
approaches $a=1$ from below
or from above. 
If the exponent $a$ of the equation of state
approaches $a=1$ from below then the
$R=r_+$ configuration is
marginally stable with $\chi_{p,T}
\rightarrow +\infty$, which means that changes in the electric charge
of the configuration
will not have any impact on the electric potential.
If the exponent $a$ of the equation of state
approaches $a=1$ from above then the
$R=r_+$ configuration is unstable, $\chi_{p,T}
\rightarrow -\infty$.
Moreover, in the region of $\frac{d-3}{d-2}<a<1$, shells with more
electric charge show more difficulty in having positive
isothermal
electric susceptibility $\chi_{p,T}$, a property
that can be deduced
from
Fig.~\ref{fig:S7} by the decreasing amount of stable configurations
with electric charge.

The details of all the calculations presented in this
section are shown in
Appendix~\ref{app:ddentropyvariables}.

\section{Conclusions}
\label{conc}

We have used the thin shell formalism to determine the mechanics of a
static charged spherical thin shell in $d$ dimensions in general
relativity and studied
its thermodynamics by
imposing the first law. The fact
that the
rest mass density and so the rest mass behaves
as a thermal quasilocal energy
and that the
pressure is
determined just by general relativity indicates there is a relation
between general relativity and thermodynamics as one equation of state
of the shell becomes fixed. We computed the entropy in terms of the
thermodynamic quantities of the shell, namely, the rest mass $M$, the area
$A$ and the electric charge $Q$ of the shell.  The derivatives of the
entropy are directly related to the temperature and the electric
potential. Indeed, with the first law of thermodynamics and general
relativity alone, we were able to restrict the expressions for the
equations of state for
the temperature and the thermodynamic electric potential.
These equations of state in turn
imply that the entropy $S$ depends solely on the two natural
radius of the
Reissner-Nordstr\"om shell spacetime, the gravitational radius
$r_+$ and the Cauchy radius $r_-$, which in turn
depend on $M$, $A$ and $Q$.
That the entropy $S$ of the $d$ dimensional shell
spacetime does not depend on the radius of the
shell, $R$, is a remarkable fact, which nevertheless
has been found for other thin shell spacetimes.

To calculate
an exact
expression for
the entropy of the shell,
one still needs full expressions for the
temperature and the thermodynamic electric potential equations of
state. We used a power law in $r_+$
with exponent $a$ for the temperature,
and opted for the
characteristic electric potential
of a Reissner-Nordstr\"om shell spacetime,
to obtain that the entropy $S$ of the shell
is proportional to $A_+^a$, where $A_+$ is the gravitational
area corresponding to
$r_+$. Shells with such entropy are of
great interest
as it is possible to obtain the black hole limit and
recover the thermodynamics of black holes.

We have then studied the thermodynamic intrinsic stability of thin
shells with such an entropy equation. The shell is stable if the
Hessian of the entropy is negative semidefinite. We analyzed the
Hessian for seven possible types of fluctuations that can occur in
the shell. Fluctuations of the shell with one free and two fixed
thermodynamic
quantities are of three types, fluctuations of the shell
with two free and one fixed  quantities are also of three
types, and fluctuations of the shell with three free
quantities, i.e.,
no fixed quantities,
are of one type.
The most important and general
type of fluctuations are the ones with no fixed 
quantities.
In the case of our entropy equation,
we have found that only one condition is
sufficient for the shell to be stable. This condition establishes that
for $0<a\leq \frac{d-3}{d-2}$ all the configurations of the shell are
thermodynamic stable, for $\frac{d-3}{d-2}<a<1$ stability depends on
the mass and electric charge, for $a=1$ the configurations are
unstable, unless the shell is at its own gravitational radius, i.e.,
at the black hole limit, in which case it is marginally stable, and
that for $1<a<\infty$ all configurations are unstable.

The physical interpretation for the stability is analyzed through
new
thermodynamic variables that can be measured in the laboratory.  One
finds that, generically, stable shells have positive heat capacity,
positive isothermal compressibility, and positive
isothermal electric
susceptibility.  For the specific equations of state that we gave,
and so for our entropy equation, it
is found that every shell with positive electric susceptibility has
positive heat capacity and positive isothermal compressibility, which
is an interesting property of the shells with the considered entropy.
In addition, marginal stability means that the electric susceptibility
is infinite, which seems to be the case in the black hole
limit. Shells with negative electric susceptibility depart from the
initial state and they should rearrange themselves until the
susceptibility becomes positive or the shell breaks down.

This work has derived some thermodynamic properties for electrically
charged spherical matter shells in higher dimensions and complements a
set of works in the thermodynamics of thin shells. Still, more future
work should be done in the investigation of the link between
thermodynamics and general relativity, and hopefully contribute to the
understanding of black hole physics and with it trying to grasp
gravitation at the tiniest possible scales.

\section*{Acknowledgments}
We thank financial support from Funda\c c\~ao para a Ci\^encia e
Tecnologia - FCT through the project~No.~UIDB/00099/2020.

\appendix

\section{Temperature as a power law in the
ADM mass and corresponding thermodynamic
electric potential as alternative to the
equations of state of Sec.~\ref{sec:firstlaw}}
\label{eos}

The reduced
equations of state for the inverse temperature
and thermodynamic electric potential
given in Eqs.~\eqref{eq:tempeqstate}
and ~\eqref{eq:electeqstate}
of Sec.~\ref{sec:firstlaw} are a good choice
for
reduced equations of state as they yield naturally
the black hole equations of state.
But there are alternatives to these equations of state.

In $d=4$, an alternative was provided
in~\cite{Lemos:2015a}, where the
reduced inverse temperature
was chosen to be
$
    b(r_+,r_-) = 2 a (r_+ + r_-)^\alpha
$,
for some constant $a$ and exponent $\alpha$.
Thus, $b$, which represents
the inverse 
temperature at infinity, is given as a power law depending on
the ADM mass, since
in $d=4$, one has
$r_+ + r_-=2Gm$.
The solution for
the reduced thermodynamic electric potential
$c(r_+,r_-)$
is then 
$c(r_+,r_-) = 2 \gamma
    \frac{ (r_+  r_-)^\delta }
{ (r_+ + r_-)^\alpha}$
for some $\gamma$, $\delta$, and so is a power law
in the charge $Q$ combined with a power law in
$m$~\cite{Lemos:2015a}.
For this $d=4$ case, it is possible analytically to
study its stability.

Motivated by this choice for a reduced
equation of state
for $d=4$, we consider for a
reduced equation of state for the
temperature in generic $d$ dimensions,
the following expression
\vskip -0.5cm
\begin{align}
    b(r_+,r_-) = a (r_+^{d-3} + r_-^{d-3})^\alpha\,,
    \label{choice1}
\end{align}
where $a$ and $\alpha$ are arbitrary real numbers,
and such a choice means that $b$ is indeed a power law
in the $d$-dimensional ADM mass $m$.
With the chosen
reduced
equation of state for the
inverse temperature, the condition in
Eq.~\eqref{eq:inteC} can be used to restrict the form
of $c(r_+,r_-)$,
which is given in this case by
$
\frac{\partial c}{\partial r_-}\frac{r_-}{d-3} -
    \frac{\partial c}{\partial r_+}\frac{r_+}{d-3} =
    \alpha c \frac{r_+^{d-3} - r_-^{d-3}}{r_+^{d-3} +
    r_-^{d-3}}
$.
The general solution for $c(r_+,r_-)$ is
\vskip -0.5cm
\begin{align}
    c(r_+,r_-) = \frac{f(r_+r_-)}{(r_+^{d-3} +
    r_-^{d-3})^\alpha}\,,
 \label{choice2}
 \end{align}
where $f(r_+r_-)$ is an arbitrary function depending only on
the charge of the shell. This generalizes
to $d$ dimensions the $c(r_+,r_-)$
obtained in~\cite{Lemos:2015a}.
The choices for the reduced equations of state
provided, namely,
Eqs.~\eqref{choice1}
and~\eqref{choice2},
permit to treat overcharged shells, i.e.,
shells with $Q>m$.
The thermodynamic stability can be performed, but we will
not do it here.

\section{Thermodynamic stability: Generic and specific
considerations to be added to Sec.~\ref{sec:stability}}
\label{app:Stability}

\subsection{Seminegative definite Hessian}

Here we establish the rules that lead to the stability conditions in
Eq.~\eqref{eq:StabCond} of Sec.~\ref{sec:stability}.
A thermodynamic system is in equilibrium if it reaches a maximum of
entropy, i.e., $dS=0$, and
it obeys
$d^2S \leq 0$.

Let us assume that the system is in the
state $dS=0$.
To understand what
$d^2S \leq 0$ leads to, it is
important to write 
the
Hessian matrix of the entropy, $S_{ij}$, i.e.,
$S_{ij} = \frac{\partial^2 S}{\partial
h_i \partial h_j}$,
$h_i$
being a set of unfixed independent parameters.
Then $d^2S \leq 0$ means that 
$S_{ij}$
has to be semidefinite negative, and so
for any arbitrary vector $\bm{v}$,
one has
\vskip -0.5cm
\begin{equation}
\sum_{ij} S_{ij}v_i v_j \leq 0\,.
    \label{genericc}
\end{equation}

\subsection{1-parameter case}

In the 1-parameter case, the entropy $S$ is a function of one unfixed
parameters $h_1$, i.e., the system is allowed to change very small
amounts in $h_1$. The
thermodynamic
stability condition,
Eq.~\eqref{genericc}, turns into 
$S_{h_1 h_1} v_1^2\leq 0$,
for an arbitrary vector $\bm{v}=(v_1)$, and so
one finds 
\vskip -0.5cm
\begin{align}
    S_{h_1 h_1} \leq 0\,.\label{eq:1paramCond}
\end{align}

\subsection{2-parameter case}

In the 2-parameter case, the entropy $S$ is a function of two unfixed
parameters $h_1$ and $h_2$. Thus, the generic
stability condition,
Eq.~\eqref{genericc}, states
$S_{h_1 h_1} v_1^2 + 2 S_{h_1 h_2} v_1 v_2 + S_{h_2 h_2} v_2^2 \leq
    0$,
for an arbitrary vector $\bm{v}=(v_1,v_2)$.
One can choose the vector $\bm{v} = (v_1,0)$, which
yields $S_{h_1 h_1} v_1^2\leq 0$,
for any $v_1$, and so a 
necessary condition is
\begin{align}
S_{h_1 h_1} \leq 0\,.
\end{align}
One can choose the vector $\bm{v} = (0,v_2)$, which
for any $v_2$, yields $S_{h_2 h_2} v_2^2\leq 0$, and so another 
necessary condition is
\vskip -0.5cm
\begin{align}
S_{h_2 h_2} \leq 0\,.
\end{align}
The third condition comes from completing the square
of the stability condition for this 2-parameter case,
yielding
$\frac{(S_{h_1 h_1} v_1 + S_{h_1 h_2} v_2)^2}{S_{h_1 h_1}} +
    \Big(S_{h_2 h_2} - \frac{S_{h_1 h_2}^2}{S_{h_1 h_1}}\Big)v_2^2
    \leq 0 $.
Thus, since $S_{h_1 h_1} \leq 0$, the third
necessary condition is
\vskip -0.5cm
\begin{align}
    S_{h_1 h_1} S_{h_2 h_2} - S_{h_1 h_2}^2 \geq 0\,.
    \label{eq:2paramCond}
\end{align}

\subsection{3-parameter case}

In the 3-parameter case, the entropy $S$ is a function of three unfixed
parameters $h_1$, $h_2$, and $h_3$. Thus, the generic
stability condition,
Eq.~\eqref{genericc},  states
$ S_{h_1 h_1} v_1^2 + S_{h_2 h_2} v_2^2 + S_{h_3 h_3} v_3^2 + 2
    S_{h_1 h_2} v_1 v_2 + 2 S_{h_1 h_3} v_1 v_3 + 2 S_{h_3
    h_2} v_2 v_3 \leq 0$,
for an arbitrary vector $\bm{v}=(v_1,v_2,v_3)$.
Analogous to the 2-parameter
case above, one can set components of $\bm{v}$
to be zero. This will give the 2-parameter conditions for each pair of
parameters. Thus, one has
\vskip -1.0cm
\begin{align}
    &S_{h_i h_i} \leq 0\,,\\
    &S_{h_i h_i} S_{h_j h_j} - S_{h_i h_j}^2
    \geq 0\,,\,\,i\neq j\,,
\end{align}
for $i,j = 1,2,3$, i.e., there are six conditions. The seventh
condition can be obtained by multiplying
the 
 generic
stability condition
$S_{h_1 h_1}$ and once again completing the square in the following
way,
$
(S_{h_1 h_1} v_1 + S_{h_1 h_2} v_2 + S_{h_1 h_3} v_3)^2 + (S_{h_2
    h_2}S_{h_1 h_1} - S_{h_2 h_1}^2)v_2^2 + (S_{h_3
    h_3}S_{h_1 h_1} - S_{h_1 h_3}^2)v_3^2
    + 2(S_{h_3 h_2}
    S_{h_1 h_1} - S_{h_1 h_2} S_{h_1 h_3})v_3 v_2 \geq 0
$.
Now one can again perform the square completion but it is more simple
to consider the roots of the polynomial $P(v_2,v_3)$ composed by the
second, third and fourth terms. Since $P(v_2,v_3) \geq 0$ and its
second derivatives are positive, the polynomial must have only one
root or no roots.  Thus, the following inequality holds
\vskip -0.5cm
\begin{align}
     &(S_{h_3 h_2} S_{h_1 h_1} - S_{h_1 h_2} S_{h_1 h_3})^2
\notag\\&- (S_{h_2 h_2}S_{h_1 h_1} - S_{h_2 h_1}^2) (S_{h_3 h_3}S_{h_1
h_1} - S_{h_1 h_3}^2)
\leq0
 \,.
 \label{eq:3paramCond}
 \end{align}

\subsection{Redundancy \label{app:redundancy}}

By the method above, it seems one needs 3 conditions for the
2-parameter case and 7 conditions for the 3-parameter case. Since the
number of conditions is higher than the number of eigenvalues of the
Hessian, there are redundant conditions. 
For any specific choice of $h_i$, the sufficient
conditions for the 2-parameter case are
Eq.~\eqref{eq:1paramCond} and Eq.~\eqref{eq:2paramCond}. For the
3-parameter case, the sufficient conditions are
Eq.~\eqref{eq:1paramCond}, Eq.~\eqref{eq:2paramCond} and
Eq.~\eqref{eq:3paramCond}.
Another method to determine
the conditions is to consider only the pivots of the Hessian.

\subsection{The case studied}

There is still
the freedom to choose the order of parameters in the
construction of the Hessian. For example, one can pick $h_1 = M$, $h_2
= A$ and $h_3 = Q$, which has been our choice,
but of course
any permutation of parameters is allowed.

For the static thin shell system we studied,
the first condition for stability, $dS=0$,
imposes that $r_+
= {\rm const}$
case, as
the entropy is proportional to a power of
$r_+$, $S = \frac{\gamma}{16\pi G}A_+^a$,
see Eq.~\eqref{eq:suggestedentropy}, with
$r_+$
itself being a function of $M$, $A$, and $Q$. 
The other condition 
$d^2S \leq 0$ leads to
Eqs.~\eqref{eq:1paramCond}-\eqref{eq:3paramCond}
which are essential the set of equations given in
Eq.~\eqref{eq:StabCond} with the choice $h_1 = M$, $h_2
= A$ and $h_3 = Q$.
The stability analysis itself is then performed
through Eqs.~\eqref{secondderivatives}-\eqref{eq:SMMAAQQa}.

\section{Graphics relevant to Sec.~\ref{sec:stability}}
\label{Graphics}

\begin{figure}[h]
\vskip -0.0cm
\centering
    \begin{subfigure}[h]{0.25\textwidth}
    \centering
    \includegraphics[width=\textwidth]{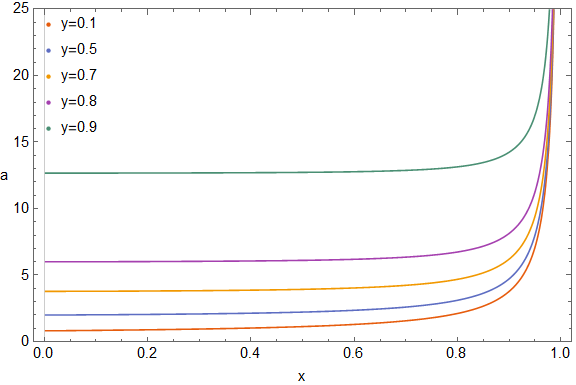}
    \caption{$d=5$, $y$ fixed}
    \label{fig:S1y}
    \end{subfigure}%
    \begin{subfigure}[h]{0.25\textwidth} 
    \centering
    \includegraphics[width=\textwidth]{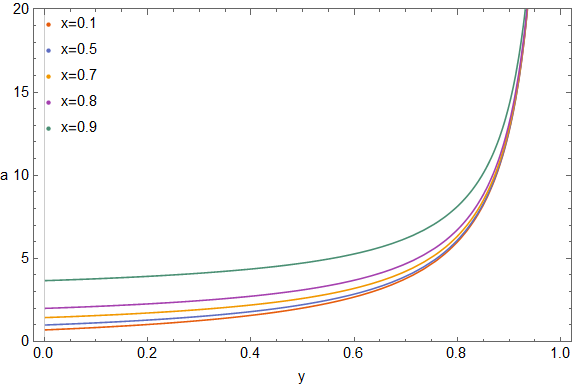}
    \caption{$d=5$, $x$ fixed}
    \label{fig:S1x}
    \end{subfigure}%
    \begin{subfigure}[h]{0.25\textwidth}
    \centering
    \includegraphics[width=\textwidth]{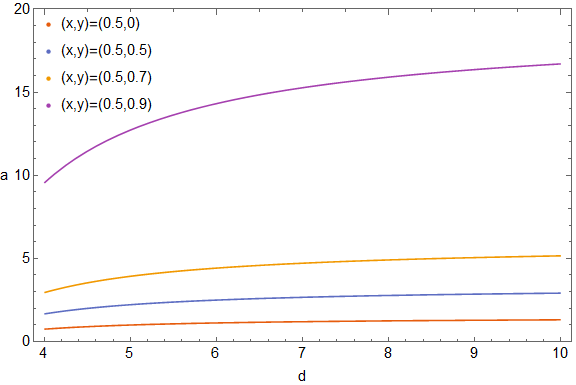}
    \caption{$(x,y)$ fixed}
    \label{fig:S1xy}
    \end{subfigure}
    \begin{subfigure}[h]{0.233\textwidth}
    \centering
    \includegraphics[width=\textwidth]{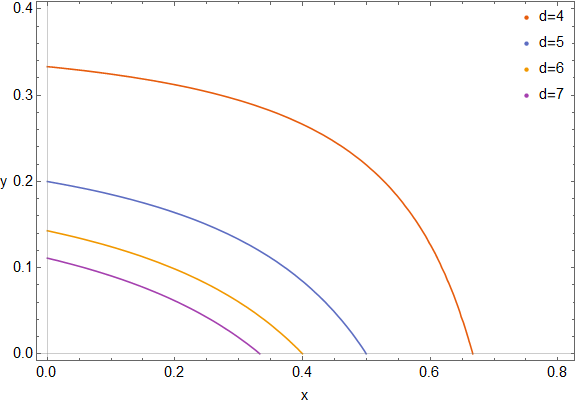}
        \centering
        \includegraphics[width=\textwidth]{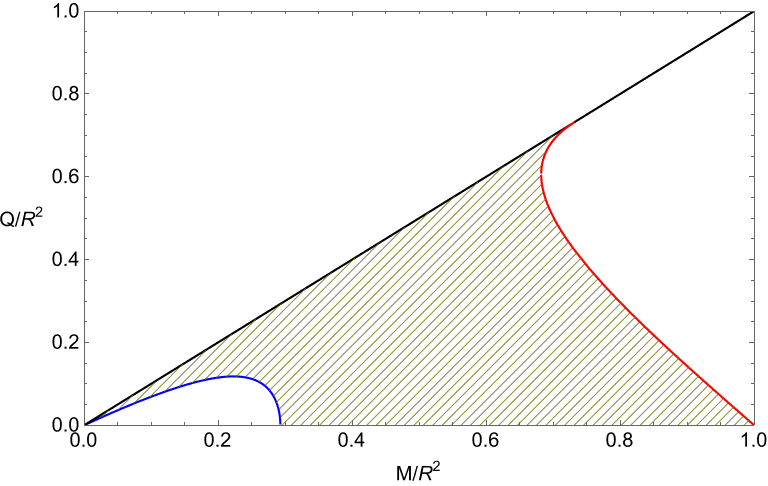}
        \label{fig:S1a1}
        \caption{$a=1$}
        \label{fig:S1a1MQ}
    \end{subfigure}
\caption{\small{
Thermodynamic stability of the shell considering
only mass fluctuations is described, see Sec.~\ref{massfluconly} and
Eqs.~\eqref{eq:S1cond},~\eqref{eq:SMMa} and~\eqref{eq:Sa1MQ}.  Plots
of $S_1\left(d,a,x,y\right) = 0$ with $d$ being the number of
spacetime dimensions, $a$ being the exponent of the equation of state
of the shell, $x=\frac{r_+^{d-3}}{R^{d-3}}$, and
$y=\frac{r_-^{d-3}}{r_+^{d-3}}$, where $r_+$ and $r_-$ are the
gravitational radius and the Cauchy radius of the Reissner-Nordstr\"om
shell spacetime, respectively, and
$R$ is the radius of the shell, are
shown. The equality $S_1\left(d,a,x,y\right) = 0$ describes marginal
stability of the shell considering only mass fluctuations. (a) The
function is plotted as $a(x)$ with $d=5$ and for fixed values of $y$,
with points with lower $a$ than the function corresponding to stable
configurations; (b) The function is plotted as $a(y)$ with $d=5$ and
for fixed values of $x$, with points with lower $a$ corresponding to
stable configurations; (c) The function is plotted as $a(d)$ for fixed
values of the pair $(x,y)$, with points with lower $a$ corresponding
to stable configurations; (d) In the
top plot, for
$a=1$ with fixed values of $d$
the function is plotted as $y(x)$, with configurations with higher $x$
than the curve being stable;
in the bottom plot, one uses the the function
in the form $S_1\left(d=5,a=1,
\frac{M}{R},\frac{Q}{R}\right) = 0$, where $M$ is the rest mass of the
shell and $Q$ is the charge of the shell, see
Sec.~\ref{furthercomments} and Eq.~\eqref{eq:Sa1MQ},
to plot, for $a=1$ with $\mu=1$
and $d=5$, the function
$Q(M)$ in the blue curve, the region of stability in
the yellow line filled area,
$Q = M$ in the black line, and $r_+ = R$
in the red curve.
}}
\label{fig:S1}
\end{figure}

\begin{figure}[h]
\vskip 0.2cm
    \centering
    \begin{subfigure}[h]{0.25\textwidth}
    \centering
    \includegraphics[width=\textwidth]{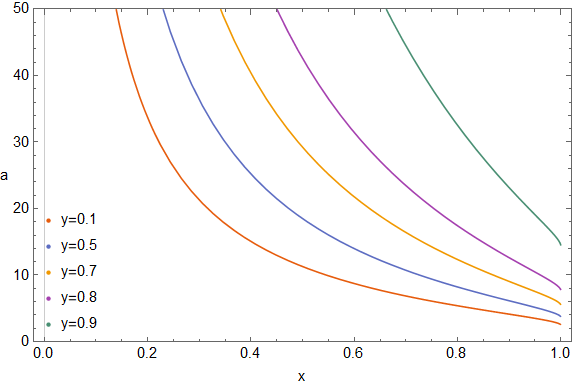}
    \caption{$d=5$, $y$ fixed}
    \label{fig:S2y}
    \end{subfigure}
    \begin{subfigure}[h]{0.25\textwidth}
    \centering
    \includegraphics[width=\textwidth]{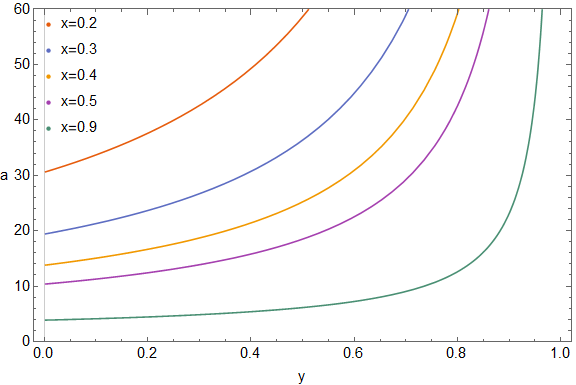}
    \caption{$d=5$, $x$ fixed}
    \label{fig:S2x}
    \end{subfigure}
    \begin{subfigure}[h]{0.25\textwidth}
    \centering
    \includegraphics[width=\textwidth]{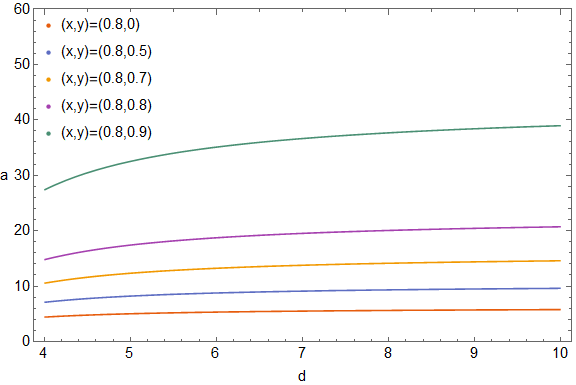}
    \caption{$(x,y)$ fixed}
    \label{fig:S2xy}
    \end{subfigure}
    \caption{\small{
Thermodynamic stability of the shell considering only area
fluctuations is described, see Sec.~\ref{areafluconly} and
Eqs.~\eqref{eq:S2cond} and \eqref{eq:SAAa}.  Plots of
$S_2\left(d,a,x,y\right) = 0$, with $d$ being the number of spacetime
dimensions, $a$ being the exponent of the equation of state of the
shell, $x=\frac{r_+^{d-3}}{R^{d-3}}$, and
$y=\frac{r_-^{d-3}}{r_+^{d-3}}$,
where $r_+$ and $r_-$ are the gravitational radius and the Cauchy
radius of the Reissner-Nordstr\"om shell spacetime, respectively, and
$R$ is the radius of the shell, are shown.
The
equality $S_2\left(d,a,x,y\right) = 0$ describes marginal stability of
the shell considering area fluctuations only. (a) The function is
plotted as $a(x)$ with $d=5$ and for fixed values of $y$; (b) The
function is plotted as $a(y)$ with $d=5$ and for fixed values of $x$;
(c) The function is plotted as $a(d)$ for fixed values of the pair
$(x,y)$.  Points with lower $a$ than the given
function correspond to stable
configurations in all three plots.  All the configurations with $a=1$
are below the surface of marginal stability, therefore they are stable
so there is no need for a plot (d).}}
\label{fig:S2}
\end{figure}

\begin{figure}[h]
\vskip 2.5cm
    \centering
    \begin{subfigure}[h]{0.25\textwidth}
    \centering
    \includegraphics[width=\textwidth]{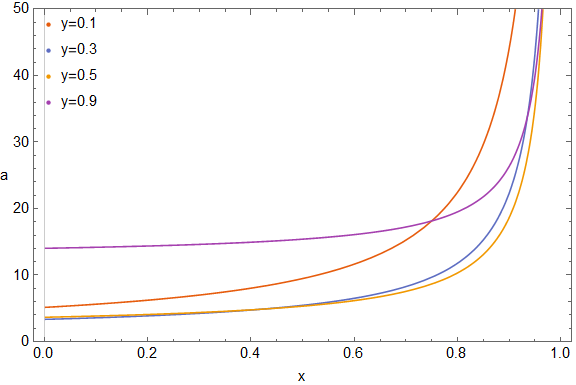}
    \caption{$d=5$, $y$ fixed}
    \label{fig:S3y}
    \end{subfigure}
    \begin{subfigure}[h]{0.25\textwidth}
    \centering
    \includegraphics[width=\textwidth]{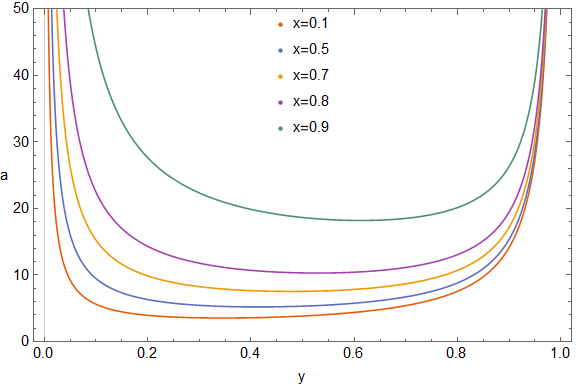}
    \caption{$d=5$, $x$ fixed}
    \label{fig:S3x}
    \end{subfigure}
    \begin{subfigure}[h]{0.25\textwidth}
    \centering
    \includegraphics[width=\textwidth]{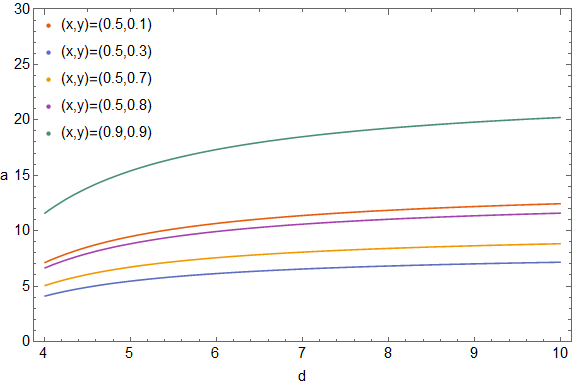}
    \caption{$(x,y)$ fixed}
    \label{fig:S3xy}
    \end{subfigure}
    \caption{\small{
Thermodynamic stability of the shell considering only charge
fluctuations is described, see Sec.~\ref{chargefluconly} and
Eqs.~\eqref{eq:S3cond} and \eqref{eq:SQQa}.  Plots of
$S_3\left(d,a,x,y\right) = 0$, with $d$ being the number of spacetime
dimensions, $a$ being the exponent of the equation of state of the
shell, $x=\frac{r_+^{d-3}}{R^{d-3}}$, and
$y=\frac{r_-^{d-3}}{r_+^{d-3}}$,
where $r_+$ and $r_-$ are the gravitational radius and the Cauchy
radius of the Reissner-Nordstr\"om shell spacetime, respectively, and
$R$ is the radius of the shell, are shown.
The
equality $S_3\left(d,a,x,y\right) = 0$ describes marginal stability of
the shell considering charge fluctuations only. (a) The function is
plotted as $a(x)$ with $d=5$ and for fixed values of $y$; (b) The
function is plotted as $a(y)$ with $d=5$ and for fixed values of $x$;
(c) The function is plotted as $a(d)$ for fixed values of the pair
$(x,y)$.  Points with lower $a$ than the 
given function correspond to stable
configurations in all three plots.  All the configurations with $a=1$
are below the surface of marginal stability, therefore they are stable
so there is no need for a plot (d).}}
    \label{fig:S3}
\end{figure}

\begin{figure}
\vskip 2.0cm
    \centering
    \begin{subfigure}[b]{0.25\textwidth}
    \centering
    \includegraphics[width=\textwidth]{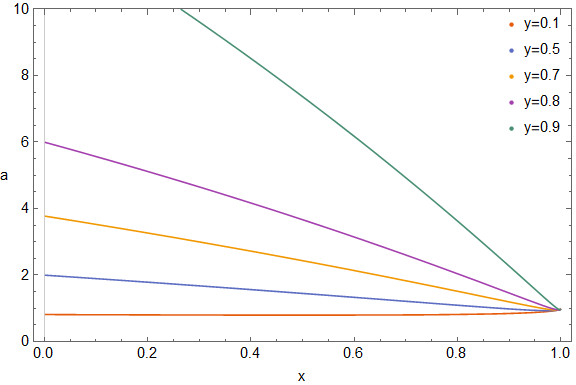}
    \caption{$d=5$, $y$ fixed}
    \label{fig:S4y}
    \end{subfigure}%
    \begin{subfigure}[b]{0.25\textwidth}
    \centering
    \includegraphics[width=\textwidth]{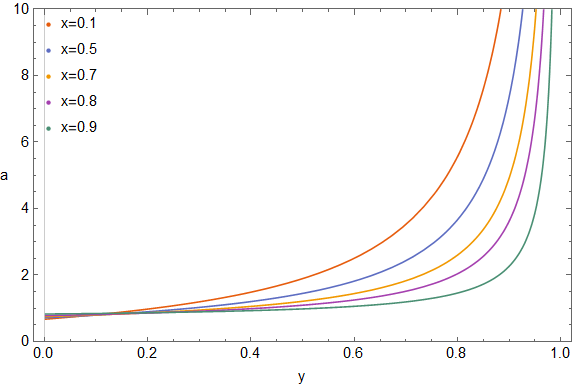}
    \caption{$d=5$, $x$ fixed}
    \label{fig:S4x}
    \end{subfigure}%
    \begin{subfigure}[b]{0.25\textwidth}
    \centering
    \includegraphics[width=\textwidth]{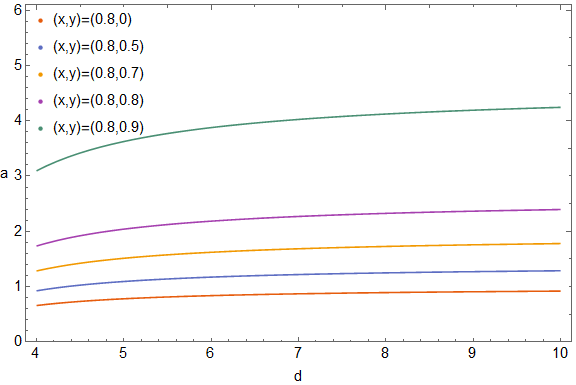}
    \caption{$(x,y)$ fixed}
    \label{fig:S4xy}
    \end{subfigure}%
    \begin{subfigure}[b]{0.235\textwidth}
    \centering
    \includegraphics[width=\textwidth]{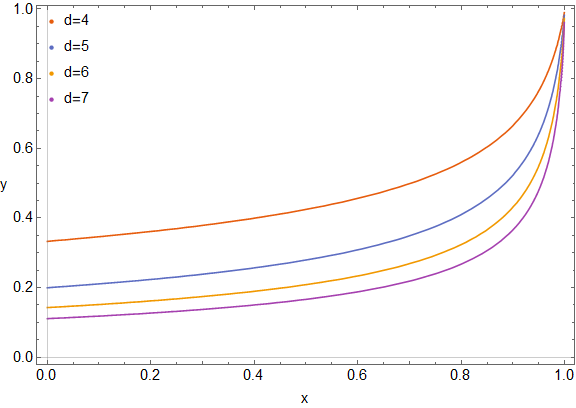}
    \caption{$a=1$}
    \label{fig:S4a1}
    \end{subfigure}
    \caption{\small{
Thermodynamic stability of the shell considering mass and area
fluctuations together is described, see
Sec.~\ref{massandareafluctogether} and Eqs.~\eqref{eq:S4cond} and
\eqref{eq:SMMAAa}.  Plots of $S_4\left(d,a,x,y\right) = 0$, with $d$
being the number of spacetime dimensions, $a$ being the exponent of
the equation of state of the shell, $x=\frac{r_+^{d-3}}{R^{d-3}}$, and
$y=\frac{r_-^{d-3}}{r_+^{d-3}}$,
where $r_+$ and $r_-$ are the gravitational radius and the Cauchy
radius of the Reissner-Nordstr\"om shell spacetime, respectively, and
$R$ is the radius of the shell, are shown.
The
equality $S_4\left(d,a,x,y\right) = 0$ describes marginal stability of
the shell considering together mass and area fluctuations. (a) The
function is plotted as $a(x)$ with $d=5$ and for fixed values of $y$,
with points with lower $a$ than the function corresponding to stable
configurations; (b) The function is plotted as $a(y)$ with $d=5$ and
for fixed values of $x$, with points with lower $a$ corresponding to
stable configurations; (c) The function is plotted as $a(d)$ for fixed
values of the pair $(x,y)$, with points with lower $a$ corresponding
to stable configurations; (d) The function is plotted as $y(x)$ for
$a=1$ with fixed values of $d$, with configurations with higher $x$
than the curve being stable.
}}
    \label{fig:S4}
\end{figure}

\begin{figure}
\vskip 2.2cm
    \centering
    \begin{subfigure}[b]{0.25\textwidth}
    \centering
    \includegraphics[width=\textwidth]{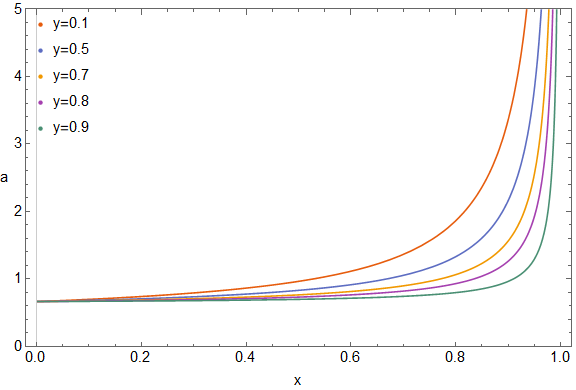}
    \caption{$d=5$, $y$ fixed}
    \label{fig:S5y}
    \end{subfigure}%
    \begin{subfigure}[b]{0.25\textwidth}
    \centering
    \includegraphics[width=\textwidth]{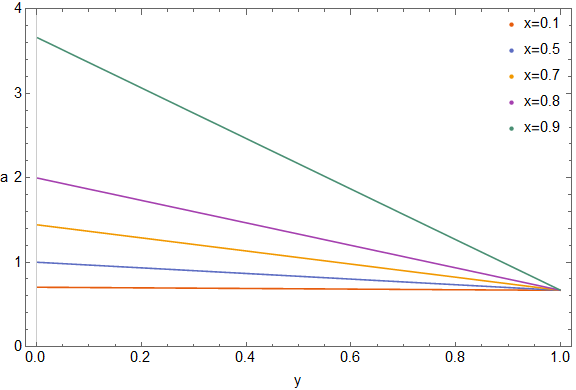}
    \caption{$d=5$, $x$ fixed}
    \label{fig:S5x}
    \end{subfigure}%
    \begin{subfigure}[b]{0.25\textwidth}
    \centering
    \includegraphics[width=\textwidth]{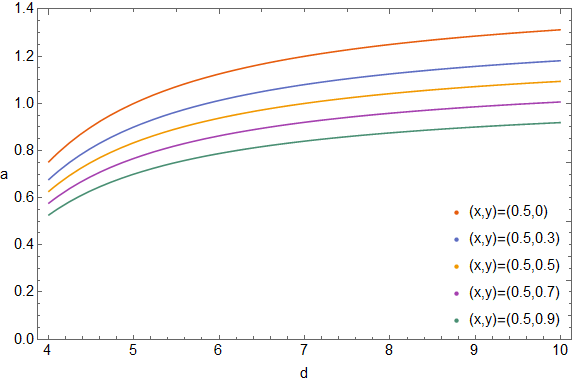}
    \caption{$(x,y)$ fixed}
    \label{fig:S5xy}
    \end{subfigure}%
    \begin{subfigure}[b]{0.235\textwidth}
    \centering
    \includegraphics[width=\textwidth]{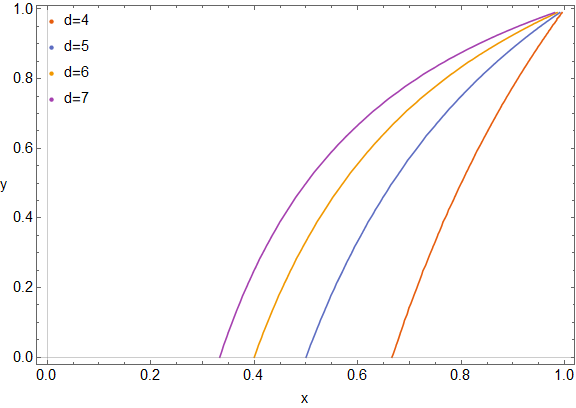}
    \caption{$a=1$}
    \label{fig:S5a1}
    \end{subfigure}
    \caption{\small{
Thermodynamic stability of the shell considering mass and charge
fluctuations together is described, see
Sec.~\ref{massandchargefluctogether} and Eqs.~\eqref{eq:S5cond} and
\eqref{eq:SMMQQa}.  Plots of $S_5\left(d,a,x,y\right) = 0$, with $d$
being the number of spacetime dimensions, $a$ being the exponent of
the equation of state of the shell, $x=\frac{r_+^{d-3}}{R^{d-3}}$, and
$y=\frac{r_-^{d-3}}{r_+^{d-3}}$, where
$r_+$ and $r_-$ are the gravitational radius and the Cauchy radius of
the Reissner-Nordstr\"om shell spacetime, respectively,
and 
$R$ is the radius of the shell, are shown. The
equality $S_5\left(d,a,x,y\right) = 0$ describes marginal stability of
the shell considering together mass and charge fluctuations. (a) The
function is plotted as $a(x)$ with $d=5$ and for fixed values of $y$,
with points with lower $a$ than the function corresponding to stable
configurations; (b) The function is plotted as $a(y)$ with $d=5$ and
for fixed values of $x$, with points with lower $a$ corresponding to
stable configurations; (c) The function is plotted as $a(d)$ for fixed
values of the pair $(x,y)$, with points with lower $a$ corresponding
to stable configurations; (d) The function is plotted as $y(x)$ for
$a=1$ with fixed values of $d$, with configurations with higher $x$
than the curve being stable.}
}
    \label{fig:S5}
\end{figure}

\begin{figure}
\vskip 2.0cm
    \centering
    \begin{subfigure}[b]{0.25\textwidth}
    \centering
    \includegraphics[width=\textwidth]{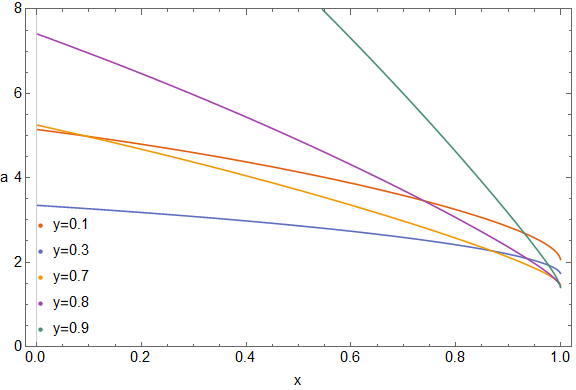}
    \caption{$d=5$, $y$ fixed}
    \label{fig:S6y}
    \end{subfigure}
    \begin{subfigure}[b]{0.25\textwidth}
    \centering
    \includegraphics[width=\textwidth]{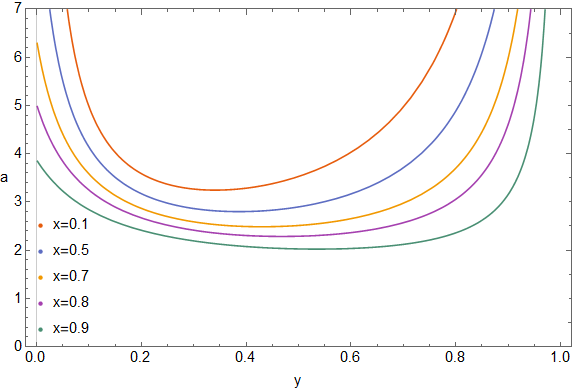}
    \caption{$d=5$, $x$ fixed}
    \label{fig:S6x}
    \end{subfigure}
    \begin{subfigure}[b]{0.25\textwidth}
    \centering
    \includegraphics[width=\textwidth]{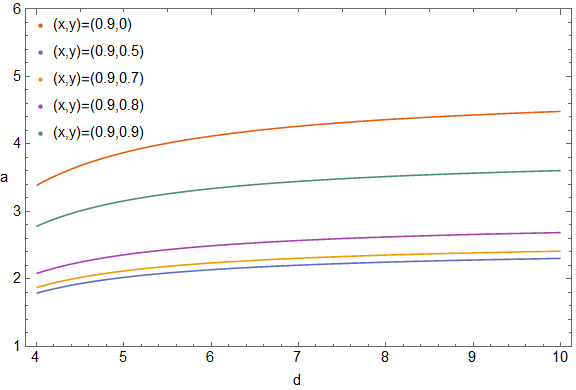}
    \caption{$(x,y)$ fixed}
    \label{fig:S6xy}
    \end{subfigure}
    \caption{\small{
Thermodynamic stability of the shell considering area and charge
fluctuations  together is described, see
Sec.~\ref{areaandchargefluctogether} and
Eqs.~\eqref{eq:S6cond} and \eqref{eq:SAAQQa}.  Plots of
$S_6\left(d,a,x,y\right) = 0$, with $d$ being the number of spacetime
dimensions, $a$ being the exponent of the equation of state of the
shell, $x=\frac{r_+^{d-3}}{R^{d-3}}$, and
$y=\frac{r_-^{d-3}}{r_+^{d-3}}$,
where $r_+$ and $r_-$ are the gravitational radius and the Cauchy
radius of the Reissner-Nordstr\"om shell spacetime, respectively, and
$R$ is the radius of the shell, are shown.
The
equality $S_6\left(d,a,x,y\right) = 0$ describes marginal stability of
the shell considering together area and charge fluctuations. 
(a) The function is
plotted as $a(x)$ with $d=5$ and for fixed values of $y$; (b) The
function is plotted as $a(y)$ with $d=5$ and for fixed values of $x$;
(c) The function is plotted as $a(d)$ for fixed values of the pair
$(x,y)$.  Points with lower $a$ than the 
given function correspond to stable
configurations in all three plots.  All the configurations with $a=1$
are below the surface of marginal stability, therefore they are stable
and there is no need for a plot (d).}}
    \label{fig:S6}
\end{figure}

\begin{figure}
\vskip 2.5cm
    \centering
    \begin{subfigure}[b]{0.25\textwidth}
    \centering
    \includegraphics[width=\textwidth]{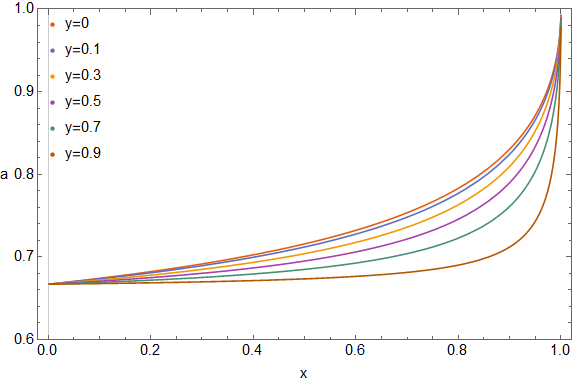}
    \caption{$d=5$, $y$ fixed}
    \label{fig:S7y}
    \end{subfigure}
    \begin{subfigure}[b]{0.25\textwidth}
    \centering
    \includegraphics[width=\textwidth]{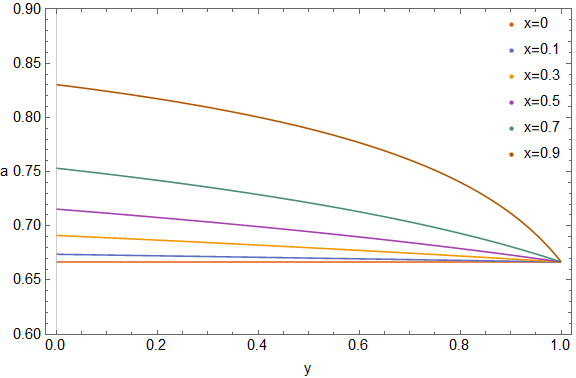}
    \caption{$d=5$, $x$ fixed}
    \label{fig:S7x}
    \end{subfigure}
    \begin{subfigure}[b]{0.25\textwidth}
    \centering
    \includegraphics[width=\textwidth]{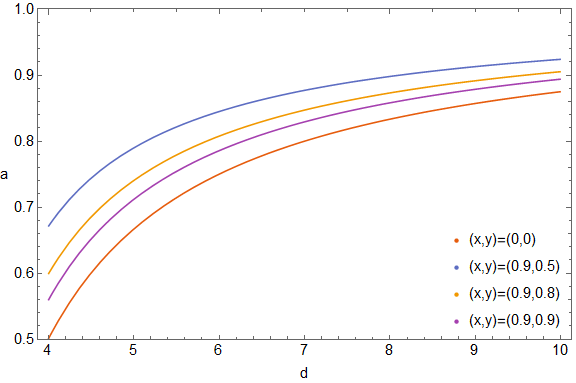}
    \caption{$(x,y)$ fixed}
    \label{fig:S7xy}
    \end{subfigure}
    \caption{\small{
Thermodynamic stability of the shell considering mass, area, and
charge fluctuations altogether is described, see
Sec.~\ref{massareaandchargeflucaltogether} and Eqs.~\eqref{eq:S7cond}
and \eqref{eq:SMMAAQQa}.  Plots of $S_7\left(d,a,x,y\right) = 0$, with
$d$ being the number of spacetime dimensions, $a$ being the exponent
of the equation of state of the shell, $x=\frac{r_+^{d-3}}{R^{d-3}}$,
and $y=\frac{r_-^{d-3}}{r_+^{d-3}}$,
where $r_+$ and $r_-$ are the gravitational radius and the Cauchy
radius of the Reissner-Nordstr\"om shell spacetime, respectively, and
$R$ is the radius of the shell, are shown.
The equality $S_7\left(d,a,x,y\right) = 0$ describes marginal
stability of the shell considering mass, area, and charge fluctuations
altogether.  (a) The function is plotted as $a(x)$ with $d=5$ and for
fixed values of $y$; (b) The function is plotted as $a(y)$ with $d=5$
and for fixed values of $x$; (c) The function is plotted as $a(d)$ for
fixed values of the pair $(x,y)$. Points with lower $a$ than the
given function correspond to stable configurations in all three plots.
All the configurations with $a=1$ are above the surface of marginal
stability, hence they are unstable, except for the points with $x=1$
which lie on the limit of the surface, hence they are marginally
stable, therefore they are stable and there is no need for a plot (d).
}}
\label{fig:S7}
\end{figure}

This appendix is dedicated to the display of the plots of the marginal
stability of the shell with the entropy given by
Eq.~\eqref{eq:suggestedentropy} that were discussed in
Sec.~\ref{sec:stability}. Each figure has plots of the marginal
condition for each seven types of fluctuations, which are all the
possible combinations of mass, area and charge fluctuations. The
correspondence is the following:
Fig.~\ref{fig:S1} corresponds to Sec.~\ref{massfluconly}
and to Sec.~\ref{furthercomments}, 
Fig.~\ref{fig:S2} to Sec.~\ref{areafluconly},
Fig.~\ref{fig:S3} to Sec.~\ref{chargefluconly}, 
Fig.~\ref{fig:S4} to Sec.~\ref{massandareafluctogether}, 
Fig.~\ref{fig:S5} to Sec.~\ref{massandchargefluctogether}, 
Fig.~\ref{fig:S6} to Sec.~\ref{areaandchargefluctogether}, and 
Fig.~\ref{fig:S7} to Sec.~\ref{massareaandchargeflucaltogether}.
The surfaces of marginal stability are 3-surfaces lying in
$\mathbb{R}^4$ with coordinates $(x,y,a,d)$, with
$x=\frac{r_+^{d-3}}{R^{d-3}}$ and
$y=\frac{r_-^{d-3}}{r_+^{d-3}}$,
where $r_+$ and $r_-$ are the
gravitational radius and the
Cauchy radius of the Reissner-Nordstr\"om shell spacetime,
$R$ is the radius of the shell, $a$ is the exponent of the equation of
state of the shell
that appears in Eq.~\eqref{eq:tempeqstate}, and $d$ is the number
of spacetime dimensions. Each point in $\mathbb{R}^4$
corresponds to a configuration of the shell for a fixed value of
$r_+$. The subset of $\mathbb{R}^4$ that corresponds to physical
configurations is described by $x\in \,\,]0,1[\,$, $y \in \,\,]0,1 [
\,$, $a \in \,\,[0,\infty[\,$ and $d \in \,\,[4,\infty[\,$. The
intersection of the 3-surface with this subset separates the subset of
physical configurations that are stable from the ones that are
unstable.

For reasons of presentation, we display cuts of the 3-surface of
marginal stability. In each figure, we display a plot of the 3-surface
as $a(x)$ for different values of $y$ with $d=5$, as $a(y)$ for
different values of $x$ with $d=5$ and as $a(d)$ for different values
of $x$ and $y$. We choose $d=5$ since it is the closest generalization
of the 4-dimensional case and has implications in holography and
unified theories. Also, we only display the curve of marginal
stability with $a=1$ in three cases: the case of mass fluctuations in
Fig.~\ref{fig:S1}, the case of mass and area fluctuations in
Fig.~\ref{fig:S4} and the case of mass and charge fluctuations in
Fig.~\ref{fig:S5}. The plot of the 3-surface with $a=1$ has physical
importance since these configurations correspond to a shell with
black hole features, in particular, a shell with
the
same entropy as a black hole with same mass and charge. However,
displaying this plot for the other cases would have no interest for the
following reasons. For the cases of Figs.~\ref{fig:S2},~\ref{fig:S3}
and~\ref{fig:S6}, there would be simply no curve. All the
configurations with $a=1$ are below the 3-surface and therefore are
stable. For the case of Fig.~\ref{fig:S7}, the plot would be of a
straight line in $x=1$, hence only the configurations with $x=1$ are
marginally stable with the others being unstable.

From the analysis in Sec.~\ref{sec:stability}, we must note that at
least Fig.~\ref{fig:S1}, Fig.~\ref{fig:S4} and Fig.~\ref{fig:S7}
have a physical interpretation. The surface of marginal stability in
Fig.~\ref{fig:S1} describes configurations with an infinite heat
capacity, whereas stable configurations have positive heat capacity
and unstable ones have negative heat capacity. The analogous happens
in Fig.~\ref{fig:S4} and Fig.~\ref{fig:S7}, that describe respectively
infinite isothermal compressibility and isothermal electric
susceptibility.

\onecolumngrid

\twocolumngrid

\newpage
\centerline{}
\newpage
\centerline{}
\newpage
\centerline{}
\newpage
\centerline{}
\newpage
\centerline{}
\newpage
\centerline{}
\newpage
\centerline{}
\newpage
\centerline{}
\newpage


\section{Laboratory variables: Details for Sec.~\ref{sec:Intrinsic}}
\label{app:lab}

\subsection{Mass and charge fluctuations together:
Second derivatives of the entropy in terms of laboratory
quantities}
\label{app:masscharge}

Here we show the detailed calculations of the stability conditions in
terms of laboratory quantities presented in
Sec.~\ref{sec:LvarMassCharge}
for the study
of mass and charge fluctuations together.
The idea is that
the second derivatives of the entropy can be written in terms of
variables that one can measure in the laboratory, i.e.,
thermodynamic
coefficients.
We start from the two equations of state
$dT$ and $d\Phi$,
seen as functions of $S$, $A$, and $Q$. 
They are 
given by
\begin{align}
    &dT = \frac{T}{C_{A,Q}}dS -
    T\frac{\lambda_{T,Q}}{C_{A,Q}} dA
    -\frac{T \lambda_{T,A}}{C_{A,Q}} dQ\,,
    \label{eq:appdTMassCharge}\\
    &d\Phi = - \frac{T \lambda_{T,A}}{C_{A,Q}} dS
    + P_{S,Q}
    dA +
    \frac{1}{\chi_{S,A}}dQ\,,\label{eq:appdphiMassCharge}
\end{align}
where $C_{A,Q}$ is the specific heat capacity at constant area and
electric
charge,
$\lambda_{T,Q}$ is the
latent heat capacity at constant temperature and charge associated to
the area,
$\lambda_{T,A}$ is the latent heat capacity at constant
temperature and area associated to the charge,
$P_{S,Q}$ is the electric pressure at constant
entropy and charge,
and $\chi_{S,A}$ is the
adiabatic electric susceptibility at
constant area.
We then write the differentials $dT(M,A,Q)$ and
$d\Phi(M,A,Q)$ by using Eq.~\eqref{1stlaw1}, i.e.,
$TdS=dM+pdA-\Phi dQ$, in
Eqs.~\eqref{eq:appdTMassCharge} and~\eqref{eq:appdphiMassCharge},
yielding
\begin{align}
    dT = &\frac{1}{C_{A,Q}}dM - \frac{T \lambda_{T,A}+\Phi}{C_{A,Q}}
    dQ - \frac{T\lambda_{T,Q} - p}{C_{A,Q}}
    dA\,,\label{eq:appdTMassCharge2}\\
    d\Phi = &- \frac{\lambda_{T,A}}{C_{A,Q}} dM +
    \left(\frac{1}{\chi_{S,A}} + \frac{\Phi
    \lambda_{T,A}}{C_{A,Q}}\right)dQ \nonumber\\&+
    \left(\frac{\kappa_{p,S}}{\kappa_{S,Q}} -
    p\frac{\lambda_{T,A}}{C_{A,Q}}\right)dA\,.
    \label{eq:appdphiMassCharge2}
\end{align}
Now,
Eq.~\eqref{1stlaw1}, i.e.,
$TdS=dM+pdA-\Phi dQ$, and Eq.~\eqref{eq:appdTMassCharge2},
yield $ S_{MM} = -\beta^2
\left(\frac{\partial T}{\partial M}\right)_{A,Q} =
- \beta^2 C_{A,Q}^{-1}$.
Thus, the first condition for stability, $S_{MM}\leq 0$,
can be written as $C_{A,Q}^{-1} \geq 0$, or equivalently,
\begin{align}
  C_{A,Q} \geq 0\,.
  \label{1stcs}
\end{align}
In addition, one can write 
$S_{MM}S_{QQ} - S_{MQ}^2$
as
$S_{MM}S_{QQ} - S_{MQ}^2=-\left(\frac{\partial
\beta}{\partial M}\right)_{A,Q}
\left(\frac{\partial \beta \Phi}{\partial Q}\right)_{M,A} -
\left(\frac{\partial \beta}{\partial Q}\right)_{M,A}^2 = -\beta^4
\left(\frac{\partial T}{\partial M}\right)_{A,Q}\left(\frac{\partial
T}{\partial Q}\right)_{M,A}\Phi -
\beta^3 \left(\frac{\partial
T}{\partial M}\right)_{A,Q} \left(\frac{\partial \Phi}{\partial
Q}\right)_{M,A} - \beta^4 \left(\frac{\partial T}{\partial
Q}\right)_{M,A}^2$, where 
Eq.~\eqref{1stlaw1}, i.e.,
$TdS=dM+pdA-\Phi dQ$, has been of help.
Then, putting
Eqs.~\eqref{eq:appdTMassCharge2} and~\eqref{eq:appdphiMassCharge2}
into
this latter equation,
one finds 
$S_{MM}S_{QQ} - S_{MQ}^2=
\beta^3\left(\frac{1}{C_{A,Q} \chi_{S,A}} -
\frac{T \lambda_{T,A}^2}{C_{A,Q}^2}\right)$.
Thus,
the second and necessary condition
for stability, namely,
$S_{MM}S_{QQ} - S_{MQ}^2\geq 0$, can be written as
\begin{align}
    \beta^3\left(\frac{1}{C_{A,Q} \chi_{S,A}} -
    \frac{T \lambda_{T,A}^2}{C_{A,Q}^2}\right)
    \geq 0\,.
    \label{eq:appCondMassCharge}
\end{align}
This can be
further
simplified.
If we introduce the heat capacity at constant electric potential and
constant area, $C_{A,\Phi} = T \left(\frac{\partial S}{\partial
T}\right)_{\Phi, A}$. This coefficient can be written as
$C_{A,\Phi}^{-1} = \beta \left(\frac{\partial T}{\partial
    S}\right)_{A,\Phi}
    = \beta \left(\left(\frac{\partial T}{\partial S}\right)_{Q,A} +
    \left(\frac{\partial T}{\partial
    Q}\right)_{S,A}\left(\frac{\partial Q}{\partial
    S}\right)_{\Phi,A}\right)
    = \beta C_{A,Q}^{-1} \left(1 - \frac{T
    \lambda_{T,A}^2}{C_{A,Q}}\chi_{S,A}\right)
    =\beta \chi_{S,A}\left(\frac{1}{C_{A,Q}\chi_{S,A}}
    - \frac{T\lambda_{T,A}^2}{C_{A,Q}^2}\right)$,
where we have used Eqs.~\eqref{eq:appdTMassCharge}
and~\eqref{eq:appdphiMassCharge} to compute the derivatives. Also, we
used that $\left(\frac{\partial Q}{\partial S}\right)_{\Phi,A} =
T\chi_{S,A}\frac{\lambda_{T,A}}{C_{A,Q}}$, that comes from inverting
Eq.~\eqref{eq:appdphiMassCharge} to obtain $dQ(S,A,\Phi)$. 
Thus,
$\beta^3\left(\frac{1}{C_{A,Q} \chi_{S,A}} -
    \frac{T \lambda_{T,A}^2}{C_{A,Q}^2}\right)
=\beta^2\frac{1}{C_{A,\Phi}\chi_{S,A}}
$, and Eq.~\eqref{eq:appCondMassCharge}
can be rewritten as $\frac{1}{C_{A,\Phi}\chi_{S,A}}\geq 0$,
or, equivalently, as
\begin{align}
    C_{A,\Phi}\chi_{S,A}\geq 0\,,
    \label{2ndcs}
\end{align}
which is the upshot of the second condition for stability
$S_{MM}S_{QQ} - S_{MQ}^2\geq 0$ in the physical variables.

Now, we consider the particular case of the thin shell with the
equations of state given by Eqs.~\eqref{eq:tempeqstate}
and~\eqref{eq:electeqstate}.
The first condition for stability, Eq.~\eqref{1stcs},
holds without further ado. 
The second condition for stability, Eq.~\eqref{2ndcs}
can be improved for the equations of state used.
The coefficient $\chi_{S,A}$ can be
calculated using the differential in
Eqs.~\eqref{eq:appdphiMassCharge2} through
$\chi_{S,A}^{-1} = \left(\frac{\partial
    \Phi}{\partial Q}\right)_{M,A} +
    \Phi \left(\frac{\partial \Phi}{\partial M}\right)_{A,Q}$.
The equation of state in consideration is
given by Eqs.~\eqref{eq:Phi} and~\eqref{eq:electeqstate},   
$\Phi = Q\left[\frac{r_+^{-(d-3)} -
    R^{-(d-3)}}{1-\frac{\mu M}{R^{d-3}}}\right]
$,
with $r_+ = r_+(M,A,Q)$ defined by Eqs.~\eqref{eq:rpm}
and~\eqref{eq:m}. Thus, the coefficient $\chi_{S,A}^{-1}$
for this case is
$
\chi_{S,A}^{-1} = \Phi^2 \frac{\mu}{R^{d-3}\left(1-\mu
    \frac{M}{R^{d-3}}\right)} + \frac{\Phi}{Q}
$,
which is positive for values of $(M,A,Q)$ that are physical. This
means that the second condition for stability
given in Eq.~\eqref{2ndcs} is
reduced to
\begin{align}
    C_{A,\Phi}\geq 0\,,
    \label{specific3}
\end{align}
for the equations of state we have used.

Here we have deduced in detail the stability conditions in physical
variables for mass and charged fluctuations presented in the main text
in Sec.~\ref{sec:LvarMassCharge}.  So,
Eqs.~\eqref{eq:appdTMassCharge2} and \eqref{eq:appdphiMassCharge2} are
Eqs.~\eqref{eq:dTMassCharge} and \eqref{eq:dphiMassCharge} of the main
text, respectively, Eqs.~\eqref{1stcs} and \eqref{2ndcs} are the two
equations given in Eq.~\eqref{eq:LabMQ} of the main text, and
Eq.~\eqref{specific3} is Eq.~\eqref{eq:SMMQQa2} of the main text.

\subsection {Mass, area, and charge fluctuations
altogether: Second derivatives
of the entropy in terms of laboratory quantities}
\label{app:ddentropyvariables}

Here we show the detailed calculations of the stability conditions in
terms of laboratory quantities presented in
Sec.~\ref{mac}
for the study
of mass, area, and charge fluctuations altogether.
The idea is that
the second derivatives of the entropy can be written in terms of
variables one can measure in the laboratory, i.e.,
thermodynamic
coefficients.
We start from the three equations of state $dS(T,p,Q)$,
$dA(T,p,Q)$ and $d\Phi(T,p,Q)$,
seen as functions of $T$, $p$, and $Q$. 
They are 
given by
\begin{align}
     dS = &\left( \frac{C_{A,Q}}{T} +
     A\frac{\alpha_{p,Q}^2}{\kappa_{T,Q}}\right)dT
     \nonumber \\&- A \alpha_{p,Q} dp + \lambda_{p,T}
     dQ\,,\label{eq:appS}\\
     \frac{dA}{A} =& \alpha_{p,Q} dT - \kappa_{T,Q} dp - \kappa_{p,T}
     dQ\,,\label{eq:appA}\\
     d\Phi =& - \lambda_{p,Q} dT -A \kappa_{p,T} dp+ \frac{1}{
    \chi_{T,p}} dQ\,,\label{eq:appphi}
\end{align}
where
 $C_{A,Q}$ is the heat capacity at
constant area and charge,
$\alpha_{p,Q}$ is the thermal expansion coefficient,
$\kappa_{T,Q}$ is the isothermal compressibility,
 $\lambda_{p,T}$ is a latent heat capacity,
 $\kappa_{p,T}$ is
the electric compressibility, $\lambda_{p,Q}$ is another
latent heat capacity,
and $\chi_{T,p}$ is the
isothermal electric susceptibility.
We can invert Eqs.~\eqref{eq:appS} and~\eqref{eq:appA} to obtain
$dT(S,A,Q)$ and $dp(S,A,Q)$. This is accomplished by rewriting
Eqs.~\eqref{eq:appS} and~\eqref{eq:appA} in matrix form, i.e.,
\begin{align}
    \begin{pmatrix}
    dS - \lambda_{p,T} dQ \\
    dA + A \kappa_{p,T} dQ
    \end{pmatrix}=
    \mathcal{M} \begin{pmatrix}
    dT \\
    dp
    \end{pmatrix}
    \,,\label{eq:appIntermediodpdT}
\end{align}
where
\begin{align}
    \mathcal{M} = \begin{pmatrix}
    \left( \frac{C_{A,Q}}{T} +
    A\frac{\alpha_{p,Q}^2}{\kappa_{T,Q}}\right) & -
    A \alpha_{p,T} \\
    A \alpha_{p,Q} & - A \kappa_{T,Q}
    \end{pmatrix}\,.
\end{align}
The matrix $\mathcal{M}$ can be inverted to yield
\begin{align}
    \mathcal{M}^{-1} = \begin{pmatrix}
    \frac{T}{C_{A,Q}}  &  -
    T\frac{\alpha_{p,Q}}{C_{A,Q}\kappa_{T,Q}} \\
    T\frac{\alpha_{p,Q}}{C_{A,Q}\kappa_{T,Q}} & -
    \left(\frac{1}{A \kappa_{T,Q}} +
    T \frac{\alpha^2_{p,Q}}{\kappa_{T,Q}^2 C_{A,Q}}\right) 
    \end{pmatrix}\,.
\end{align}
Applying $\mathcal{M}^{-1}$ on both sides of
Eq.~\eqref{eq:appIntermediodpdT}, we obtain $dT(S,A,Q)$ and
$dp(S,A,Q)$ as
\begin{align}
    dT = & \frac{T}{C_{A,Q}}dS -
    T\frac{\alpha_{p,Q}}{C_{A,Q}\kappa_{T,Q}}dA - T \mathcal{B}
    dQ
    \label{T1}\,,\\
    dp =& \frac{T \alpha_{p,Q}}{C_{A,Q}\kappa_{T,Q}}dS -
    \left(\frac{1}{A \kappa_{T,Q}} +
    T\frac{\alpha_{p,Q}^2}{\kappa_{T,Q}^2 C_{A,Q}} \right)
    dA \nonumber\\
    & -\left( \frac{\kappa_{p,T}}{\kappa_{T,Q}} +
    T\frac{\alpha_{p,Q}}{\kappa_{T,Q}}\mathcal{B}\right)dQ
    \label{p1}\,,
\end{align}
where $\mathcal{B} = \frac{\lambda_{p,T}}{C_{A,Q}} +
A\frac{\alpha_{p,Q}\kappa_{p,T}}{C_{A,Q}\kappa_{T,Q}}$. We can then
compute directly $dT(M,A,Q)$ and $dp(M,A,Q)$ by substituting the
differential $dS$ given in Eq.~\eqref{1stlaw1},
i.e., $TdS=dM+pdA-\Phi dQ$, into Eqs.~\eqref{T1} and \eqref{p1},
yielding
\begin{align}
    dT =& \frac{dM}{C_{A,Q}} + \left(\frac{p}{C_{A,Q}} -
    T\frac{\alpha_{p,Q}}{C_{A,Q}\kappa_{T,Q}}\right)dA \nonumber\\&-
    \left(\frac{\Phi}{C_{A,Q}} + T\frac{\lambda_{p,T}}{C_{A,Q}} + A
    \frac{\alpha_{p,T} \kappa_{p,T}}{\kappa_{T,Q} C_{A,Q}}
    \right)dQ\,,\label{eq:dTMAQAPP}\\
    dp =& \frac{\alpha_{p,Q}}{C_{A,Q}\kappa_{T,Q}} dM\nonumber\\& -
    \left[ \frac{1}{A \kappa_{T,Q}} - \frac{\alpha_{p,Q}}{C_{A,Q}
    \kappa_{T,Q}}\left( p - T \frac{\alpha_{p,Q}}{\kappa_{T,Q}}\right)
    \right] dA \nonumber\\& - \left(\frac{\kappa_{p,T}}{\kappa_{T,Q}}
    - \frac{\alpha_{p,Q}}{C_{A,Q} \kappa_{T,Q}}\mathcal{C}\right)
    dQ\,,\label{eq:dpMAQAPP}
\end{align}
where 
 $\mathcal{C} =
T A
\frac{\kappa_{p,T}\alpha_{p,Q}}{\kappa_{T,Q}} + T\lambda_{p,T} +
\Phi$.
Then, substituting $dT$
and $dp$ given in Eqs.~\eqref{T1} and \eqref{p1}, respectively,
into
$d\Phi(T,p,Q)$, written in Eq.~\eqref{eq:appphi}, we obtain
\begin{align}
    d\Phi = & - \mathcal{B} dM +
    \left[\frac{\kappa_{p,T}}{\kappa_{T,Q}} - \left(p -
    T\frac{\alpha_{p,Q}}{\kappa_{T,Q}}\right)\mathcal{B}
    \right]dA\nonumber\\& + \left(\mathcal{B}\mathcal{C} +
    \frac{1}{\chi_{p,T}} + A\frac{\kappa_{p,T}^2}{\kappa_{T,Q}}
    \right)dQ\,.\label{eq:dPhiMAQAPP}
\end{align}
With $dT(M,A,Q)$, $dp(M,A,Q)$ and $d\Phi(M,A,Q)$, we can write the
second derivatives of the entropy since these are derivatives of
$\beta(M,A,Q)$, $\beta p(M,A,Q)$ and $\beta \Phi(M,A,Q)$. The second
derivatives are explicitly
$S_{MM} = - \frac{\beta^2}{C_{A,Q}}$,
$S_{AA} = - \left[ \frac{1}{A \kappa_{T,Q}} +
\frac{\beta^2}{C_{A,Q}}\left( p -
T \frac{\alpha_{p,Q}}{\kappa_{T,Q}}\right)^2\right]$,
$S_{QQ} = -\beta \left[ \frac{\mathcal{C}^2\beta}{C_{A,Q}}
+ \frac{1}{\chi_{p,T}} +
A \frac{\kappa_{p,T}^2}{\kappa_{T,Q}}\right]$,
$S_{MA} = \frac{\beta^2}{C_{A,Q}}
\left(T\frac{\alpha_{p,Q}}{\kappa_{T,Q}} - p \right)$,
$S_{MQ} = \frac{\beta^2}{C_{A,Q}}\mathcal{C}$,
$S_{AQ} = - \beta\left[ \frac{\kappa_{p,T}}{\kappa_{T,Q}}
- \frac{\beta^2}{C_{A,Q}}\left(p
- T\frac{\alpha_{p,Q}}{\kappa_{T,Q}}
\right)\mathcal{C}\right]$.
Thus, the relevant combinations of the second
derivatives for the stability conditions are
$S_{MM}= - \frac{\beta^2}{C_{A,Q}}$,
$S_{MM}S_{AA} - S_{MA}^2 = \frac{\beta^3}{A C_{A,Q}
    \kappa_{T,Q}}$,
and
$(S_{MM} S_{AQ} - S_{MA}S_{MQ})^2
- (S_{MM}S_{AA} - S_{MA}^2)(S_{MM}S_{QQ} -
    S_{MQ}^2)= -
    \frac{\beta^6}{A C_{A,Q}^2 \kappa_{T,Q}
    \chi_{p,T}}$.
For stability  one has 
$S_{MM}\leq0$,
$S_{MM}S_{AA} - S_{MA}^2\geq0$, and
$(S_{MM} S_{AQ} - S_{MA}S_{MQ})^2
- (S_{MM}S_{AA} - S_{MA}^2)(S_{MM}S_{QQ} -
    S_{MQ}^2)\leq 0$, which translates into
\begin{align}
    &C_{A,Q}\geq 0,\,\label{eq:stabcondbetterAPP1}\\
    &\kappa_{T,Q}\geq 0\,,\label{eq:stabcondbetterAPP2}\\
    & \chi_{p,T}\geq 0\,.\label{eq:stabcondbetterAPP3}
\end{align}
These are the stability conditions for mass, area, and charge
fluctuations. In words, the heat capacity at constant area
and charge, the isothermal compressibility associated to the pressure,
and the
isothermal electric susceptibility must be positive.

The particular case of the thin shell with the equations of state
given by Eqs.~\eqref{eq:tempeqstate}, \eqref{eq:electeqstate},
and~\eqref{eq:p} can be also worked out in detail, we will not do it
here.

Here, we have deduced in detail the stability conditions in physical
variables for mass, area, and charged fluctuations presented in the
main text in Sec.~\ref{mac}.  So, Eqs.~\eqref{eq:dTMAQAPP}
and~\eqref{eq:dpMAQAPP} are Eqs.~\eqref{eq:dTMAQ} and~\eqref{eq:dpMAQ}
of the main text, respectively, Eq.~\eqref{eq:dPhiMAQAPP} is
Eq.~\eqref{eq:dPhiMAQ} of the main text, and
Eqs.~\eqref{eq:stabcondbetterAPP1}, \eqref{eq:stabcondbetterAPP2},
\eqref{eq:stabcondbetterAPP3} are the three equations given in
Eq.~\eqref{eq:stabcondbetter} of the main text.

\end{document}